\documentclass{article}
\usepackage[a4paper, total={17cm, 25.7cm}]{geometry}
\usepackage[utf8]{inputenc} 
\usepackage[T1]{fontenc} 
\usepackage{hyperref} 
\usepackage{url} 
\usepackage[numbers,square, sort&compress]{natbib}
\usepackage{booktabs} 
\usepackage{amsfonts} 
\usepackage{nicefrac} 
\usepackage{microtype} 
\usepackage{lipsum}		
\usepackage{graphicx}
\usepackage{doi}
\usepackage{dsfont}
\usepackage{rotating}
\usepackage[section]{placeins}

\usepackage{soul, color}
\usepackage{todonotes}
\usepackage{amsmath}
\usepackage{cleveref}
\usepackage{listings}
\usepackage{xcolor}
\usepackage{amsthm} 
\usepackage{algpseudocode}
\usepackage{algorithm}
\usepackage{longtable}
\usepackage{setspace}
\usepackage{bm}
\usepackage{float}

\usepackage{hyperref}
\usepackage{amssymb}
\usepackage{array}
\newcommand{\nc}{\newcommand}
\nc{\ntitle}[1]{
 \begin{center}
   \fbox{\textbf{\Large #1}}
  \end{center}         }

\nc{\tred}[1]{\textcolor[rgb]{0.00,0.00,0.00}{#1}}
\nc{\tb}[1]{\textcolor[rgb]{0.00,0.00,0.00}{#1}}

\nc{\slideline}{\smallskip \hrule\hrule \smallskip}
\nc{\stitle}[1]{
\textbf{\large #1}
\slideline
\slideline
          }

\nc{\nn}{\nonumber}
\nc{\fns}{\footnotesize}

\nc{\revisionline}{\vspace{.1in} \today \vspace{.1in} \hrule\hrule\hrule\vspace{.1in}}
\nc{\newpp}{\vspace{.1in} \noindent}

\nc{\wh}{\widehat}

\nc{\Ef}{ {\rm E}_{\infty} }
\nc{\Ex}{ {\rm E} }
\nc{\Ec}{ {\rm E}_1 }
\nc{\Pf}{ {\rm P}_{\infty} }
\nc{\Pc}{ {\rm P}_{1} }
\nc{\Prb}{ {\rm P} }
\nc{\sd}{\pm \hat{\sigma} }

\nc{\indep}{{\, \perp \! \! \! \perp  \,} }
\nc{\tsps}{^{ {\rm T} } }

\nc{\pu}{\pi_{\rm U}}
\nc{\pbi}{\pi_{\rm B}}
\nc{\pnb}{\pi_{\rm NB}}
\nc{\prp}{\propto}
\nc{\pr}{ {\rm pr} }

\nc{\al}{\alpha}
\nc{\dl}{\delta}
\nc{\la}{\lambda}
\nc{\om}{\omega}
\nc{\vep}{\varepsilon}

\nc{\snf}{\sum_{n=1}^{\infty}}
\nc{\skf}{\sum_{k=1}^{\infty}}
\nc{\sner}{\sum_{n=1}^{86}}
\nc{\sjn}{\sum_{j=1}^{n}}
\nc{\skn}{\sum_{k=1}^{n}}
\nc{\sumim}{\sum_{i=1}^m}
\nc{\sumjn}{\sum_{j=1}^n}
\nc{\sumlL}{\sum_{l=1}^{L}}
\nc{\sumL}{\sum_{l=1}^{L}}
\nc{\sumkK}{\sum_{k=1}^{K_i}}
\nc{\sumrR}{\sum_{r=1}^R}

\nc{\hivp}{\sum_{ {\rm HIV}^+ } }
\nc{\sumiN}{ \sum_{i=1}^N }
\nc{\summM}{ \sum_{m=1}^M }
\nc{\sumjM}{ \sum_{j=1}^M }

\nc{\lsq}{\left[}
\nc{\rsq}{\right]}
\nc{\lbc}{\left \{ }
\nc{\rbc}{\right \} }
\nc{\lp}{\left(}
\nc{\rp}{\right)}

\nc{\imp}{\Rightarrow}
\nc{\lbf}{\lim_{b \rightarrow \infty}}
\nc{\limNinf}{\lim_{N \rightarrow \infty}}
\nc{\limminf}{\lim_{m \rightarrow \infty}}
\nc{\limninf}{\lim_{n \rightarrow \infty}}
\nc{\convd}{\stackrel{D}{\longrightarrow}}
\nc{\convp}{\stackrel{P}{\longrightarrow}}
\nc{\eqd}{\stackrel{{\EuScript D}}{=}}

\nc{\trans}{^{\text T}}
\nc{\ol}{\overline}
\nc{\logit}{\text{logit}\,}

\nc{\rl}{ {\rm {\bf R} } }
\nc{\zah}{ {\rm {\bf Z} } }

\nc{\lkn}{\Lambda^n_k}
\nc{\stp}{ {\cal C}_b }
\nc{\istp}{ {\cal I}_A }
\nc{\snb}{S_{N_b}}
\nc{\stb}{S_{T_b}}
\nc{\ixlog}{I_{ \{ 0 \leq x \leq \log \al \} } }
\nc{\iulog}{I_{ \{ 0 \leq u  \leq \log \al \} } }
\nc{\rgn}{ \Upsilon_n }
\nc{\var}{{\rm var}}
\nc{\cov}{{\rm cov}}
\nc{\corr}{{\rm corr}}
\nc{\dpl}{\partial}
\nc{\half}{ {\textstyle \frac{1}{2}} }
\nc{\tr}{{\rm trace}}
\nc{\real}{\mathbb{R}}
\nc{\bbC}{\mathbb{C}}

\def\boxit#1{\vbox{\hrule\hbox{\vrule\kern6pt\vbox{\kern6pt#1\kern6pt}\kern6pt\vrule}\hrule}}

\nc{\calb}{ {\cal B} }
\nc{\calc}{ {\cal C} }
\nc{\bcalc}{ \mbox{\boldmath{${\cal C}$}}}
\nc{\cald}{ {\cal D} }
\nc{\cale}{ {\cal E} }
\nc{\cali}{ {\cal I} }
\nc{\call}{ {\cal L} }
\nc{\calm}{ {\cal M} }
\nc{\caln}{ {\cal N} }
\nc{\cals}{ {\cal S} }
\nc{\calo}{ {\cal O} }
\nc{\bcalo}{ \mbox{\boldmath{${\cal O}$}}}
\nc{\calt}{ {\cal T} }
\nc{\calv}{ {\cal V} }
\nc{\bcalu}{ \mbox{\boldmath{${\cal U}$}}}
\nc{\calu}{ {\cal U} }
\nc{\calw}{ {\cal W} }
\nc{\calx}{ {\cal X} }

\nc{\sca}{ {\EuScript A} }
\nc{\scb}{ {\EuScript B} }
\nc{\scc}{ {\EuScript C} }
\nc{\scd}{ {\EuScript D} }
\nc{\sce}{ {\EuScript E} }
\nc{\scf}{ {\EuScript F} }
\nc{\scF}{ {\EuScript f} }
\nc{\scg}{ {\EuScript G} }
\nc{\sch}{ {\EuScript H} }
\nc{\sci}{ {\EuScript I} }
\nc{\scj}{ {\EuScript J} }
\nc{\sck}{ {\EuScript K} }
\nc{\scl}{ {\EuScript L} }
\nc{\sclic}{ \scl_i^{\rm c} }
\nc{\scm}{ {\EuScript M} }
\nc{\scn}{ {\EuScript N} }
\nc{\sco}{ {\EuScript O} }
\nc{\scp}{ {\EuScript P} }
\nc{\scq}{ {\EuScript Q} }
\nc{\scr}{ {\EuScript R} }
\nc{\scs}{ {\EuScript S} }
\nc{\sct}{ {\EuScript T} }
\nc{\scu}{ {\EuScript U} }
\nc{\scv}{ {\EuScript V} }
\nc{\scw}{ {\EuScript W} }
\nc{\scx}{ {\EuScript X} }
\nc{\scy}{ {\EuScript Y} }
\nc{\scz}{ {\EuScript Z} }
\nc{\scxo}{ {\EuScript X}_{\rm obs} }
\nc{\Xobs}{ \pmb{\scx}_{\rm obs} }
\nc{\Xcom}{ \pmb{\scx} }
\nc{\Xmis}{ \pmb{\scx}_{\rm mis} }

\nc{\bsci}{ \mbox{\boldmath{$\sci$}}}
\nc{\bscj}{ \mbox{\boldmath{$\scj$}}}

\nc{\sumlic}{\sum_{l \in sclic}}

\nc{\scyo}{ {\EuScript Y}_{\rm obs} }

\nc{\bga}{\begin{array}{c}}
\nc{\ena}{\end{array}}

\nc{\mhat}{ {\hat{p}}_M }
\nc{\fhat}{ {\hat{p}}_F }
\nc{\ph} { \hat{p} }

\nc{\ta}{ {\tilde{a}} }
\nc{\tc}{ {\tilde{c}} }

\nc{\bal}{\mbox{\boldmath{$\alpha$}}}
\nc{\balpha}{\mbox{\boldmath{$\alpha$}}}
\nc{\bone}{\mbox{\boldmath{$1$}}}
\nc{\bbet}{\mbox{\boldmath{$\beta$}}}
\nc{\bbeta}{\mbox{\boldmath{$\beta$}}}
\nc{\bDel}{\mbox{\boldmath{$\Delta$}}}
\nc{\bDelta}{\mbox{\boldmath{$\Delta$}}}
\nc{\bdel}{\mbox{\boldmath{$\delta$}}}
\nc{\bdelta}{\mbox{\boldmath{$\delta$}}}
\nc{\bet}{\mbox{\boldmath{$\eta$}}}
\nc{\beps}{\mbox{\boldmath{$\epsilon$}}}
\nc{\bvep}{\mbox{\boldmath{$\vep$}}}
\nc{\bgam}{\mbox{\boldmath{$\gamma$}}}
\nc{\bgamma}{\mbox{\boldmath{$\gamma$}}}
\nc{\bGamma}{\mbox{\boldmath{$\Gamma$}}}
\nc{\bLam}{\mbox{\boldmath{$\Lambda$}}}
\nc{\bLambda}{\mbox{\boldmath{$\Lambda$}}}
\nc{\blambda}{\mbox{\boldmath{$\lambda$}}}
\nc{\bmu}{ \mbox{\boldmath{$\mu$}}}
\nc{\bOm}{ \mbox{\boldmath{$\Omega$}}}
\nc{\bOmega}{ \mbox{\boldmath{$\Omega$}}}
\nc{\bom}{ \mbox{\boldmath{$\omega$}}}
\nc{\bomega}{ \mbox{\boldmath{$\omega$}}}
\nc{\bpi}{ \mbox{\boldmath{$\pi$}}}
\nc{\bPi}{ \mbox{\boldmath{$\Pi$}}}
\nc{\bpsi}{ \mbox{\boldmath{$\psi$}}}
\nc{\bPsi}{ \mbox{\boldmath{$\Psi$}}}
\nc{\bphi}{ \mbox{\boldmath{$\phi$}}}
\nc{\bPhi}{ \mbox{\boldmath{$\Phi$}}}
\nc{\bxi}{ \mbox{\boldmath{$\xi$}}}
\nc{\bXi}{ \mbox{\boldmath{$\Xi$}}}
\nc{\bSig}{\mbox{\boldmath{$\Sigma$}}}
\nc{\bSigma}{\mbox{\boldmath{$\Sigma$}}}
\nc{\bsig}{\mbox{\boldmath{$\sigma$}}}
\nc{\btau}{\mbox{\boldmath{$\tau$}}}
\nc{\bThe}{\mbox{\boldmath{$\Theta$}}}
\nc{\bTheta}{\mbox{\boldmath{$\Theta$}}}
\nc{\bthe}{\mbox{\boldmath{$\theta$}}}
\nc{\btheta}{\mbox{\boldmath{$\theta$}}}
\nc{\bzeta}{\mbox{\boldmath{$\zeta$}}}
\nc{\bIm}{\mbox{\boldmath{$\Im$}}}

\nc{\ba}{ { \bf a }}
\nc{\bA}{ { \bf A }}
\nc{\bB}{ { \bf B }}
\nc{\bb}{ { \bf b }}
\nc{\bc}{ { \bf c }}
\nc{\bC}{ { \bf C }}
\nc{\bD}{ { \bf D }}
\nc{\bd}{ { \bf d }}
\nc{\be}{ { \bf e }}
\nc{\bF}{ { \bf F }}
\nc{\bG}{ { \bf G }}
\nc{\bh}{ { \bf h }}
\nc{\bH}{ { \bf H }}
\nc{\bI}{ { \bf I }}
\nc{\bJ}{ { \bf J }}
\nc{\bK}{ { \bf K }}
\nc{\bL}{ { \bf L }}
\nc{\bM}{ { \bf M }}
\nc{\bn}{ { \bf n }}
\nc{\bO}{ { \bf O }}
\nc{\bP}{ { \bf P }}
\nc{\br}{ { \bf r }}
\nc{\bR}{ { \bf R }}
\nc{\bs}{ { \bf s }}
\nc{\bS}{ { \bf S }}
\nc{\bT}{ { \bf T }}
\nc{\bt}{ { \bf t }}
\nc{\bu}{ { \bf u }}
\nc{\bU}{ { \bf U }}
\nc{\bv}{ { \bf v }}
\nc{\bV}{ { \bf V }}
\nc{\bW}{ { \bf W }}
\nc{\bw}{ { \bf w }}
\nc{\bx}{ { \bf x }}
\nc{\bX}{ { \bf X }}
\nc{\by}{ { \bf y }}
\nc{\bY}{ { \bf Y }}
\nc{\bz}{ { \bf z }}
\nc{\bZ}{ { \bf Z }}

\nc{\YR}{[\bY,R]}
\nc{\YgivenR}{[\bY \mid R]}
\nc{\RgivenY}{[R \mid \bY]}
\nc{\Y}{[\bY]}
\nc{\R}{[R]}

\nc{\dio}{d_i^o}
\nc{\timi}{t_{i,m_i}}
\nc{\betahat}{\hat{\bbet}}
\nc{\mui}{\bmu_{\rm I}}
\nc{\mue}{\bmu^{\rm E}}
\nc{\mup}{\bmu^{\rm P}}
\nc{\muihat}{\hat{\bmu}_{\rm I}}
\nc{\muehat}{\hat{\bmu}^{\rm E}}
\nc{\muphat}{\hat{\bmu}^{\rm P}}
\nc{\delhat}{\hat{\bdel}}
\nc{\muhat}{\hat{\bmu}}

\nc{\iid}{\stackrel{\rm iid}{\sim}}
\nc{\law}{\stackrel{\scl}{=}}

\nc{\phiij}{ \phi_{ij}( \Delta_0) }
\nc{\phiiprmj}{ \phi_{i'j}( \Delta_0) }
\nc{\phiijprm}{ \phi_{ij'}( \Delta_0) }
\nc{\phixy}{ \phi( X_i(S_{ik}), Y_j(T_{jl}) ) }
\nc{\phixydo}{ \phi( X_i(S_{ik}), Y_j(T_{jl})-\Delta_0 ) }
\nc{\phixyd}{ \phi( X_i(S_{ik}), Y_j(T_{jl})-\Delta) }
\nc{\phixydstar}{ \phi^*( X_i(S_{ik}), Y_j(T_{jl})-\Delta) }
\nc{\phixystdttil}{ \tilde{\phi}( X_i(s), Y_j(t)-\Delta, \theta) }
\nc{\phixydttil}{ \tilde{\phi}( X_i(S_{ik}), Y_j(T_{jl})-\Delta, \theta) }
\nc{\Nmn}{{\sqrt{N} \over mn}}

\nc{\Xis}{X_i(s)}
\nc{\Yjt}{Y_j(t)}

\nc{\bthehat}{\hat{\bthe}}

\nc{\Ritil}{\tilde{R}_i}

\nc{\Ybar}{\overline{Y}}
\nc{\Rbar}{\overline{R}}
\nc{\Nbar}{\overline{N}}
\nc{\intzeroinf}{\int_0^\infty}

\nc{\Fhat}{\hat{F}}
\nc{\Ghat}{\hat{G}}

\nc{\FhatS}{\hat{F}(S_{ik})}
\nc{\GhatT}{\hat{G}(T_{jl})}

\nc{\Fhatik}{\hat{F}_{ik}}
\nc{\Ghatjl}{\hat{G}_{jl}}
\nc{\Fik}{F_{ik}}
\nc{\Gjl}{G_{jl}}
\nc{\phiijkl}{\phi_{ik,jl}(\Delta)}
\nc{\phiijkltil}{\tilde{\phi}_{ik,jl}(\Delta_0,\theta_0)}
\nc{\ord}{N^{-3/2}}           
\nc{\sumijkl}{\sum_{ijkl}}

\nc{\Citil}{\tilde{C}_i}
\nc{\Crtil}{\tilde{C}_r}
\nc{\Djtil}{\tilde{D}_j}
\nc{\Ditil}{\tilde{D}_i}

\nc{\Cithe}{\tilde{C}^{\theta}_i}
\nc{\Djthe}{\tilde{D}^{\theta}_j}

\nc{\Sikthe}{S_{ik}^{\theta}}
\nc{\Tjlthe}{T_{jl}^{\theta}}

\nc{\Zi}{ \bZ_{-i}}
\nc{\zic}{ \lbc z(\bs_j) \: : \: i \neq j \rbc }
\nc{\zkap}{ \bz_{\kappa} }
\nc{\sumi}{ \sum_i }
\nc{\sumj}{ \sum_j }
\nc{\sumij}{ \sum_{i < j} }
\nc{\sumiandj}{ \sum_{i, j} }
\nc{\zsi}{ z(\bs_i) }
\nc{\Zsi}{ Z(\bs_i) }
\nc{\zsj}{ z(\bs_j) }
\nc{\zsn}{ z(\bs_n) }
\nc{\zsone}{ z(\bs_1) }
\nc{\pZ}{ \Pr \lbc \bZ \rbc }
\nc{\qz}{ Q( \bz ) }
\nc{\qZ}{ Q( \bZ ) }

\nc{\thetaYD}{\theta_{Y\mid D}}
\nc{\thetaD}{\theta_D}
\nc{\psiDY}{\psi_{D\mid Y}}
\nc{\psiY}{\psi_Y}

\nc{\tn}{\Theta^{\nu}}
\nc{\Etn}{E_{\theta^{\nu}}}
\nc{\tnone}{\Theta^{\nu+1}}
\nc{\Lm}{L_{\text{m}}}
\nc{\Lo}{L_{\text{o}}}
\nc{\Ym}{Y_{\text{m}}}
\nc{\Yo}{Y_{\text{o}}}
\nc{\ym}{y_{\text{m}}}
\nc{\yo}{y_{\text{o}}}

\nc{\vijb}{v_{ij} - \bX_{i(j)}  \bbet}
\nc{\vikb}{v_{ik} - \bX_{i(k)}  \bbet}
\nc{\vilb}{v_{il} - \bX_{i(l)}  \bbet}
\nc{\betart}{ \bbet^{(r)}_{t_i} }
\nc{\betarj}{ \bbet^{(r)}_j }
\nc{\yij}{y_{ij}}
\nc{\Xmisi}{ {\bX_{ i{\rm (mis)} }} }
\nc{\Xobsi}{ {\bX_{ i{\rm (obs)} }} }
\nc{\Zobsi}{ {\bZ_{ i{\rm (obs)} }} }
\nc{\bSigobs}{ \bSig_{  {\rm obs} } }
\nc{\bSigmis}{ \bSig_{  {\rm mis} } }
\nc{\bSigmo}{ \bSig_{  {\rm mis,obs} } }
\nc{\bSigom}{ \bSig_{  {\rm obs,mis} } }
\nc{\Xil}{{\bX}_{il}}
\nc{\Zil}{{\bZ}_{il} }
\nc{\omilr}{\omega_{il}^{(r)}}
\nc{\delio}{\bdel_i^{{\rm obs}} }

\nc{\yio}{ {y_i^{\rm o} }}
\nc{\Yio}{ {Y_i^{\rm o }} }
\nc{\Yim}{ {Y_i^{\rm m} }}
\nc{\yim}{ {y_i^{\rm m} }}
\nc{\Yc}{Y^{\rm c}}
\nc{\Yic}{Y_i^{\rm c}}
\nc{\yc}{y^{\rm c}}
\nc{\yic}{y_i^{\rm c}}

\nc{\yi}{y_i}
\nc{\Yi}{Y_i}

\nc{\fyic}{f ( \yic ; \; \psiY )}
\nc{\fyi}{f ( y_i ; \; \psiY ) }
\nc{\fdigivenyic}{f ( d_i  \mid  \yic ; \; \psiDY )}
\nc{\fditilgivenyic}{f ( \tilde{d}_i  \mid  \yic ; \; \psiDY )}
\nc{\fditilgivenyi}{f ( \tilde{d}_i  \mid  \yi ; \; \psiDY )}
\nc{\Fditilgivenyic}{F ( \tilde{d}_i  \mid  \yic ; \; \psiDY )}
\nc{\Fditilgivenyi}{F ( \tilde{d}_i  \mid  \yi ; \; \psiDY )}
\nc{\fdigivenyi}{f (d_i \mid y_i ; \;  \psiDY  )}
\nc{\fyicdi}{f \left( \yic, d_i \right)}
\nc{\fyidi}{f \left( \yi, d_i \right)}

\nc{\fymidr}{f_{Y \mid R}}
\nc{\fyr}{f_{Y,R}}
\nc{\frmidy}{f_{R \mid Y}}
\nc{\fy}{f_Y}
\nc{\fr}{f_R}

\nc{\fyicgivendi}{f (\yic \mid d_i; \; \thetaYD )}
\nc{\fyigivendi}{f (\yi \mid d_i; \; \thetaYD )}
\nc{\fyicgivens}{f (\yic \mid s; \; \thetaYD )}
\nc{\fyigivens}{f (\yi \mid s; \; \thetaYD )}
\nc{\fdi}{f ( d_i; \; \thetaD )}

\nc{\fyicX}{f ( \yic \mid X_i; \; \psiY )}
\nc{\fyiX}{f ( y_i \mid X_i; \; \psiY ) }
\nc{\fdigivenyicX}{f ( d_i  \mid  \yic, X_i ; \; \psiDY )}
\nc{\fdigivenyiX}{f (d_i \mid y_i, X_i ; \;  \psiDY  )}
\nc{\fyicdiX}{f \left( \yic, d_i \mid X_i \right)}

\nc{\fyicgivendiX}{f (\yic \mid d_i, X_i; \; \thetaYD )}
\nc{\fyigivendiX}{f (y_i \mid d_i, X_i; \; \thetaYD )}
\nc{\fdiX}{f ( d_i \mid X_i; \; \thetaD )}

\nc{\Yistar}{\bY_i^*}

\nc{\Dio}{D_i^{\rm obs}}
\nc{\bdelio}{\bdel_{ i \, {\rm (obs)}} }

\nc{\fygivend}{f_{Y \mid \delta}}
\nc{\fyd}{f_{Y, \delta}}
\nc{\fd}{f_\delta}
\nc{\FD}{F_D}
\nc{\fygivenbd}{f_{Y\mid b, \delta}}

\nc{\alphahat}{\hat{\bal}}
\nc{\phihat}{\hat{\bphi}}
\nc{\thetahat}{\hat{\bthe}}
\nc{\thetatilde}{\tilde{\bthe}}
\nc{\scoretheta}{\bS(\bthe; \, \scc)}
\nc{\hesstheta}{\bH(\bthe; \, \scc)}
\nc{\infotheta}{\sci(\bthe; \, \scc)}
\nc{\sitheta}{\bs_i(\bthe; \, \scc_i)}
\nc{\sithetahat}{\bs_i(\thetahat; \, \scc_i)}

\nc{\loglikobs}{\ell_{{\rm o}}(\bthe; \, \sco)}
\nc{\scoreobs}{\bS_{{\rm o}}(\bthe; \, \sco)}
\nc{\hessobs}{\bH_{{\rm o}}(\bthe; \, \sco)}
\nc{\infoobs}{\scj_{{\rm o}}(\bthe; \, \sco)}

\nc{\Cil}{\scc_{il}}
\nc{\olog}{\lambda^*(\bthe, \Xobs)}
\nc{\LthetaC}{\scl(\bthe; \, \scc)}
\nc{\LthetaCi}{\scl_i(\bthe; \, \scc_i)}
\nc{\LthetaCil}{\scl_i (\bthe; \, \scc_{il}) }
\nc{\lthetaC}{\ell(\bthe; \, \scc)}
\nc{\lthetaCi}{\ell_i(\bthe; \, \scc_i)}
\nc{\lthetaCil}{\ell_i (\bthe; \, \scc_{il}) }
\nc{\Qtheta}{\scq \left( \bthe \, \left| \,  \bthe^{(r)} \right. \right)}
\nc{\thetar}{\bthe^{(r)}}
\nc{\thetas}{\bthe^{(s)}}
\nc{\alphas}{\bal^{(s)}}
\nc{\psis}{\psi^{(s)}}
\nc{\alphasplusone}{\bal^{(s+1)}}
\nc{\psisplusone}{\bpsi^{(s+1)}}
\nc{\alphapsis}{\left( \alphas, \psis \right)}

\nc{\thetarplusone}{\bthe^{(r+1)}}
\nc{\ologi}{\lambda^*_i(\bthe, \Xobs)}
\nc{\llogi}{\lambda_i \left( \bthe, \tilde{\Xcom}_{il} \right) }

\nc{\scxil}{\tilde{\Xcom}_{il}}
\nc{\siginv}{\bSig_i^{-1}}

\nc{\fofym}{ f \left( \by_i \mid \bbet_m, \bSig \right) }
\nc{\mphim}{ \phi_M \lsq \bSig^{-1/2}(\by_i - \bX_i \bbet_m) \rsq }
\nc{\mphit}{ \phi_M \lsq \bSig^{-1/2}(\by_i - \bX_i \bbet_{t_i}) \rsq }
\nc{\mphij}{ \phi_M \lsq \bSig^{-1/2}(\by_i - \bX_i \bbet_j) \rsq }
\nc{\mphik}{ \phi_M \lsq \bSig^{-1/2}(\by_i - \bX_i \bbet_k) \rsq }
\nc{\expkerm}{ \exp  \lbc -\half \bu_i(\bbet_m)' \bSig^{-1} \bu_i(\bbet_m)
  \rbc }
\nc{\expkerk}{ \exp  \lbc -\half \bu_i(\bbet_k)' \bSig^{-1} \bu_i(\bbet_k)
  \rbc }
\nc{\expkerj}{ \exp  \lbc -\half \bu_i(\bbet_j)' \bSig^{-1} \bu_i(\bbet_j)
  \rbc }
\nc{\normscorem}{\left( \bX_i' \bSig^{-1} \bX_i \bbet_m - \bX_i' \bSig^{-1}
  \by_i \right) }
\nc{\normscorej}{\left( \bX_i' \bSig^{-1} \bX_i \bbet_j - \bX_i' \bSig^{-1}
  \by_i \right) }
\nc{\piti}{ \pi \left( t_i, \bal, \bZ_i\bgam \right) }
\nc{\omij}{ \om_{ij} \left( t_i, \bal, \bZ_i\bgam \right) }
\nc{\phibetak}{ \phi_M(\bbet_k) }
\nc{\phibetaj}{ \phi_M(\bbet_j) }
\nc{\dphidbetak}{ \left. \dpl \phibetak \right/ \dpl \bbet_k }
\nc{\dphidbetakf}{ \frac{ \dpl \phibetak }{ \dpl \bbet_k } }
\nc{\uik}{\bu_i \left( \bbet_k  \right)}
\nc{\mset}{ \{ 0, 1, \ldots, M \} }
\nc{\betasigma}{ \left( \lbc \bbet^{(r)}_t \rbc, \bSig^{(r)} \right) }
\nc{\Thetar}{ \bThe^{(r)} }

\nc{\shatkm}{\hat{S}_{\rm KM}}

\nc{\ds}{\displaystyle}

\nc{\beq}{\begin{eqnarray*}}
\nc{\eeq}{\end{eqnarray*}}

\nc{\beqna}{\begin{eqnarray}}
\nc{\eeqna}{\end{eqnarray}}

\nc{\bct}{\begin{center}}
\nc{\ect}{\end{center}}

\nc{\bds}{\begin{description}}
\nc{\eds}{\end{description}}

\nc{\bit}{\begin{itemize}}
\nc{\eit}{\end{itemize}}

\nc{\bnu}{\begin{enumerate}}
\nc{\enu}{\end{enumerate}}

\nc{\bgt}{\begin{table}}
\nc{\bgtb}{\begin{center} \begin{tabular}}
\nc{\entb}{\end{tabular} \end{center} }
\nc{\ent}{\end{table}}

\nc{\ts}{\textstyle}

\nc{\trd}[1]{\textcolor[rgb]{0.00,0.00,0.00}{#1}}
\nc{\tp}[1]{\textcolor[rgb]{0,0.00,0}{#1}}

\title{Improving Longitudinal Targeted Maximum Likelihood Estimation in Target Trial Emulation using Joint Calibrated Weights \\ \large{Preprint}}

\date{}
\author{Juliette Limozin, Shaun R. Seaman, Li Su\\
	MRC Biostatistics Unit\\
	University of Cambridge \\
    \small{\texttt{juliette.limozin@mrc-bsu.cam.ac.uk}}
}
\begin{document}
\maketitle
\begin{abstract} 
\small{In target trial emulation (TTE), marginal structural models (MSMs) can be used to characterise per-protocol treatment effects over time. The MSM parameters are often estimated by inverse probability weighting (IPW), with weights estimated by maximum likelihood. However, IPW-based estimators can be unstable in small samples and are sensitive to misspecification of the weight models. An alternative method for estimating the MSM parameters is longitudinal targeted maximum likelihood estimation (LTMLE). LTMLE is double robust and potentially more efficient than IPW. Nevertheless, LTMLE also relies on inverse probability weights and may therefore share the instability of IPW-based estimators. We propose joint calibrated LTMLE, which integrates LTMLE with joint calibrated weights tailored for per-protocol effect estimation in TTE. This calibration of weights improves finite-sample performance by enforcing covariate balance in both the treatment and censoring processes simultaneously. Simulations show that the proposed method has improved efficiency and robustness to weight model misspecification, compared to standard LTMLE. We illustrate the method using a case study to evaluate the effect of highly active antiretroviral therapy on CD4 cell count among HIV-positive women.
}
\end{abstract}


\section{Introduction} \label{intro}

\subsection{Target trial emulation and per-protocol effect estimation}
Target trial emulation (TTE) is an increasingly popular approach for estimating causal treatment effects using observational data \cite{hansford_reporting_2023}. When implemented successfully, TTE can provide causal effect estimates that closely approximate those that would have been obtained from the target randomised controlled trial (RCTs) being emulated \cite{hernan_using_2016, matthews_comparing_2021}. TTE is particularly helpful when an RCT cannot be conducted due to time, financial or ethical constraints \cite{matthews_target_2022}. 

In TTEs, practitioners may be interested in evaluating the causal relationship between sustained treatment strategies and the outcome trajectories over time. The corresponding causal estimand of interest is the per-protocol effect, defined as a contrast between the mean counterfactual outcome trajectories under pre-specified treatment strategies, such as the \textit{always-treated} strategy, in which treatment is received at every visit, and the \textit{never-treated} strategy. In the presence of time-varying confounding of treatment, g-methods such as inverse probability weighting (IPW) and g-computation are commonly used to estimate mean counterfactual outcomes under static treatment strategies and thereby obtain per-protocol effect estimates \cite{daniel_methods_2013, robins_marginal_2000, hernan_marginal_2001}. However, most of these methods rely on having enough support in the observed data for {\textit{all}} treatment strategies under comparison. In per-protocol analyses typical of TTE, support for certain sustained treatment strategies is often limited, particularly over long follow-up periods. {This lack of support} can lead to {weakly identified estimands and result in} unstable, or highly variable estimates of the mean counterfactual outcomes under the treatment strategies of interest.

To address this challenge, marginal structural models (MSMs) \cite{robins_marginal_2000} \tp{can be specified to characterise} the \tred{mean} counterfactual outcome \tred{trajectory under a} treatment strategy, \tb{with baseline covariates optionally included to model potential effect modification.} For example, an MSM for the mean counterfactual CD4 cell count trajectory in HIV-positive patients can \tb{model cumulative treatment effects stratified by baseline CD4 cell count groups} \cite{ko_estimating_2003, yiu_joint_2022}. By pooling information across all follow-up visits, MSMs enable estimation of mean counterfactual outcome trajectories even when support for some treatment strategies is limited at specific time points in the observed data \cite{petersen_targeted_2014}. 
\tred{In particular, unsaturated MSMs are well-suited to TTE settings with poor support for some treatment strategies, as they impose a parsimonious structure on the mean counterfactual outcome trajectory over time. This contrasts with saturated MSMs, which estimate a separate mean at each time point without imposing structural constraints. \tp{Unsaturated} MSMs may be defined over the full set of possible treatment strategies or restricted to a subset of treatment strategies of interest. \tp{In this article, we focus on \textit{unsaturated}, \textit{restricted} MSMs, modelling only the mean counterfactual outcomes over time for the always-treated strategy and the never-treated strategy, and imposing no restrictions on the mean counterfactual outcomes for other treatment strategies}.}


\subsection{Estimation of marginal structural model parameters in target trial emulation}
\tred{In TTE}, MSM parameters are most commonly estimated using IPW \cite{robins_marginal_2000,hernan_marginal_2001,daniel_methods_2013}, where the weights are constructed as \tb{the product of two sets of time-varying weights, the first accounting for baseline and time-varying confounding of treatment assignment, and the second for censoring due to loss to follow-up.} By weighting the observed data, IPW aims to create a \tred{pseudo-population} that is representative of the target population of interest in TTE, \tred{i.e., a population in which there is no loss to follow-up and in which all patients adhere to their assigned treatment} strategies from baseline \tred{until the end of the trial}. \tred{This weighting procedure seeks to balance} \tb{the distributions of observed covariates between adhering and non-adhering patients, and between censored and uncensored patients over time \cite{jackson_diagnosing_2019, yiu_joint_2022}. Such covariate balance ensures the absence of systematic differences between patients who are artificially censored due to treatment protocol non-adherence and those who are not, as well as between patients who are censored due to loss to follow-up and those who remain in the trial,} \tred{thereby enabling estimation of the per-protocol effect.}

Although IPW estimation of MSM parameters (IPW-MSM) remains one of the most widely used approaches for treatment effect estimation in longitudinal settings \cite{robins_marginal_2000, clare_causal_2019,murray_causal_2021}, an important limitation is that IPW-MSM is single robust. In practice, inverse probability weights are typically obtained using maximum likelihood parameter estimates of the treatment and censoring process models, which, if misspecified, can result in biased treatment effect estimates. These concerns have motivated growing interest in estimators with improved robustness and efficiency for longitudinal treatment effects; see \citet{tran_double_2019} for a review. Several such estimators have been proposed, including the augmented inverse probability weighting estimator \cite{robins_robust_2000} and the weighted iterated conditional expectation estimator \cite{bang_doubly_2005}. 

Building on this line of work, targeted maximum likelihood estimation (TMLE) provides a general framework for constructing doubly robust and asymptotically efficient estimators of mean counterfactual outcomes and has been extended to longitudinal settings \cite{stitelman_general_2012, schnitzer_modeling_2014,schomaker_using_2019,schnitzer_dataadaptive_2020, benkeser_doubly-robust_2023,shirakawa_longitudinal_2024}. TMLE estimates causal effects via a two-step procedure in which an initial outcome regression model is first fitted, and subsequently updated using inverse probability weights to target the causal estimand of interest \cite{tran_double_2019}. In particular, the longitudinal TMLE (LTMLE) developed by \citet{petersen_targeted_2014} builds on the work of \citet{robins_robust_2000} and \citet{bang_doubly_2005}, providing a doubly robust substitution estimator for the parameters of an MSM in longitudinal settings. This allows us to remain within the MSM framework for TTE, with off-the-shelf implementation available in \texttt{R} \cite{lendle_ltmle_2017}. 

\tb{When the MSM is restricted to specific treatment strategies, as is typical in TTE, LTMLE offers an important advantage over standard IPW-based estimators, where} patients' follow-up is usually artificially censored at the time they deviate from the treatment strategies assigned \cite{hernan_observational_2008, murray_causal_2021, daniel_methods_2013}. \tb{Specifically}, LTMLE fits outcome regression models using data from the \textit{entire} study cohort, thereby retaining outcome information from patients who \textit{partially} adhere to treatment strategies of interest.
This more efficient use of the observed data can be particularly beneficial in settings with limited support for sustained treatment strategies.

Despite these advantages, a key limitation of LTMLE is its reliance on inverse probability weights, commonly estimated via maximum likelihood. {It is well known that IPW-based estimators can be unstable \trd{in settings with limited covariate overlap} and are sensitive to misspecification of the weight models. 
Although LTMLE is doubly robust, it uses the inverse of the estimated probabilities from the treatment and censoring models in its targeting step and, consequently, can \trd{share} the instability of IPW-based estimators, particularly when \textit{the outcome regression model is misspecified}. Evidence on simpler doubly robust estimators in point treatment settings shows that \trd{these estimators are sensitive to misspecification of the
propensity score model and can perform poorly when some estimated propensity scores are small, because small estimation errors in propensity scores lead to large errors in the inverse propensity score weights. \cite{kang_demystifying_2007,robins_performance_2007}.}

Achieving correct specification is especially difficult in longitudinal settings due to the complexity of time-varying treatment and censoring processes. Although data-adaptive methods such as SuperLearner \cite{polley_superlearner_2025} may mitigate the risk of model misspecification, they typically require sufficiently large sample sizes to perform well. As discussed by \citet{schomaker_using_2019}, the use of data-adaptive models within LTMLE can be challenging in settings with long follow-up periods, declining sample sizes \trd{due to loss to follow-up} or limited support for some treatment strategies of interest.

{The above considerations motivate us to propose an alternative method for estimating the inverse probability weights used by LTMLE, specifically a method that targets covariate balance \trd{in the} treatment and censoring processes \cite{yiu_joint_2022}.}

\subsection{Joint calibration of inverse probability treatment and censoring weights for covariate balance}

Covariate-balancing weights methods \trd{seek to address} the challenges of IPW by directly optimising the weights to achieve covariate balance in the weighted observed data, typically via imposing covariate moment conditions. Compared to standard IPW, these methods increase the likelihood of achieving covariate balance in finite samples and can provide robustness to misspecification of the weight models. Some covariate-balancing weights methods additionally stabilise weights by formulating convex minimisation problems that penalise extreme weight dispersion \cite{chattopadhyay_balancing_2020}. An overview of notable contributions is provided in Table~1 in the Supplementary Materials.

Existing covariate-balancing weights methods have largely focused on point interventions and/or single outcomes, with relatively limited research in longitudinal settings. The \textit{joint calibrated estimation} of inverse probability of treatment and censoring weights for MSMs, proposed by \citet{yiu_joint_2022}, was the first of its kind to combine covariate-balancing techniques to address both time-varying confounding of treatment assignment and censoring simultaneously. Although the widely used Covariate Balancing Propensity Score (CBPS) \trd{method} \cite{imai_covariate_2014} has been adapted to longitudinal settings with binary time-varying treatments \cite{imai_robust_2015} and extended to MSMs with continuous time-varying treatments \cite{huffman_covariate_2018}, these approaches do not incorporate joint calibration to balance covariates for both treatment and censoring processes \textit{simultaneously}. Moreover, joint calibration is expected to offer superior computational efficiency compared to CBPS-based approaches in per-protocol analyses, as \trd{CBPS-based approaches} require solving a much larger set of balancing constraints that consider all possible treatment strategies a patient could follow over time \cite{imai_robust_2015}. Despite the progress, there remains a gap in developing covariate-balancing weights methods specifically tailored to TTE.

Incorporating joint calibration of inverse probability weights into LTMLE offers several potential advantages. Firstly, \tred{although inverse probability weights estimated from correctly specified treatment and censoring models achieve covariate balance asymptotically, residual imbalance may still persist in finite samples. Because LTMLE incorporates these weights within its targeting step, we speculate that reducing or eliminating this residual imbalance \trd{by employing joint calibrated weights} could improve the finite-sample performance of LTMLE for MSM parameter estimation}.

Secondly, integrating joint calibrated weights within LTMLE may provide additional protection against model misspecification. Although LTMLE is doubly robust to misspecification of either the models for \trd{inverse probability weights} or the outcome regression model, this guarantee is \textit{asymptotic} and does not ensure good finite-sample performance. In particular, when the treatment \tp{and/or} censoring models are misspecified, the resulting poorly estimated inverse probability weights may inadequately balance covariates \tp{and can be highly variable, which would} adversely affect the finite-sample performance of LTMLE. \tp{This impact could be further exacerbated by misspecification of the outcome regression model.} By directly improving covariate balance, joint calibrated weights \tb{may} mitigate the impact of \tp{misspecifying the treatment \tp{and/or} censoring models}, thereby providing an additional layer of robustness and potentially reducing mean squared error (MSE) of LTMLE \cite{yiu_joint_2022}. 

\trd{Thirdly}, although joint calibration does not resolve practical violations of the positivity assumption, 
\trd{failure of the joint calibration procedure to converge in such settings may serve as a diagnostic indicator of practical positivity violations and insufficient support in the observed data \cite{chattopadhyay_balancing_2020}. } 

\trd{Finally}, joint calibration is applied to an initial set of estimated weights and is therefore compatible with both parametric and data-adaptive approaches for estimating the treatment and censoring models, which can be used to obtain the initial weights prior to calibration. 

\subsection{Overview of contributions}

In this article, we propose a joint calibrated LTMLE for per-protocol effect estimation over time in TTE. Specifically, we adapt the joint calibrated weights approach of \citet{yiu_joint_2022} to the TTE setting and integrate it with LTMLE to estimate MSM parameters that characterise the per-protocol effect over time. We focus on estimating the per-protocol effect of a binary treatment strategy on a continuous longitudinal outcome; \tb{extensions to survival outcomes and multiple treatment strategies are discussed in Section~\ref{discussion}.} \tb{Our simulation studies demonstrate that the proposed joint calibrated LTMLE reduces} the mean squared error of MSM parameter estimation compared to standard LTMLE \tred{under model misspecification}, providing an additional layer of robustness to LTMLE.

To adapt the joint calibrated weights approach of \citet{yiu_joint_2022} to TTE, we account for the distinct features of causal estimands in per-protocol analyses. Specifically, we propose calibration restrictions that target two corresponding hypothetical populations: (1) all trial-eligible patients following the \textit{always-treated} strategy \tred{until the end of the trial},
and (2) all trial-eligible patients following the \textit{never-treated} strategy \tred{until the end of the trial}.
This setting differs from the estimation of treatment weights for standard MSMs \tb{that model the effect of all possible treatment strategies}, such as \tb{was done} by CBPS and related methods \cite{imai_robust_2015,huffman_covariate_2018,kallus_kernel_2019,kallus_optimal_2021}. 
\tb{In our setting, two censoring mechanisms arise: artificial censoring due to treatment protocol non-adherence and censoring due to loss to follow-up. Thus, weighting for the treatment process can be viewed as a form of (artificial) censoring weighting.} Because joint calibration explicitly accommodates the structure of censoring weights \citep{yiu_joint_2022}, it is particularly well suited to the TTE framework. 

Since the \texttt{R} package \texttt{ltmle} \cite{lendle_ltmle_2017,r_core_team_r_2023} for implementing the LTMLE by \citet{petersen_targeted_2014} does not permit replacing the default inverse probability weights with joint calibrated weights, we develop our own implementation.

The remainder of the article is structured as follows. In Section~\ref{HERSintro}, we introduce the HIV Epidemiology Research Study (HERS) data as a running example and describe the corresponding target trial protocol. Section~\ref{methods} describes the notation, causal estimand, causal assumptions and estimation procedure for per-protocol effect estimation using LTMLE with joint calibrated weights, along with the construction of confidence intervals. Section~\ref{simulations} evaluates the finite-sample performance of the proposed method through a set of simulation studies. In Section~\ref{application}, we apply the proposed method to the HERS data. We conclude in Section~\ref{discussion} with a discussion.

\section{HIV Epidemiology Research Study}\label{HERSintro}

The HIV Epidemiology Research Study (HERS) was a natural history study that enrolled 1,310 women with, or at high risk of HIV infection, across four U.S. sites (Baltimore, Detroit, New York, and Providence) between 1993 and 1995, with follow-up continuing until 2000 \citep{ko_estimating_2003}. Participants attended up to 12 scheduled visits approximately six months apart, during which clinical, behavioural, and sociological data were collected, including self-reported information on antiretroviral treatment.

Building on previous analyses of the HERS cohort \citep{ko_estimating_2003, yiu_joint_2022,limozin_inference_2025}, we apply the TTE framework to estimate the effect of sustained Highly Active Antiretroviral Therapy (HAART) on CD4 cell count over a two-year period among HIV-positive participants. We focus on visits 8 through 12 of the HERS cohort, corresponding to the period during which HAART became more widely available. We therefore emulate a hypothetical trial with its baseline defined at visit 8 and follow-up through visit 12. 
The corresponding trial protocol is summarised in Table~2 of the Supplementary Materials. 

A total of 491 women met the trial eligibility criterion of no history of HAART use by visit 8. Among them, 150 participants were subsequently lost to follow-up, leaving 341 women remaining in the trial at visit 12. 
Data on potential confounders were available throughout follow-up, including prior CD4 cell counts, HIV viral load, self-reported HIV-related symptoms, race, and study site. 

In the HERS cohort, the probability of receiving HAART at each visit depends on a complex set of baseline and time-varying covariates. As noted by \citet{yiu_joint_2022}, this complexity increases the risk of treatment model misspecification and makes covariate balance across treatment groups difficult to achieve when inverse probability of treatment weights are estimated by maximum likelihood. To address these issues, \citet{yiu_joint_2022} proposed joint calibrated weights within a standard MSM framework to improve covariate balance.

\Cref{HERS data summary} shows the number of patients who followed the always-treated and never-treated strategies and remained under follow-up by each visit. As shown, only 76 out of 491 eligible patients initiated the always-treated strategy, and 29 of them adhered to this strategy and remained under follow-up until the end of the trial. The very small number of patients following the always-treated strategy indicates poor support for this treatment strategy, which can lead to unstable inverse probability of treatment weights estimated by maximum likelihood. The sample size of the eligible cohort is also relatively small (491 patients), creating further finite-sample challenges.

These issues motivate us to (1) focus on an unsaturated MSM to characterise the per-protocol effect over time, and (2) extend the IPW-MSM approach used in the analysis of Yiu and Su \cite{yiu_joint_2022} by integrating joint calibrated weights into LTMLE for MSM parameter estimation. This allows us to retain the benefits of calibration while additionally exploiting the double robustness and potential efficiency gain of LTMLE. 

\begin{table}[htbp]
 \centering
 \caption[Data tabulation of the emulated trial from the HERS cohort by treatment strategy and visit.]{Data tabulation of the emulated trial from the HERS cohort by treatment strategy and visit. The numbers in a column represent the number of patients adhering to the always-treated or never-treated strategy and remaining under follow-up up to the corresponding visit in the HERS cohort.}
 \begin{tabular}{*{1}{l}*{5}{c}}
 \toprule
 & \multicolumn{5}{c}{Visit} \\
 \cmidrule(lr){2-6}
 Treatment strategy& 8 & 9 & 10 & 11 & 12\\
 \midrule
Always-treated & 76 &52& 41& 39& 29 \\ 
Never-treated & 415 &338 &269& 228& 159 \\
 \bottomrule
 \end{tabular}
 \label{HERS data summary}
\end{table}


\section{Methods}\label{methods}

\subsection{Notation}

Consider an emulated target trial on a discrete time scale, with $n$ patients meeting the eligibility criteria at the trial baseline $t = 0$. Eligible patients start the follow-up at baseline and are subsequently assessed at follow-up visits ($ t=1,2,\ldots, T$), until the earliest of loss to follow-up and the end of the trial at visit $T$. Each patient's time-independent variables are measured at baseline ($t = 0$), whereas time-varying variables are measured at baseline and each follow-up visit ($ t=1,2,\ldots, T$).

We consider an independent and identically distributed data structure $O_1,\ldots, O_n$, where for each patient $i= 1,\ldots,n$, we define the following variables in $O_i$.
\begin{itemize} 
\item $Y_{t}$ is the patient's continuous outcome variable measured at visit $t$, where $t=0, 1, 2, \ldots, T$.
\item $\boldsymbol{V}$ is the patient's vector of the time-independent covariates measured at $t = 0$.
\item $A_{t}$ is a binary variable that indicates the patient's treatment status at visit $t$. $A_t = 1$ if the patient is receiving treatment at visit $t$, and $A_t = 0$ if not.
\item $\boldsymbol L_{t}$ is the vector of the patient's time-varying covariates measured at visit $t$.
\item $C_{t}$ indicates whether the patient is censored due to loss to follow-up between visits $t$ and $t+1$. $C_0=0$ (i.e., baseline visit assessments are complete for all patients) and $C_{t-1}=1 \implies C_{t}=1$. 
\end{itemize}

We assume the temporal ordering $(\boldsymbol V, \boldsymbol L_{0}, A_{0}, Y_{0}, C_{0}, \ldots, \boldsymbol L_{t}, A_{t},$ $Y_{t}, C_{t},\ldots,\boldsymbol L_{T}, A_{T},$ $Y_{T})$, for each of $O_i$, $i = 1,\ldots,n$.

Throughout this article we will use the overbar to denote patient variable history, for example: $\overline A_{t} = (A_{0}, A_{1},\ldots, A_{t})$ and $\overline{\boldsymbol L}_{t} = (\boldsymbol L_{0}, Y_0, \boldsymbol L_{1}, Y_1,\ldots,\boldsymbol L_{t-1}, Y_{t-1}, \boldsymbol L_{t})$, where the observed outcome history up to $t-1$ is absorbed into $\overline{\boldsymbol L}_{t}$.

Let $Y^{\overline a_t}_t$ and ${\boldsymbol L}^{\overline a_t}_{t+1}$ be a patient's counterfactual outcome and counterfactual time-varying covariates, respectively, if, possibly contrary to fact, the patient had received the treatment sequence $\overline A_t=\overline a_t$. 

\subsection{Causal estimand and assumptions}

The causal estimand of interest is the per-protocol effect of treatment on the outcome over time, defined as
\begin{align} \label{estimand}
& \mathbb{E}\left(Y^{\overline a_t = \overline 1}_t\right) - \mathbb{E}\left(Y^{\overline a_t = \overline 0}_t\right), \qquad t = 0,\ldots,T,
\end{align}
that is, the difference in the mean counterfactual outcome at visit $t$ if, possibly contrary to fact, all patients in the population eligible for the target trial at baseline were always treated from baseline up to visit $t$, and the mean counterfactual outcome at visit $t$ if, possibly contrary to fact, they were never treated from baseline up to visit $t$.

Typically, in an emulated trial, the estimand of interest is the estimand in Equation~\eqref{estimand} evaluated at $t = T$, at the end of the trial. However, as explained in \Cref{intro}, due to \tred{lack of support and }data sparsity issues, focusing on this time-specific contrast is challenging. Therefore, we use MSMs to summarise the per-protocol effects over time as follows.

We assume that these counterfactual outcomes 
can be characterised using a marginal structural working model of the form
\begin{equation}\label{workingMSM}
 \mathbb{E}\left(Y^{\overline a_t = \overline a}_t | \boldsymbol V;\bm \beta\right) = m(a,t, \bm V;\boldsymbol{\beta}),
\end{equation}
which models the mean counterfactual outcomes under treatment protocol with $\overline a = \overline 1$ or $\overline a =\overline0$ as a function of the treatment strategy $a$, visit $t$, possibly some baseline covariates $\boldsymbol V$ (with baseline values of time-varying covariates $\boldsymbol L_0$ absorbed into $\boldsymbol V$), and a parameter vector $\boldsymbol{\beta}$ \cite{yiu_joint_2022}. The per-protocol effect for the entire trial population (when $\boldsymbol V = \varnothing$) and those for the subgroups defined by $\boldsymbol V$ can be obtained from this MSM. Note that the model in~\Cref{workingMSM} may be specified as a saturated MSM, \tred{in which 
a separate parameter is used for the mean counterfactual outcome of each treatment strategy at each visit, or as an unsaturated MSM that imposes additional structure to improve efficiency.}

For the HERS data example, we consider a working MSM motivated by previous analyses that examined a cumulative effect of HAART stratified by baseline CD4 cell count groups \cite{ko_estimating_2003,yiu_joint_2022}:
\begin{align} \label{HERS msm}
 \mathbb{E}\left(Y^{\overline a_t = \overline a}_t | \boldsymbol V;\bm \beta\right)& = m(a,t, \bm V;\boldsymbol{\beta}) \nonumber\\
 & =\beta_{0t} + \sum_{l = 1} ^2\beta_lI(G = l)+\sum_{l = 0}^2 \beta_{\alpha,l}\,I(G = l)\, a \tb{\times} (t+1), \\
 & \qquad t= 0 ,\ldots,4, ~~a\in \{1,0\}. \nonumber
\end{align} 
Here, $t = 0$ corresponds to visit 8, $t = 1$ to visit 9 and so on. $Y_{t}^{\bar a_t = \bar a}$ is the counterfactual CD4 cell count at visit $t$ under either the always-treated or never-treated strategy. $\beta_{0t}$ represents visit-specific intercept terms that capture time effects. The coefficients $\beta_{\alpha,0}$, $\beta_{\alpha,1}$ and $\beta_{\alpha,2}$ represent stratum-specific cumulative effects of sustained HAART treatment vs no treatment, with strata defined by baseline CD4 cell count at visit 7. Specifically, the stratum indicator $G$ is defined as $G= 0$ if $Y_7 < 200$, $G = 1$ if $200 \leq Y_7 \leq 500$, and $G = 2$ if $Y_7 > 500$, where $Y_7$ is the CD4 count at visit 7. The coefficients $\beta_1, \beta_2$ represent the main effects of baseline CD4 strata relative to the reference group $G=0$.

One can also consider a working model with fewer modelling assumptions, such as the following model that only includes a cumulative treatment effect and no stratification by baseline covariates:
\begin{equation} \label{linear msm}
 \mathbb{E}(Y^{\overline a_t = \overline a}_t; \bm \beta) = m( a,t;\boldsymbol{\beta}) = \beta_0 + \beta_{1}\, a \times (t+1),~~~ t = 0,\ldots,T, ~a \in \{1,0\}.
\end{equation} 

The identification and estimation of the MSM parameters $\boldsymbol{\beta}$ requires the following assumptions.\\

\textit{No Interference.}

For $i \ne j$, the treatment received by patient $i$ has no effect on the counterfactual outcomes of patient $j$ \cite{daniel_methods_2013}. \\

\textit{Consistency.}

$\forall \:t, \:\overline Y_{t} = \overline Y_{t}^{\overline A_{t}}$ and $\overline{\boldsymbol L}_{t+1} = \overline{\boldsymbol L}_{t+1} ^{\overline A_{t}}$, meaning that the observed outcomes $\overline Y_{t}$ and covariates $\overline{\boldsymbol L}_{t+1}$ are equal to their corresponding counterfactual outcomes and covariates under the treatment history actually received \cite{daniel_methods_2013,young_simulation_2014}.\\

\textit{Positivity of treatment assignment, treatment adherence and censoring.}
\begin{align*}
&\Pr(A_{0} = a | \boldsymbol V, \boldsymbol L_{0}) > 0; \\
 \forall \:t>0,~~~~ &\Pr(A_{t} = a \mid \overline A_{t-1} = \overline a, C_{t-1} = 0, \boldsymbol V, \overline{\boldsymbol L}_{t}) >0; \\
\forall \:t>0,~~~~ &\Pr(C_{t} = 0 \mid C_{t-1} = 0, \overline A_{t} = \overline a, \boldsymbol V, \overline{\boldsymbol L}_{t}) >0
\end{align*}
for $ a \in \{1,0\}$, i.e. a patient has non-zero probability of being assigned to either treatment at trial baseline, adhering to the treatment assigned and remaining in the study at all times conditional on the observed histories of treatment and covariates \cite{young_simulation_2014}.\\

\textit{Sequentially ignorable treatment assignment.}

$\forall \:t, \: Y_{t}^{\overline a}
\indep A_{t} \mid \boldsymbol V, \overline{\boldsymbol L}_{t}, \overline A_{t-1}= \overline a, C_{t-1} = 0 $ for $ a \in \{1,0\}$, i.e, conditional on the treatment and covariate histories, there is no unmeasured confounding between the counterfactual outcome and the current treatment received \cite{daniel_methods_2013}.\\

\textit{Sequentially ignorable loss to follow up.}

$\forall \:t>0, \: C_{t} \indep Y_{t+1}^{\overline a}, Y_{t+2}^{\overline a}, \ldots, Y_{T}^{\overline a} \mid C_{t-1}=0$, $\overline A_{t}= \overline a$, $\boldsymbol V$, $\overline{\boldsymbol L}_{t}$, for $ a \in \{1,0\}$ \cite{daniel_methods_2013}, i.e. at a given visit, a patient's probability of remaining in the trial does not depend on their future counterfactual outcomes, conditional on the treatment and covariate history up to that visit.

\subsection{Longitudinal targeted maximum likelihood estimation for marginal structural models} \label{LTMLE}


The working MSM in~\Cref{workingMSM} can be estimated by LTMLE \cite{petersen_targeted_2014},
which requires specification of nuisance models for the treatment, censoring, and outcome processes conditional on observed histories. 
We summarise the corresponding LTMLE estimation procedure as follows: 
\paragraph{LTMLE estimation procedure for MSM}\begin{enumerate}
 \item \textit{Estimate the inverse probability of treatment and censoring weights for both treatment strategies} 
 
To account for artificial censoring due to treatment protocol non-adherence, a patient's \textit{inverse probability of treatment weight} \cite{robins_marginal_2000} \tb{for treatment strategy $a$} at visit $t$ is 
\begin{align}\label{IPTW}
 W^{A}_{a,t} = \frac{1}{\prod_{j = 0}^t \Pr( A_{j}= a\mid\overline A_{ j-1}=\overline a, \boldsymbol V, \overline{\boldsymbol L}_{j}, C_{j-1} = 0)}, ~~~~~~~~~~a \in \{1,0\}.
\end{align}
\tb{The inverse probability of treatment weights $W^A_{a,t}$ in~\Cref{IPTW} applied to patients who adhere to always-treated or never-treated strategies aims to create a pseudo-population representative of the target population in the absence of artificial censoring. These weights can therefore be interpreted as inverse probability of \textit{artificial censoring} weights due to treatment protocol non-adherence.}

Here, $\Pr(A_{j}=a\mid\overline A_{j-1}=\overline a, \boldsymbol V, \overline{\boldsymbol L}_{j}, C_{j-1} = 0)$ is the conditional probability that the patient receives treatment $a$ at visit $j$, given that they have adhered to the treatment strategy $a$ through visit $j-1$, their observed covariate history to visit $j$, and that they have not been censored by visit $j$.

The probability $\Pr(A_{j}=a\mid\overline A_{j-1}=\overline a, \boldsymbol V, \overline{\boldsymbol L}_{j}, C_{j-1} = 0)$, for $j = 0,\ldots,T$, is \textit{often} estimated by $\Pr(A_{j}=a\mid\overline A_{j-1}=\overline a, \boldsymbol V, \overline{\boldsymbol L}_{j}, C_{j-1} = 0; \hat {\boldsymbol \alpha})$ \tb{with $\overline A_{j-1}$ set at $\overline a$}, where $\hat {\boldsymbol \alpha}$ denotes the maximum likelihood estimate of a parameter vector ${\boldsymbol \alpha}$ from a \tb{pooled} logistic regression model for the \tb{observed} treatment process $\Pr(A_{j}=a\mid\overline A_{j-1}, \boldsymbol V, \overline{\boldsymbol L}_{j}, C_{j-1} = 0; {\boldsymbol \alpha})$. 

To address censoring due to loss to follow-up, the \textit{inverse probability of censoring weight} \cite{robins_marginal_2000} \tred{at visit $t$ for a patient who has received treatment $a$ at all previous visits is defined as follows,} 
\begin{align}\label{IPCW}
 W^{C}_{a,t} = \frac{1}{\Pi_{j = 0}^{t-1}\Pr(C_{j} =0 \mid C_{j-1} = 0,\overline A_{j}= \overline a, \boldsymbol V, \overline{\boldsymbol L}_{j})}, ~~~~~~~~~~~~~~~~~a \in \{1,0\},
\end{align}
with $W^{C}_{a,0} = 1$.
Here, $\Pr(C_{j} =0\mid C_{j-1} = 0,\overline A_{j} = \overline a, \boldsymbol V, \overline{\boldsymbol L}_{j})$ is the conditional probability that the patient remains under follow-up between visit $j$ and visit $j+1$, given that they have remained in the trial and have adhered to the treatment strategy $a$ up to visit $j$, as well as their observed covariate history up to visit $j$.

The probability $\Pr(C_{j} =0\mid C_{j-1} = 0,\overline A_{j} = \overline a, \boldsymbol V, \overline{\boldsymbol L}_{j})$, for $j = 0,\ldots,T$, is often estimated by $\Pr(C_{j} =0\mid C_{j-1} = 0,\overline A_{j} = \overline a, \boldsymbol V, \overline{\boldsymbol L}_{j}; \hat {\boldsymbol \gamma})$ \tb{with $\overline A_{j-1}$ set at $\overline a$}, where $\hat {\boldsymbol \gamma}$ is the maximum likelihood estimate of a parameter vector ${\boldsymbol \gamma}$ from a \tb{pooled} logistic regression model for the censoring process $\Pr(C_{j} =0\mid C_{j-1} = 0,\overline A_{j}, \boldsymbol V, \overline{\boldsymbol L}_{j}; {\boldsymbol \gamma})$.

Finally, the estimated \textit{inverse probability of treatment and censoring weight} for each patient who has been following the treatment strategy $a$ and uncensored up to visit $t$ is
\begin{align} \label{MLE Weights}
 W^{AC}_{a,t}(\hat{\boldsymbol \alpha},\hat{\boldsymbol \gamma}) = W^A_{a,t}(\hat{\boldsymbol \alpha})\times W^C_{a,t}(\hat{\boldsymbol \gamma}). 
\end{align}
 \item Rescale the observed longitudinal outcomes $Y_0,\ldots, Y_T$ to lie in the interval $(0,1)$.
 \item For $t = T,\ldots,0$:
 \begin{itemize}
 \item For $k = t,\ldots,0$:
 \begin{enumerate}
 \item \textit{Construct initial outcome predictions.} \tb{Define $I_k = \{ i \in \{1, \dots, n\}: C_{i, k-1} = 0 \}$ as the set of patients still under follow-up at visit $k$, of size $n_k=n-\sum_{i=1}^n C_{i,k-1}$. }
 \begin{itemize}
 \item If $k = t$, fit a quasi-binomial logistic regression of \tb{$Y_{t}$ on $\overline A_{t}$, $\overline{\boldsymbol L}_{t}$} \tp{to patients in $I_t$}. Define $\bm{\widehat Q}_{k}^{1,t} $ and $\bm{\widehat Q}_{k}^{0,t} $ as the $n_k$-vectors of \tred{predicted outcome values} obtained from this regression model by setting $\overline A_{i,t} = \overline 1$ and $\overline A_{i,t} = \overline 0$ respectively \tp{to patients in} $I_t$. \\

 \item If $k < t$, fit \tb{a quasi-binomial logistic regression of $\bm{\widehat Q}_{k+1}^{1,t*} $ on $\overline A_{k}$, $\overline{\boldsymbol L}_{k}$ \tp{to patients in $I_{k+1}$}. Define $\bm{\widehat Q}_{k}^{1,t}$ as the $n_k$-vector of predicted outcome values obtained from this model by setting $\overline A_{i,k} = \overline 1$ for $i \in I_k$.
 Similarly, fit a quasi-binomial logistic regression of $\bm{\widehat Q}_{k+1}^{0,t*} $ on $\overline A_{k}$, $\overline{\boldsymbol L}_{k}$ \tp{to patients in $I_{k+1}$}. Define $\bm{\widehat Q}_{k}^{0,t}$ as the $n_k$-vector of predicted outcome values obtained by setting $\overline A_{i,k} = \overline 0$ for $i \in I_k$.}\\ 
 \end{itemize}
 Define the \tred{\textit{initial} predicted outcome} vector of \tb{length $n_k \times 2$} as $\bm{\widehat Q}_{k}^{t} = (\bm{\widehat Q}_{k}^{1,t} , \bm{\widehat Q}_{k}^{0,t} )$.
 \item \textit{Construct the clever covariates.} Construct the \tb{$(n_k \times 2) \times \dim(\boldsymbol{\beta})$ covariate matrix $\bcalc$. Rows 1 to $n_k$ and rows $n_k+1$ to $2n_k$ of $\bcalc$ correspond to, respectively, the $n_k \times dim(\bm \beta)$ submatrices $\bm c(1,t,k;\boldsymbol V)$ and $\bm c(0,t,k;\boldsymbol V)$. The $i$th row of each \tred{of these two $n_k \times {\rm dim}(\beta)$ submatrices} corresponds to patient $i$ and is given by }
 \[
 \boldsymbol{I}(\overline A_{i,k} = \overline a,\, C_{i,k-1}=0)
 \times
 \frac{\partial m(a,t,\boldsymbol V_i;\boldsymbol{\beta})}
 {\partial \boldsymbol{\beta}},
 \qquad a \in \{0,1\} , \qquad i \in I_k.
 \]
 For the working MSM in \Cref{linear msm}, \tred{the $i$th row of $\bm c(1,t,k;\boldsymbol V)$ corresponds to 
 \begin{align*}
 \left(1\times\boldsymbol{I}(\overline A_{i,k} = \overline 1, C_{i,k-1} = 0), (t+1)\times \boldsymbol{I}(\overline A_{i,k} = \overline 1, C_{i,k-1} = 0)\right),
 \end{align*}
and the $i$th row of $\bm c(0,t,k;\boldsymbol V)$ corresponds to
 \begin{align*}
 \left(1 \times \boldsymbol{I}(\overline A_{i,k} = \overline 0, C_{i,k-1} = 0), 0\times \boldsymbol{I}(\overline A_{i,k} = \overline 0, C_{i,k-1} = 0)\right),
 \end{align*}
 for each patient $i\in I_k$.}
 \item \textit{Target the initial predictions.} 
 \begin{itemize}
 \item \tred{If $k = t$, set $\bm{\widehat Q}_{k+1}^{t*} = (Y_{1,t}, \ldots Y_{n_t,t}, Y_{1,t}, \ldots Y_{n_t,t})$.}
 \item Fit \tb{the targeting model}, an intercept-free weighted quasi-binomial logistic regression of $\bm{\widehat Q}_{k+1}^{t*} = (\bm{\widehat Q}_{k+1}^{1,t*} , \bm{\widehat Q}_{k+1}^{0,t*} )$ on the covariate matrix $\bcalc$ with offset $\logit (\bm{\widehat Q}_{k}^{t})$ and \tb{weighted by the stacked $n_k\times2$ vector of weights $\bm \omega_{k}^{t} = (\bm\omega_{k}^{1,t},\bm\omega_{k}^{0,t})$, with each element $i = 1,...,n_k$ of $\bm \omega_{k}^{a,t}$ defined as
 $h(a,t,\bm V_i)\times W^{AC}_{a,i,k}(\hat{\bm \alpha}, \hat{\bm \gamma})$ for $a\in\{1,0\}$. \tred{The function} $h(a,t,\bm V_i)$ is user-specified.} 
 
 For the MSM in \Cref{linear msm}, the targeting model is
 \begin{equation}
 \logit\{\mathbb{E}(\bm{\widehat Q}_{k+1}^{t*} )\} = \logit (\bm{\widehat Q}_{k}^{t})+ \epsilon_0 + \epsilon_1 \, \bm a \times (t+1), \nonumber
 \end{equation} 
 where $\boldsymbol{\epsilon} = (\epsilon_0, \epsilon_1)$ is a parameter vector and \tb{$\bm a = (1,...,1,0,...,0)$ is a $n_k\times 2$ vector}.
 For the MSM of the HERS data example in \Cref{HERS msm}, the targeting model is
 \begin{align}
\logit&\{\mathbb{E}(\bm{\widehat Q}_{k+1}^{t*} )\} \nonumber \\
& = \logit(\bm{\widehat Q}_{k}^{t})+ \epsilon_{0t} + \sum_{l = 1} ^2\epsilon_lI(\bm G = l) + \sum_{l = 0}^2I(\bm G = l)\, \epsilon_{\alpha,l}\, \bm a \times (t+1), \nonumber
 \end{align} \tb{where $\epsilon_{0t}$, $\epsilon_{1}$, $\epsilon_{2}$, $\epsilon_{\alpha,0}$, $\epsilon_{\alpha,1}$ and $\epsilon_{\alpha,2}$ are regression parameters and $\bm G$ denotes the stacked vector of baseline CD4 cell count strata, of length $n_k \times 2$, and $I(\bm G = l)$ is the corresponding vector of indicator functions.}
 \end{itemize}
 
 \item \textit{Update the predictions.} Define $\bm{\widehat Q}_{k}^{t*} = (\bm{\widehat Q}_{k}^{1,t*}, \bm{\widehat Q}_{k}^{0,t*})$ as the predicted outcomes from the fitted targeting model.
 \end{enumerate}
 \end{itemize}
 \item Rescale $\bm{\widehat Q}^{0*}_0, \bm{\widehat Q}^{1*}_0,\ldots, \bm{\widehat Q}^{T*}_0 $ back to the original scale of $Y_0,\ldots,Y_T$
 \item \tb{Fit a weighted linear regression of the stacked vector of targeted predictions $(\bm{\widehat Q}^{0*}_0, \bm{\widehat Q}^{1*}_0,\ldots, \bm{\widehat Q}^{T*}_0)$ on the covariates specified by the working MSM $m(a,t, \bm V;\boldsymbol{\beta})$, using weights $h(a,t,\bm V)$. The covariate matrix for this model is a $(2 \times \sum_{t=0}^T n_t )\times dim(\bm\beta)$ matrix defined row-wise as
\[
\begin{bmatrix}
\dfrac{\partial m(a,t,\boldsymbol V_i;\boldsymbol\beta)}{\partial \boldsymbol\beta}
\end{bmatrix}_{(t,a,i)},
\]
where the rows are stacked over $t=0,\ldots,T$, $a\in\{1,0\}$, and $i=1,\ldots,n_t$, in the same order as the stacked targeted predictions.
 This gives the LTMLE estimate $\bm{\hat\beta}$ of $\boldsymbol{\beta}$.}
 \end{enumerate}
 
\paragraph{Choice of function $h(a,t,\bm V)$ }The choice of function $h(a,t,\bm V)$ is discussed extensively by \citet{petersen_targeted_2014}. 
We note that setting $h(a,t,\bm V) = \prod_{j = 0}^t \Pr( A_{j}= a\mid\overline A_{ j-1}=\overline a, \boldsymbol V, C_{j-1} = 0; \hat{\boldsymbol \eta}) \prod_{j = 0}^{t-1}\Pr(C_{j} =0 \mid C_{j-1} = 0,\overline A_{j}= \overline a, \boldsymbol V ; \hat{\boldsymbol \mu})$ is equivalent to using \textit{stabilised} inverse probability of treatment and censoring weights estimated via maximum likelihood in Step 3(c), where $\hat {\boldsymbol \eta}$ and $\hat {\boldsymbol \mu}$ are the maximum likelihood estimates of ${\boldsymbol \eta}$ and ${\boldsymbol \mu}$ in some parametric models $\Pr(A_{j}=a\mid\overline A_{j-1}, \boldsymbol V, C_{j-1} = 0; {\boldsymbol \eta})$ and $\Pr(C_{j} =0\mid\overline C_{j-1} = 0,\overline A_{j}, \boldsymbol V ; {\boldsymbol \mu})$, respectively; see Section~3.1 of the Supplementary Materials for more details on how the stabilised weights are calculated. Assuming the MSM is correctly specified, this choice of $h(a,t,\bm V)$ results in stabilised weights that place more weight on treatment strategies, trial visits, and baseline strata defined by $\bm V$ with more support in the data \cite{petersen_targeted_2014}, which could improve MSM estimation in TTE settings with poor support for certain treatment strategies. \citet{robins_marginal_2000} also recommend the use of stabilised weights in the presence of time-varying confounding to reduce the variability of inverse probability weights. If baseline covariates $\bm V$ are included in $h(\cdot)$, they must also be included \tb{as covariates in the specified MSM} \cite{petersen_targeted_2014}. 

\subsection{Joint calibrated weights for target trial emulation}

The joint calibrated weights proposed by \citet{yiu_joint_2022} provide an alternative to inverse probability weighting by creating a pseudo-population that more closely represents the target population through calibration restrictions that enforce covariate moments conditions in finite samples \cite{su_sensitivity_2022}. In this section, we adapt the approach of \citet{yiu_joint_2022} to the TTE framework for estimating the per-protocol effect estimand defined in \Cref{estimand}.

Let us introduce some additional notation: 
\begin{itemize}
 \item $R_{t}$ denotes whether a patient remains under follow-up at visit $t$, as determined by the censoring indicator $C_{t}$. Specifically, $R_{t} = 0 \iff C_{t-1} = 1$, meaning that the patient was censored between visits $t-1$ and $t$. 
\item $S^{(1)}_{t}$ and $S^{(0)}_{t}$ indicate adherence to the always-treated or never-treated strategies up to visit $t$, respectively. Thus, $\overline A_t = \bar 1 \iff S^{(1)}_{t} = 1$ (and $0$ otherwise), and $\overline A_t = \overline 0 \iff S^{(0)}_{t} = 1$ (and $0$ otherwise). $S^{(1)}_{t}$ and $S^{(0)}_{t}$ therefore act as indicators of artificial censoring due to treatment protocol non-adherence. $S^{(a)}_{t-1} = 0 \implies S^{(a)}_{t} = 0$ for $t= 0,\ldots,T$ and $a \in \{1,0\}$, with the convention $S^{(a)}_{-1} = 1$.
\item \tb{$\boldsymbol{W}^{\star}_{a,t}(\bm \lambda^{(a)})$ denotes the vector of joint calibrated weights $W^{\star}_{i,a,t}(\bm \lambda^{(a)})$ for $i = 1,\ldots,n$, $a \in \{1,0\}$ at visit $t$ ($t = 0,\ldots,T$), indexed by some calibration parameter vector ${\boldsymbol \lambda}^{(a)}$}. We use the superscript $\star$ to indicate that these weights are calibrated.
\end{itemize}

 \tb{We first introduce the proposed calibration restrictions for artificial censoring due to treatment protocol non-adherence and for censoring due to loss to follow-up. We then describe the form of calibrated weights and outline how to implement the joint calibration.}

\subsubsection{Calibration restrictions for artificial censoring due to treatment protocol non-adherence} \label{cal_res_tx}

Following the derivation of calibration restrictions for inverse probability of \textit{censoring} weights in \citet{yiu_joint_2022}, we obtain the following calibration restrictions for the inverse probability of treatment weights in the per-protocol analysis of TTE:
\begin{align} \label{calibration restriction artificial by time with censoring}
 \sum_{i = 1}^n R_{i,t}\left[S^{(a)}_{i,t}W^{\star}_{i,a,t}({\boldsymbol \lambda}^{(a)})-S^{(a)}_{i,t-1}W^{\star}_{i,a,t-1}({\boldsymbol \lambda}^{(a)})\right]\widetilde{\bm X}_{i,t}= 0, \\
 \qquad t = 0,\ldots,T, \quad a\in \{1,0\},\quad\text{ with } S^{(a)}_{i,-1} = W^{\star}_{i,a,-1} = 1 \ \forall (i,a), \notag 
\end{align}
where $\widetilde{\bm X}_{i,t}$ is a vector of functionals of covariates $(\bm V_i, \bm L_{i,t})$ including 1. Including the term $R_{i,t}$ in the left-hand side of \Cref{calibration restriction artificial by time with censoring} ensures that the calibration restrictions for artificial censoring due to treatment protocol non-adherence apply only to patients who remain under follow-up at visit $t$. The joint calibration then aims to achieve that, after weighting, covariate associations with both treatment protocol non-adherence and censoring due to loss to follow-up are eliminated in finite samples \citep{yiu_joint_2022}.

To facilitate interpretation, suppose that $\widetilde{\bm X}_{i,t} = (1, \boldsymbol V_i, {\boldsymbol L}_{i,t})$ and there is no censoring due to
loss to follow-up. In this case, the calibration restrictions in \Cref{calibration restriction artificial by time with censoring} imply the following:
\begin{itemize}
 \item $\sum_{i = 1}^n S^{(1)}_{i,t}W^{\star}_{i,1,t}({\boldsymbol \lambda}^{(1)})= n$,
 for $t = 0, \ldots, T$, meaning that the weighted number of always-treated patients at visit $t$ equals the total number of patients at the trial baseline. This reflects the target population of the emulated trial in which no one was artificially censored due to non-adherence to the always-treated strategy. 

 \item Similarly, $\sum_{i = 1}^n S^{(1)}_{i,t}W^{\star}_{i,1,t}({\boldsymbol \lambda}^{(1)}) \boldsymbol V_i= \sum_{i = 1}^n \boldsymbol V_i$, for $t = 0, \ldots, T$, meaning that the weighted sample average of $\boldsymbol V_i$ among always-treated patients at visit $t$ equals the sample average of $\boldsymbol V_i$ at baseline in the target population.

 \item $\sum_{i = 1}^n S^{(1)}_{i,t}W^{\star}_{i,1,t}({\boldsymbol \lambda}^{(1)}){\boldsymbol L}_{i,t} = \sum_{i = 1}^n S^{(1)}_{i,t-1}W^{\star}_{i,1,t-1}({\boldsymbol \lambda}^{(1)}){\boldsymbol L}_{i,t}$,
 for $t = 0, \ldots, T$, meaning that weighting the always-treated patients recovers the sample average of ${\boldsymbol L}_{i,t}$ that would have been observed in the absence of artificial censoring due to non-adherence to the always-treated strategy at visit $t$.
\end{itemize} 
Similar interpretations can be obtained for the calibration restrictions for the never-treated strategy. 

\tb{We \tp{note} two differences between our setting and that of \citet{yiu_joint_2022}. Firstly, Yiu and Su considered calibration restrictions for treatment assignment such that, after weighting with their resulting calibrated weights, the treatment assignments up to visit $t$ are unassociated with the covariate histories conditional on the treatment histories in the observed data sample, so their calibration restrictions consider \textit{all possible treatment strategies}. In our setting, we only consider the two treatment strategies of interest, the always-treated and never-treated strategies. The goal of weighting is different: we weight (1) the observed always-treated patients to obtain a representative sample of the target population in which all patients follow the always-treated strategy, and (2) the observed never-treated patients to obtain a representative sample of the target population in which all patients follow the never-treated strategy}. \tred{Secondly, if there is no censoring, the approach of \citet{yiu_joint_2022} preserves the original sample size $n$ after weighting, whereas our approach constructs separate weighted populations for each treatment strategy, effectively resulting in a stacked sample of size $2n$ corresponding to the two strategies.}

\subsubsection{Calibration restrictions for censoring due to loss to follow-up}\label{cal_res_censor}

This section describes the calibration restrictions for censoring due to loss to follow-up, which has been extensively covered by \citet{yiu_joint_2022}. Following their work, the censoring weight calibration restrictions are given by
\begin{align} \label{calibration restriction censoring by time}
 \sum_{i = 1}^n S^{(a)}_{i,t-1}\left[R_{i,t}W^{\star}_{i,a,t}({\boldsymbol \lambda}^{(a)})-R_{i,t-1}W^{\star}_{i,a,t-1}({\boldsymbol \lambda}^{(a)})\right]\widetilde{\bm H}_{i,t-1}= 0, \\
\qquad t = 1,\ldots,T, \quad a\in \{1,0\}, \quad \text{ with } R_{i,0} = 1 \ \forall i, \notag 
\end{align}
where $\widetilde{\bm H}_{i,t-1}$ is a vector of functionals of covariates $(\bm V_i, \bm L_{i,t-1})$ including 1. We index with $t-1$ because the time ordering of the censoring indicator $R_t$ in relation to other covariates $\bm L_{t-1} \rightarrow A_{t-1} \rightarrow Y_{t-1} \rightarrow C_{t-1} \rightarrow R_{t}$ means that the calibrations restrictions for censoring weights can only include functionals of covariates up to $t-1$. \tb{Including the term $S^{(a)}_{i,t-1}$ in the left-hand side of \Cref{calibration restriction censoring by time} ensures that the calibration restrictions for censoring due to loss to follow-up apply only to patients who adhere to the treatment strategy $a$ up to visit $t-1$, i.e., with $S^{(a)}_{i,t-1}=1$.}

Suppose that $\widetilde{\bm H}_{i,t-1} = (1, \boldsymbol V_i, {\boldsymbol L}_{i,t-1})$ and that all patients adhered to the always-treated strategy, so that $S^{(1)}_{i,t} = 1$, for $i = 1, \ldots, n$ and $t = 0, \ldots, T$, and hence $W^{\star}_{i,1,0}({\boldsymbol \lambda}^{(1)}) = 1$. Under these assumptions, the calibration restrictions in \Cref{calibration restriction censoring by time} can be interpreted as follows:
\begin{itemize}
 \item $\sum_{i = 1}^n R_{i,t}W^{\star}_{i,1,t}({\boldsymbol \lambda}^{(1)}) = n$, for $t = 1,\ldots, T$, meaning that weighting the always-treated, uncensored patients at visit $t$ recovers the total number of patients (i.e., all patients observed at baseline). 
 
 \item $\sum_{i = 1}^n R_{i,t}W^{\star}_{i,1,t}({\boldsymbol \lambda}^{(1)}) \boldsymbol V_i= \sum_{i = 1}^n \boldsymbol V_i$, for $t = 1,\ldots,T$, meaning that the weighted sample average of the baseline covariates $\boldsymbol V_i$ among always-treated uncensored patients at visit $t$ is equal to the sample average of $\boldsymbol V_i$ at baseline. 
 
 \item $\sum_{i = 1}^n R_{i,t}W^{\star}_{i,1,t}({\boldsymbol \lambda}^{(1)}) {\boldsymbol L}_{i,t-1} = \sum_{i = 1}^n R_{i,t-1}W^{\star}_{i,1,t-1}({\boldsymbol \lambda}^{(1)}) {\boldsymbol L}_{i,t-1}, t = 1,\ldots, T$, meaning that weighting the always-treated uncensored patients at visit $t$ recovers the sample average of ${\boldsymbol L}_{i,t-1}$ in the always-treated had there been no patients lost to follow-up between visits $t-1$ and $t$.
 \end{itemize}

\subsubsection{Implementation of the joint calibration} \label{implementation calibration}

We now describe how to implement the joint calibration of artificial censoring and censoring weights. 

\tb{\paragraph{Form of calibrated weights} Following \citet{yiu_joint_2022}, we consider their Type (1) calibration form that relates the calibrated weights to an initial set of inverse probability of treatment and censoring weights obtained via maximum likelihood estimation (MLE). Specifically, let $\boldsymbol{W}_a^{AC}(\hat{\boldsymbol{\alpha}}, \hat{\boldsymbol{\gamma}})$ denote the vector of unstabilised MLE-based weights $W^{AC}_{i,a,t}(\hat{\boldsymbol{\alpha}}, \hat{\boldsymbol{\gamma}})$ for the \tb{treatment strategy $a$} ($i = 1, \ldots, n$ and $t = 0,\ldots, T$), as defined in \Cref{MLE Weights}. For convenience, we refer to $\boldsymbol{W}_a^{AC}(\hat{\boldsymbol{\alpha}}, \hat{\boldsymbol{\gamma}})$ as the `\textit{MLE weights}'. }
\tb{For the treatment strategy $a$, at each visit $t$, the formula for the joint calibrated weights is:
\begin{align} \label{type 1 form}
 \boldsymbol{W}^{\star}_{a,t}(\boldsymbol{\lambda}^{(a)}_{t}) = \boldsymbol{W}^{AC}_{a,t}(\hat{\bm \alpha}, \hat{\bm \gamma}) \circ \exp(\boldsymbol{K}^{(a)}_t \boldsymbol{\lambda}^{(a)}_{t}), \; a\in \{1,0\},
\end{align}
where $\boldsymbol{W}^{\star}_{a,t}(\boldsymbol{\lambda}^{(a)}_{t})$ is the vector of joint calibrated weights ${W}^{\star}_{i,a,t}(\boldsymbol{\lambda}^{(a)}_{t})$, $i = 1,\ldots,n$, based on calibration restrictions for the treatment strategy $a$, \tb{$\circ$ denotes element-wise product,} $\boldsymbol{K}^{(a)}_t$ is the $n\times p_t$ matrix with row $i = 1,\ldots,n$ corresponding to $\left(R_{i,t}S^{(a)}_{i,t}\widetilde{\bm X}_{i,t}, S^{(a)}_{i,t-1}R_{i,t}\widetilde{\bm H}_{i,t-1}\right)$, $p_t =\dim(\widetilde{\bm X}_{i,t}) + \dim(\widetilde{\bm H}_{i,t-1})$ and $\bm \lambda^{(a)}_t$ is a vector of length $p_t$ denoting calibration parameters} \tred{corresponding to visit $t$. \tp{Define} $\bm \lambda^{(a)}= (\bm \lambda^{(a)}_0, \ldots, \bm \lambda^{(a)}_T$). } 

\tb{\paragraph{Estimation of the initial weights} \tred{\tp{By following} Step 1 in the LTMLE estimation procedure}, the initial MLE weights are estimated from the observed data. \tb{In typical per-protocol effect estimation}, these weights are estimated using each patient's data only up to the time when they first deviate from the treatment strategy \tb{they are assigned at baseline}. For this article, we estimate the initial MLE weights using these data, as well as the data from patients' follow-up after treatment strategy non-adherence, in order to retain as much information as possible before the calibration step.}

\tb{\paragraph{Convex minimisation for calibration} The calibration procedure is carried out \textit{separately} for each treatment strategy $a\in\{1,0\}$ and \textit{sequentially} over visits $t= 0, \ldots, T$. For a given treatment strategy $a$, at each visit $t$, we have a collection of calibration restrictions that are linear in the calibrated weights, $W^{\star}_{i,a,t}({\boldsymbol \lambda_t}^{(a)})$, for $i = 1,\ldots,n$ and $a \in \{1,0\}$, given the \textit{known} calibrated weights from the previous visit $t-1$, $W^{\star}_{i,a,t-1}(\widehat{\bm\lambda}^{(a)}_{t-1})$. These restrictions are defined by Equations~\eqref{calibration restriction artificial by time with censoring} and~\eqref{calibration restriction censoring by time}. The implementation of joint calibration proceeds \textit{sequentially} over time by solving these calibration restrictions for $\bm\lambda^{(a)}_t$ to obtain ${\bm W}^{\star}_{a,t}({\widehat{\bm\lambda}^{(a)}_t)}$, given calibrated weights up to visit $t-1$, ${\bm W}^{\star}_{a,t-1}(\widehat{\bm\lambda}^{(a)}_{t-1})$. 
This is equivalent to minimising a convex function in $\boldsymbol{\lambda_t}^{(a)}$, 
\begin{align}
 {\boldsymbol{K}^{(a)}_t}\trans\{\boldsymbol{W}^{AC}_{a,t}(\hat{\bm \alpha}, \hat{\bm \gamma}) \circ \exp(\boldsymbol{K}^{(a)}_t \boldsymbol{\lambda}^{(a)}_{t})\} - {\boldsymbol{l}^{(a)}_t}\trans{\bm W}^{\star}_{a,t-1}(\widehat{\bm\lambda}^{(a)}_{t-1}), \; a \in \{1,0\}, \label{convex minimization}
\end{align}
where $\bm l_t^{(a)}$ is the $n\times p_t$ matrix with row $i= 1,\ldots,n$ corresponding to $\left(R_{i,t}S^{(a)}_{i,t-1}\widetilde{\bm X}_{i,t},\right.$ $\left.S^{(a)}_{i,t-1}R_{i,t-1}\widetilde{\bm H}_{i,t-1}\right)$. $\boldsymbol{K}^{(a)}_t$ and $\bm l_t^{(a)}$ are matrices collecting the covariate terms appearing in the calibration restrictions in Equations~\eqref{calibration restriction artificial by time with censoring} and~\eqref{calibration restriction censoring by time}. The vectors $\bm \lambda^{(a)}_t$ and $\widehat{\bm\lambda}^{(a)}_{t-1}$ have dimensions $p_t$ and $p_{t-1} =\dim(\widetilde{\bm X}_{i,t-1}) + \dim(\widetilde{\bm H}_{i,t-2})$, respectively. Because the objective function in~\Cref{convex minimization} is convex in $\bm \lambda^{(a)}_t$, a unique solution exists and can be solved efficiently using the \texttt{R} package \texttt{nleqslv} \cite{hasselman_nleqslv_2023}.}

We present \tb{a detailed algorithm} for obtaining the joint calibrated weights 
\tb{based on unstabilised initial MLE weights in \Cref{joint calibration algo}. If the chosen $h(a,t,\bm V)$ function in the LTMLE algorithm is equivalent to using stabilised MLE weights as discussed at the end of Section~\ref{LTMLE}, then joint calibration of stabilised MLE weights should be performed. It follows the same procedure, with $\boldsymbol{W}^{AC}_{a,t}(\hat{\bm \alpha}, \hat{\bm \gamma})$ simply replaced by the stabilised initial MLE weights. Because stabilisation reduces the variability of MLE weights \cite{robins_marginal_2000}, we expect joint calibrated weights derived from initial \textit{stabilised} MLE weights to have less variability than those derived from initial \textit{unstabilised} MLE weights.} 

\begin{algorithm}[hb!]
\caption{Iterative algorithm for obtaining the joint calibrated weights based on initial unstabilised MLE weights.} \label{joint calibration algo}
{\footnotesize
\begin{enumerate}
 \item Estimate the initial MLE weights $W^{AC}_{i,a,t}(\hat{\bm \alpha}, \hat{\bm \gamma}), \; i = 1,\ldots,n, \; t = 0,\ldots,T$, using the full observed data for the treatment and censoring processes.
 \item 
 Create protocol-specific indicators for the always-treated
($S^{(1)}_{t} = 1$) and never-treated ($S^{(0)}_{t} = 1$) treatment strategies in the observed treatment data.
 \item For each treatment strategy $a \in \{1,0\}$:
 \begin{enumerate}
 \item \textbf{Initialisation at $t = 0$}: The calibration restrictions at baseline $t = 0$ are 
 \begin{align}
 \sum_{i = 1}^n \left\{S^{(a)}_{i,0}W^{\star}_{i,a,0}({\boldsymbol \lambda^{(a)}_0})-1\right\}\widetilde{\bm X}_{i,0}= 0 \nonumber
 \end{align}
 \begin{itemize}
\item Obtain $\hat{\bm\lambda}^{(a)}_0$ by minimising 
 \begin{align}
 {\bm K^{(a)}_0}\trans\left\{\bm{W}^{AC}_{a,0}(\hat{\bm \alpha}, \hat{\bm \gamma})\circ \exp(\bm K^{(a)}_0\bm\lambda^{(a)}_0)\right\} - {\bm l_0}\trans\bm 1 \nonumber
 \end{align}
 where $\bm{W}^{AC}_{a,0}(\hat{\bm \alpha}, \hat{\bm \gamma}) = (W^{AC}_{1,a,0}(\hat{\bm \alpha}, \hat{\bm \gamma}), \ldots, W^{AC}_{n,a,0}(\hat{\bm \alpha}, \hat{\bm \gamma}))$, $\bm K^{(a)}_0$ is the matrix such that row $i$ corresponds to $S^{(a)}_{i,0}\widetilde{\bm X}_{i,0}$, and $\bm l_0$ is the matrix such that row $i$ corresponds to $\widetilde{\bm X}_{i,0}$.
 
 \item Set $\bm{W}^{\star}_{a,0}(\widehat{\bm\lambda}^{(a)}_0) = \bm{W}^{AC}_{a,0}(\hat{\bm \alpha}, \hat{\bm \gamma})\circ \exp(\bm K_0^{(a)}\widehat{\bm\lambda}^{(a)}_0)$.
 \end{itemize}
 \item \textbf{For $t = 1,\ldots,T$}: Given the minimiser $\widehat{\bm\lambda}^{(a)}_{t-1}$ from the previous step, we have calibrated weights up to visit $t-1$, $\bm{W}^{\star}_{a,t-1}(\widehat{\bm\lambda}^{(a)}_{t-1})$. The calibration restrictions at visit $t$ are
 \begin{align}
 \sum_{i = 1}^n R_{i,t}\left[S^{(a)}_{i,t}W^{\star}_{i,a,t}({\bm \lambda}^{(a)}_t)-S^{(a)}_{i,t-1}W^{\star}_{i,a,t-1}(\widehat{\bm\lambda}^{(a)}_{t-1})\right]\widetilde{\bm X}_{i,t}= 0 \nn \\
 \sum_{i = 1}^n S^{(a)}_{i,t-1}\left[R_{i,t}W^{\star}_{i,a,t}({\bm \lambda}^{(a)}_t)-R_{i,t-1}W^{\star}_{i,a,t-1}(\widehat{\bm\lambda}^{(a)}_{t-1})\right]\widetilde{\bm H}_{i,t-1}= 0 \nn
 \end{align} 
 \begin{itemize}
 \item Obtain $\widehat{\bm\lambda}^{(a)}_t$ by minimising 
 \begin{align}
 {\bm K^{(a)}_t}\trans\left\{\bm{W}^{AC}_{a,t}(\hat{\bm \alpha}, \hat{\bm \gamma})\circ \exp(\bm K^{(a)}_t\bm\lambda^{(a)}_t)\right\} - {\bm l^{(a)}_t}\trans\bm{W}^{\star}_{a,t-1}(\widehat{\bm\lambda}^{(a)}_{t-1}) \nonumber
 \end{align}
 where $\bm{W}^{AC}_{a,t}(\hat{\bm \alpha}, \hat{\bm \gamma}) = (W^{AC}_{1,a,t}(\hat{\bm \alpha}, \hat{\bm \gamma}), \ldots, W^{AC}_{n,a,t}(\hat{\bm \alpha}, \hat{\bm \gamma}))$, $\bm K^{(a)}_t$ is the matrix with row $i$ corresponding to $\left(R_{i,t}S^{(a)}_{i,t}\widetilde{\bm X}_{i,t},\; S^{(a)}_{i,t-1}R_{i,t}\widetilde{\bm H}_{i,t-1}\right)$, and $\bm l^{(a)}_t$ is the matrix with row $i$ corresponding to $\left (R_{i,t}S^{(a)}_{i,t-1}\widetilde{\bm X}_{i,t},\; S^{(a)}_{i,t-1}R_{i,t-1}\widetilde{\bm H}_{i,t-1}\right)$.
 \item Set $\bm{W}^{\star}_{a,t}(\widehat{\bm\lambda}^{(a)}_t) = \bm{W}^{AC}_{a,t}(\hat{\bm \alpha}, \hat{\bm \gamma})\circ \exp(\bm K^{(a)}_t\hat{\bm\lambda}^{(a)}_t)$.
 \end{itemize}
 \end{enumerate}
\end{enumerate}}
\end{algorithm}

\subsection{Joint calibrated LTMLE: estimation algorithm} \label{calibrated LTMLE}

By integrating the joint calibrated weights with LTMLE, we propose the following algorithm for \textit{Joint Calibrated Longitudinal Targeted Maximum Likelihood Estimation} (joint calibrated LTMLE) of the MSM parameters $\bm \beta$, \tb{estimating} the per-protocol effect estimand defined in \Cref{estimand}, \tb{based on unstabilised initial MLE weights:}
\paragraph{Joint calibrated LTMLE algorithm with unstabilised weights}\begin{enumerate}
 \item Estimate unstabilised inverse probability of treatment and censoring weights for both treatment strategies via maximum likelihood. 
 \item Jointly calibrate the unstabilised MLE weights 
 using the calibration algorithm described in \Cref{joint calibration algo}. This gives joint calibrated weights $\bm{W}^{\star}_{a,t}(\widehat{\bm\lambda}^{(a)}_t)$ for $a \in \{1,0\}$ and $t=1, \ldots, T$.
 \item Follow Steps~2--5 of the LTMLE algorithm in \Cref{LTMLE} \tb{with a user-specified function $h(a,t,V), a\in \{1,0\}$}, replacing the MLE weights in Step~3(c) with the joint calibrated weights.
 \end{enumerate}

\tb{The estimation algorithm for when using stabilised initial MLE weights is specified in Section~3.2 of the Supplementary Materials.}

Under correct specification of the treatment and censoring models in Equations~\eqref{IPTW} and~\eqref{IPCW} used to construct the initial MLE weights, the joint calibrated weights converge to the \textit{true} inverse probability weights. This is because the true weights satisfy the population version of the calibration restrictions in Equations~\eqref{calibration restriction artificial by time with censoring} and~\eqref{calibration restriction censoring by time} \cite{yiu_joint_2022,su_sensitivity_2022}.
Consequently, replacing the MLE weights in the LTMLE algorithm with joint calibrated weights preserves consistency of the estimated MSM parameters and their double robustness with respect to misspecification of either the regression models for obtaining predicted outcomes $\widehat{\bm Q}_k^t$ or the treatment and censoring models used to estimate the initial MLE weights \cite{yiu_joint_2022,su_sensitivity_2022,petersen_targeted_2014}.

\subsection{Confidence intervals} \label{confidence intervals}

\paragraph{Full nonparametric bootstrap} To obtain valid statistical inference for the MSM parameters and the per-protocol effect estimand, we can construct confidence intervals (CIs) using nonparametric bootstrap \cite{su_sensitivity_2022}. We can generate $B$ bootstrap samples by resampling the observed data with replacement, using patients as the resampling unit. For bootstrap sample \tb{$b$ ($b = 1,\ldots,B$)}, the estimation steps as described in \Cref{calibrated LTMLE} are applied to obtain the bootstrap estimate $\hat{\bm \beta}_b$ of the MSM parameters $\bm \beta$. A 95\% CI of MSM parameters $\bm \beta$ is then constructed by taking the $2.5$th and $97.5$th percentiles of $\{\hat{\bm \beta}_b : b = 1,\ldots,B\}$ as the lower and upper bounds of the CI. Similarly, a 95\% CI of the mean counterfactual outcomes is constructed by taking the $2.5$th and $97.5$th percentiles of the bootstrap distribution of the mean counterfactual outcome estimates, \tb{which are obtained by calculating the mean counterfactual outcome estimates using the bootstrap MSM parameter estimates $\{\hat{\bm \beta}_b : b = 1,\ldots,B\}$.}

\tb{As noted by \citet{su_sensitivity_2022}, estimating and calibrating the MLE weights in each bootstrap sample can be computationally expensive. In our setting, this computational burden is increased further by the need to also re-estimate $\bm{\widehat Q}_{k}^{t}$, for $t = T,\ldots, 0$ and $k = t,\ldots,0$, as required by the joint calibrated LTMLE algorithm within each bootstrap sample. To alleviate this, we borrow a computational strategy introduced by \citet{tran_robust_2018}. While their work focused on a different estimation problem (using a standard TMLE algorithm to estimate counterfactual outcome means rather than the MSM parameters targeted in our setting), their underlying solution is useful to our setting. Specifically, they proposed a `modified' TMLE algorithm that \textit{separates} the step of obtaining the initial outcome predictions from the targeting step. By estimating the MLE weights and initial outcome predictions \textit{only once}, using the original data, and \tred{performing only} the targeting step within each bootstrap sample, they obtained a computationally efficient and consistent nonparametric bootstrap variance estimator of the counterfactual outcome means variance. Since the modified TMLE has the same asymptotic behaviour as the original TMLE, the resulting bootstrap procedure yields valid inference \cite{tran_robust_2018}. We adapt this separation strategy from \citet{tran_robust_2018} to our context, the proposed joint calibrated LTMLE, in order to efficiently construct CIs for the MSM parameters. Similarly to their modified TMLE, we separate the step of obtaining the initial outcome predictions $\bm{\widehat Q}_{k}^{t}$ (Step 3(a) of the LTMLE algorithm in Section~\ref{LTMLE}) from the targeting step (Steps 3(b) and 3(c) of the LTMLE algorithm) for $\bm{\widehat Q}_{k}^{t *}$. The estimation procedure is as follows:} \\

\paragraph{Modified joint calibrated LTMLE algorithm with unstabilised weights}\begin{enumerate}
 \item Estimate unstabilised inverse probability of treatment and censoring weights for both treatment strategies via maximum likelihood. 
 \item Jointly calibrate the MLE weights using the calibration algorithm in \Cref{joint calibration algo}. This gives joint calibrated weights $\bm{W}^{\star}_{a,t}(\widehat{\bm\lambda}^{(a)}_t)$ for $a \in \{1,0\}$ and $t=1, \ldots, T$.
 \item 
Rescale the observed longitudinal outcomes $Y_0,\ldots, Y_T$ to to lie in the interval $(0,1)$.
 \item \textit{Separate initial outcome prediction step.} For $t = T,\ldots,0$:
 \begin{itemize}
 \item For $k = t,\ldots,0$: Generate a vector $\bm{\widehat Q}_{k}^{t} = (\bm{\widehat Q}_{k}^{1,t} , \bm{\widehat Q}_{k}^{0,t} )$ of length $n_k\times 2$:
 \begin{itemize}
 \item If $k = t$, fit a quasi-binomial logistic regression of $Y_t$ on $\overline A_{t}$, $\overline{\boldsymbol L}_{t}$ \tp{to patients in $I_t$}. 
 Define $\bm{\widehat Q}_{k}^{1,t} $ and $\bm{\widehat Q}_{k}^{0,t} $ as the vectors of \textit{initial} predicted \tb{outcome values from this regression model, setting} $\overline A_{t} = \overline 1$ and $\overline A_{t} = \overline 0$, respectively, \trd{for patients in $I_t$}.
 \\
 \item If $k < t$, fit quasi-binomial logistic regressions of $\bm{\widehat Q}_{k+1}^{1,t}, \bm{\widehat Q}_{k+1}^{0,t}$ on $\overline A_{k}$, $\overline{\boldsymbol L}_{k}$, respectively, \tp{to patients in $I_{k+1}$}. 
 Define $\bm{\widehat Q}_{k}^{1,t} $ and $\bm{\widehat Q}_{k}^{0,t} $ as the corresponding vectors of \textit{initial} predicted \tb{outcome values from these two models, setting} $\overline A_{k} = \overline 1$ and $\overline A_{k} = \overline 0$, respectively, \trd{for patients in $I_{k}$}. 
 
 \end{itemize}
 \end{itemize}
 \item \textit{Separate targeting step.} For $t = T,\ldots,0$:
 \begin{itemize}
 \item For $k = t,\ldots,0$: Apply Steps 3(b)-3(d) of the LTMLE algorithm in \Cref{LTMLE}, using joint calibrated weights $\bm{W}^{\star}_{a,k}(\widehat{\bm\lambda}^{(a)}_k)$ for $a \in \{1,0\}$ 
 in Step 3(c).
 \end{itemize}
 \item Rescale $\bm{\widehat Q}^{0*}_0, \bm{\widehat Q}^{1*}_0,\ldots, \bm{\widehat Q}^{T*}_0 $ to the original scale of $Y_0,\ldots,Y_T$.
 \item Fit a linear regression of $\bm{\widehat Q}^{0*}_0, \bm{\widehat Q}^{1*}_0,\ldots, \bm{\widehat Q}^{T*}_0 $ according to the working MSM $m(a,t,$ $\bm V;\boldsymbol{\beta})$, using weights $h(a,t,\bm V)$. This gives the modified LTMLE estimator of $\boldsymbol{\beta}$.
 \end{enumerate}

\paragraph{Bootstrap based on modified LTMLE} Nonparametric bootstrap CIs of the MSM parameters $\bm \beta$ can then be constructed using the modified LTMLE (with MLE weights or joint calibrated weights), by \tb{performing only the targeting and MSM fitting steps (Step~5--7 above) within the bootstrap samples}. Specifically, Steps 1-4 of the modified LTMLE algorithm are first performed using the \textit{original data}, producing point estimates of the joint calibrated weights
$\bm{W}^{\star}_{a,k}(\widehat{\bm\lambda}^{(a)}_k)$ for $a \in \{1,0\}$ 
and initial predicted outcomes $\bm{\widehat Q}_k^t$, for $t = T,\ldots,0$, $k = t,\ldots,0$. For bootstrap sample \tb{$b$ ($b = 1,..,B$)}, \tb{Steps 5--7} are then performed using the \textit{fixed} point estimates $\bm{W}^{\star}_{a,k}(\widehat{\bm\lambda}^{(a)}_k)$ for $a \in \{1,0\}$ and $\bm{\widehat Q}_k^t$ ($t = T,\ldots,0$, $k = t,\ldots,0$) to obtain the bootstrap estimate $\hat{\bm \beta}_b$ of the MSM parameters. A 95\% CI for $\bm \beta$ is constructed using the $2.5$th and $97.5$th percentile of $\{\widehat{\bm \beta}_b : b = 1,\ldots,B\}$. Corresponding CIs for the mean counterfactual outcomes are constructed similarly using the bootstrap MSM parameter estimates. 

\paragraph{Bootstrap based on modified
LTMLE with calibration performed within each bootstrap sample}
Following \citet{su_sensitivity_2022}, joint calibration can alternatively be incorporated \textit{within each bootstrap sample} to further correct for chance covariate imbalances within the bootstrap samples and possibly improve the resulting bootstrap CIs. In this approach, Steps~1, 3 and~4 of the modified LTMLE algorithm are first performed using the original data. For bootstrap sample \tb{$b$ ($b = 1,...,B$)}, \tb{Step 2 is performed to jointly calibrate the initial MLE weights estimated from the original data, using calibration restrictions applied to the bootstrap sample $b$}. \tb{Steps~5-7 are} subsequently applied using these bootstrap-sample-specific calibrated weights with the fixed initial outcome predictions $\bm{\widehat Q}_k^t$ ($t = T,\ldots,0$, $k = t,\ldots,0$) estimated from the original data. Bootstrap CIs are constructed using the resulting
$\{\widehat{\bm \beta}_b : b = 1,\ldots,B\}$ as described above.


\section{Simulations}\label{simulations}

In this section, we present four simulation studies to evaluate the finite-sample performance of the joint calibrated LTMLE by comparing it with the standard LTMLE using MLE weights across various TTE settings. In the first simulation study, we consider a target trial with two follow-up visits after baseline and no censoring. The second simulation study extends this setting by including censoring in the data-generating mechanism. In the third simulation study, we examine performance under a longer study horizon with nine follow-up visits after the trial baseline. Finally, in the fourth simulation study, we evaluate the empirical coverage of 95\% CIs constructed using the three bootstrap methods presented in \Cref{confidence intervals}: the full nonparametric bootstrap, the bootstrap based on modified LTMLE, and the bootstrap based on modified LTMLE with calibration performed within each bootstrap sample.

\subsection{Design}
\begin{table}[ht!]
\caption{Summary of data-generating mechanism for the simulation studies.}
\label{data simu mechanism}
\centering
\footnotesize
\begin{tabular}{p{0.28\textwidth} p{0.63\textwidth}} 
 \toprule
 \textbf{Data simulation settings} & $n$: number of patients \\
 & $T$: number of follow-up visits; \\
 & $\gamma$: coefficient that describes the confounding strength of time-varying covariates $X_{3,t}, X_{4,t}$ on treatment assignment at visit $t$, $t = 0,1,\ldots, T$\\
 & $\boldsymbol \alpha = (\alpha_0, \alpha_{1,1}, \alpha_{1,0},\ldots,\alpha_{T,1}, \alpha_{T,0})$: intercepts in the treatment models, representing the rate of treatment protocol adherence\\
 & \\
 \textbf{Baseline ($t = 0$)} & \\
 \quad Covariates & $ X_{1,0} = Z_{1,0} $, \\
 & $ X_{2,0} = Z_{2,0}$, \\
 & $X_{3,0} = Z_{3,0}$, \\
 & $X_{4,0} = Z_{4,0}$ \\
 & where $Z_{1,0}, Z_{2,0}, Z_{3,0}, Z_{4,0} \overset{\text{iid}}{\sim} N(0,1)$ \\
 & \\
 \quad Treatment & $A_0 \sim \text{Bernoulli}(p_0)$ \\
 & $\text{logit}(p_0) = \alpha_{0} + 0.5X_{1,0} + 0.5X_{2,0}+\gamma X_{3,0} +\gamma X_{4,0}$ \\
 & \\
 \quad Outcome & $Y_0 = 200 + 5(2A_0 + \sum_{l = 1}^4 X_{l,0})+\epsilon_0 $, where $\epsilon_0 \sim N(0,20)$\\
 & \\
 \quad Censoring & $C_0 \sim \text{Bernoulli}(p^C_0)$ \\
 & $\text{logit}(p^C_0) = -2.5 +0.5X_{1,0} -0.5X_{2,0}+ 0.2X_{3,0} - 0.2X_{4,0}$\\
 \\
 \textbf{Follow-up visits ($t = 1,\ldots,T$)} & \\
 \quad Covariates & $ X_{1,t} = U_1Z_{1,t} $, \\
 & $ X_{2,t} = U_1Z_{2,t}$, \\
 & $X_{3,t} = Z_{3,t} + 0.5\sum_{k = 0}^{t-1} A_k$, \\
 & $X_{4,t} = Z_{4,t} + 0.5\sum_{k = 0}^{t-1} A_k$ \\
 & where $U_1 = 1- 0.3A_{t-1}$, $Z_{1,t}, Z_{2,t}, Z_{3,t}, Z_{4,t} \overset{\text{iid}}{\sim} N(0,1)$ \\
 & \\
 \quad Treatment & $A_t \sim \text{Bernoulli}(p_t)$ \\
 & $\text{logit}(p_t) = \alpha_{t,1}A_{t-1}+ \alpha_{t,0}(1-A_{t-1})+ 0.5X_{1,t} + 0.5X_{2,t} +\gamma X_{3,t} +\gamma X_{4,t}$ \\
 & \\
 \quad Outcome & $Y_t = 200 + 5(2A_t + A_{t-1} + \sum_{l = 1}^4 X_{l,t} + \sum_{l = 1}^4 X_{l,t-1})+\epsilon_t $, where $\epsilon_t \sim N(0,20)$\\
 & \\
 \quad Censoring & $C_t |C_{t-1} = 0\sim \text{Bernoulli}(p^C_t)$ \\
 & $\text{logit}(p^C_t) = -2.5 +0.5X_{1,t} -0.5X_{2,t}+ 0.2X_{3,t} - 0.2X_{4,t}$\\
 & Note that $C_T = 0$ \\
 &\\
 \bottomrule
\end{tabular}
\end{table}

We adapt the data-generating mechanism of \citet{yiu_joint_2022} to the objectives of our simulation studies. The resulting data-generating mechanism is presented in \Cref{data simu mechanism}. \tb{It includes several input variables and parameters that allow us to vary the number of patients and follow-up visits, confounding strength, and the rate of treatment protocol adherence.}

\begin{table}[ht!]
\centering
\caption[Summary of the simulation study scenarios.]{Summary of the simulation study scenarios. Only parameters that vary across scenarios are listed. All other components of the data-generating mechanism are held fixed and described in Table~\ref{data simu mechanism}.}
\label{sim_scenarios}
\footnotesize
\begin{tabular}{p{0.08\textwidth} p{0.10\textwidth} p{0.10\textwidth} p{0.25\textwidth} p{0.33\textwidth}} 
\toprule
\textbf{Study} & \textbf{Follow-ups ($T$)} & \textbf{Censoring} & \textbf{Purpose} & \textbf{DGM parameters ($\gamma,\boldsymbol\alpha$)} \\
\midrule
Study 1 & 2 & No &Baseline comparison of joint calibrated LTMLE vs.\ LTMLE using MLE weights & \textit{Weak confounding:}
\[
\begin{aligned}
\gamma &= 0.2, & \alpha_0 &= 0, \\
\alpha_{1,1} &= 1, & \alpha_{1,0} &= -1.25, \\
\alpha_{2,1} &= 0.8, & \alpha_{2,0} &= -1.25
\end{aligned}
\]
\textit{Strong confounding:}
\[
\begin{aligned}
\gamma &= 5, & \alpha_0 &= 0, \\
\alpha_{1,1} &= 0.1, & \alpha_{1,0} &= -5, \\
\alpha_{2,1} &= -5, & \alpha_{2,0} &= -5.1
\end{aligned}
\]
\\
Study 2 & 2 & Yes &Assessing performance in the presence of censoring & Same as Study 1 \\
\\
Study 3 & 9 & No & Evaluating performance under a longer study horizon& \textit{Weak confounding}: $\gamma = 0.2$, $\alpha_0 = 0$, $\boldsymbol{\alpha_{\text{treat}}} = (\alpha_{1,1}, \alpha_{2,1},\ldots,\alpha_{9,1}) = (1.85,$1.65,1.45$,1.25, 1.05, 0.85$, $0.65$, $0.45, 0.25)$,
$\boldsymbol{\alpha_{\text{control}}}= (\alpha_{1,0}, \alpha_{2,0},\ldots,\alpha_{9,0})$ = $(-2.15, -2.15,\ldots, -2.15)$ \\
&&&& \textit{Strong confounding}: $\gamma = 2.5$, 
$\alpha_0 = 0$, 
$\boldsymbol{\alpha_{\text{treat}}}$ = $(2$,$-0.5$,$-3$,$-5.5$,$-8$,$-10.5$,$-13$, $-15.5$, $-18)$, 
$\boldsymbol{\alpha_{\text{control}}}$ $= (-4.55, -4.55,\ldots, -4.55)$ \\
\\
Study 4 & 2 & No &Evaluating the performance of the 95\% CIs using the three bootstrap methods& Same as Study 1\\
\bottomrule
\end{tabular}
\\[0.5em]
\end{table}

\Cref{sim_scenarios} summarises \tb{the different values of these input variables and parameters that were used to generate data for} the four simulation studies. For each simulation study, we simulate 1,000 datasets with different sample sizes ($n = 300$, $500$, $1000$, $2500$). We assess the performance of the methods under weak or strong time-varying confounding (determined by the value of the parameter $\gamma$\tb{, see Table~\ref{data simu mechanism}}). Scenarios with strong confounding are designed to induce practical near-positivity violations due to poor overlap in covariate distributions. 

The values of $\boldsymbol \alpha$ \tb{(see Table~\ref{data simu mechanism})} are chosen for each simulation study to ensure that, regardless of the confounding strength, the marginal probability of a patient being assigned to a treatment strategy at baseline is equal to 0.5. In addition, they ensure that the marginal probability that a patient adheres to the treatment strategy is approximately the same across follow-up visits regardless of the treatment strategy initiated at baseline (either always-treated or never-treated). Consequently, varying the strength of confounding does not affect the overall proportion of \tb{artificial censoring} arising from treatment protocol non-adherence.

\subsection{Estimation and inference}

The estimand is the parameters $(\beta_0,\beta_1)$ in the MSM $\mathbb{E}(Y^{\overline a_t = \overline a}_t) = \beta_0 + \beta_1\, a \times (t+1) $. The true values of $\beta_0$ and $\beta_1$ are 200 and 10, respectively.

In Simulation Studies~1--3, we estimate $(\beta_0, \beta_1)$ using \textbf{(a)} LTMLE using MLE weights as described in \Cref{LTMLE} and\textbf{ (b)} joint calibrated LTMLE as described in \Cref{calibrated LTMLE}. When estimating the initial MLE weights, correctly specified logistic models are fitted for the treatment and censoring processes. LTMLE is implemented using working models for initial predicted outcomes $\bm{\widehat Q}_k^t$ that include the correct set of time-varying covariates and treatment variables. Details can be found in Section~4 of the Supplementary Materials. For simplicity, we chose $h(a,t,\bm V) = 1, ~ \forall a,t,\bm V$ in all simulation studies. 

For the calibration restrictions, we include an intercept term and time-varying covariates so that $\widetilde{\bm X}_{i,t} = (1,X_{1,t},X_{2,t},X_{3,t}, X_{4,t})$, and, when censoring is included, $\widetilde{\bm H}_{i,t-1} = (1, X_{1,t-1},X_{2,t-1},X_{3,t-1}, X_{4,t-1})$.

We also assess the performance of both LTMLE methods under model misspecification. Specifically, we consider functional-form misspecification \cite{kang_demystifying_2007} by replacing the true covariates $X_{1,k},X_{2,k},X_{3,k}, X_{4,k}$ with transformed covariates $W_{1,k} = X_{1,k}^3/9$, $W_{2,k}=X_{1,k}X_{2,k}$, $W_{3,k}=\log(|X_{3,k}|)+4$ and $W_{4,k} = 1/(1+\exp(X_{4,k}))$ when estimating the initial MLE weights and the initial predicted outcomes $\bm{\widehat Q}_k^t$ \tb{and when} performing the joint calibration. 

We compare the performance of joint calibrated LTMLE with LTMLE using MLE weights by examining the empirical bias, empirical standard error (SE) and root mean squared error (RMSE) of $(\widehat \beta_0, \widehat \beta_1)$ under each method. 

In Simulation Studies~1-3, we additionally evaluate performance for estimating $\mathbb{E}(Y^{\bar 1}_T)$, the mean counterfactual outcome at the end of the trial under the always-treated strategy, as this estimand is often 
of interest in TTE. Note that the corresponding mean counterfactual outcome under the never-treated protocol is $\beta_0$.

In Simulation Study 4, we assess the empirical coverage, average CI width and average computation time of the three types of 95\% CIs for $(\beta_0, \beta_1)$ discussed in \Cref{confidence intervals}, using either MLE weights or joint calibrated weights. We therefore compare the following five CI construction methods:
\begin{enumerate}
 \item[(1)] \tb{Full nonparametric b}ootstrap CIs with LTMLE using MLE weights, re-estimating the MLE weights for each bootstrap sample. 
 \item[(2)] Bootstrap CIs \tb{based on} {modified} LTMLE using MLE weights, fixing the MLE weights and initial predicted outcomes $\bm{\widehat Q}_k^t$ at the values estimated from the original data. 
 \item[(3)] \tb{Full nonparametric b}ootstrap CIs with joint calibrated LTMLE, re-estimating the initial weights and re-calibrating them within each bootstrap sample.
 \item[(4)] Bootstrap CIs \tb{based on} {modified} LTMLE using calibrated weights, fixing the calibrated weights and initial predicted outcomes $\bm{\widehat Q}_k^t$ at the values estimated from the original data.
 \item[(5)] \tb{Bootstrap CIs based on modified LTMLE with calibration performed within each bootstrap sample}, fixing the initial weights and initial predicted outcomes $\bm{\widehat Q}_k^t$ at the values estimated from the original data and re-calibrating the weights within each bootstrap sample. 
\end{enumerate}
 In Simulation Study 4, we focus on the setting with correctly specified models. For all CI methods, we again set $h(a,t,\bm V) = 1, ~ \forall a,t,\bm V$ for simplicity. For each CI construction method, we generate 500 bootstrap samples with replacement, using patients as resampling units.

\subsection{Computational resources}

All simulations were conducted using \texttt{R} (version 4.5.1). The joint calibration procedure was implemented using the \texttt{nleqslv} package. Because the \texttt{ltmle} package does not allow substitution of the MLE weights with joint calibrated weights within the LTMLE algorithm, we implemented the estimation procedure using our own code. The implementation was validated by benchmarking results against the \texttt{ltmle} package output for LTMLE using MLE weights.

We used the \texttt{doParallel} and \texttt{doRNG} packages \cite{daniel_doparallel_2022,gaujoux_dorng_2026} to parallelise the simulations. Simulation Studies~1-3 required approximately 5 hours using 15 cores of the research institution’s high-performance computing cluster, while Simulation Study 4 required approximately 19 hours using 67 cores. All R scripts are available at \url{https://github.com/juliettelimozin/Joint-Calibrated-LTMLE}.

\subsection{Results}

\subsubsection{Simulation 1: Baseline comparison of joint calibrated LTMLE vs.\ LTMLE using MLE weights with two follow-up visits and no censoring}

\begin{table}[ht!]
\centering
\caption[Empirical bias, empirical standard error (SE) and root mean squared error (RMSE) for the two LTMLE estimators of $(\beta_0, \beta_1)$ in Simulation Study 1 with \textbf{\textit{two follow-up visits}} and \textbf{\textit{no censoring}}.]{\small Empirical bias, empirical standard error (SE) and root mean squared error (RMSE) for the two LTMLE estimators of $(\beta_0, \beta_1)$ in Simulation Study 1 with \textbf{\textit{two follow-up visits}} and \textbf{\textit{no censoring}}. MLE: LTMLE using MLE weights; CMLE: joint calibrated LTMLE.} \label{simu 1 results}
\footnotesize
\begin{tabular}{lcccccccccccc}
\toprule
& \multicolumn{6}{c}{Correct covariates} & \multicolumn{6}{c}{Transformed covariates} \\
\cmidrule(lr){2-7} \cmidrule(lr){8-13}
& \multicolumn{2}{c}{Bias} & \multicolumn{2}{c}{SE} & \multicolumn{2}{c}{RMSE} & \multicolumn{2}{c}{Bias} & \multicolumn{2}{c}{SE} & \multicolumn{2}{c}{RMSE} \\
& \multicolumn{2}{c}{$(\beta_0, \beta_1)$} & \multicolumn{2}{c}{$(\beta_0, \beta_1)$} & \multicolumn{2}{c}{$(\beta_0, \beta_1)$} & \multicolumn{2}{c}{$(\beta_0, \beta_1)$} & \multicolumn{2}{c}{$(\beta_0, \beta_1)$} & \multicolumn{2}{c}{$(\beta_0, \beta_1)$} \\
\midrule
\multicolumn{13}{l}{\textbf{Weak confounding}} \\
\multicolumn{13}{l}{$n = 300$} \\
 MLE & 0.00 & 0.05 & 1.47 & 0.96 & 1.47 & 0.96 & -1.75 & 1.56 & 1.52 & 1.03 & 2.31 & 1.86 \\ 
 CMLE & 0.00 & 0.05 & 1.47 & 0.95 & 1.47 & 0.96 & -1.68 & 1.50 & 1.47 & 0.99 & 2.23 & 1.80 \\ 
 \\
\multicolumn{13}{l}{$n = 500$} \\
 MLE & -0.01 & -0.00 & 1.03 & 0.71 & 1.03 & 0.71 & -1.89 & 1.61 & 1.24 & 0.81 & 2.26 & 1.80 \\ 
 CMLE & -0.01 & -0.01 & 1.03 & 0.71 & 1.03 & 0.71 & -1.78 & 1.52 & 1.07 & 0.75 & 2.07 & 1.70 \\ 
\\
\multicolumn{13}{l}{$n = 1000$} \\
MLE & 0.00 & -0.01 & 0.77 & 0.51 & 0.77 & 0.51 & -2.05 & 1.72 & 1.85 & 0.92 & 2.77 & 1.95 \\ 
 CMLE & 0.00 & -0.01 & 0.77 & 0.51 & 0.77 & 0.51 & -1.84 & 1.58 & 1.00 & 0.63 & 2.09 & 1.70 \\
 \\
\multicolumn{13}{l}{$n = 2500$} \\
 MLE & 0.01 & -0.01 & 0.48 & 0.32 & 0.48 & 0.32 & -2.27 & 1.87 & 2.25 & 1.09 & 3.20 & 2.17 \\ 
 CMLE & 0.01 & -0.01 & 0.48 & 0.32 & 0.48 & 0.32 & -1.94 & 1.66 & 0.95 & 0.58 & 2.16 & 1.76 \\
 \\
\midrule
\multicolumn{13}{l}{\textbf{Strong confounding}} \\
\multicolumn{13}{l}{$n = 300$} \\
 MLE & -0.11 & 0.06 & 3.48 & 2.35 & 3.48 & 2.35 & -4.58 & 3.80 & 1.87 & 1.35 & 4.95 & 4.03 \\ 
 CMLE & -0.01 & -0.03 & 3.63 & 2.35 & 3.63 & 2.35 & -4.09 & 3.51 & 1.72 & 1.23 & 4.44 & 3.72 \\ 
 \\
\multicolumn{13}{l}{$n = 500$} \\
 MLE & 0.09 & -0.16 & 3.40 & 2.32 & 3.40 & 2.33 & -4.68 & 3.89 & 1.38 & 1.03 & 4.88 & 4.03 \\ 
 CMLE & 0.17 & -0.14 & 3.15 & 2.12 & 3.16 & 2.13 & -4.25 & 3.62 & 1.30 & 0.99 & 4.44 & 3.75 \\ 
\\
\multicolumn{13}{l}{$n = 1000$} \\
 MLE & 0.09 & -0.02 & 3.27 & 2.32 & 3.27 & 2.32 & -4.68 & 3.85 & 0.97 & 0.72 & 4.78 & 3.92 \\ 
 CMLE & 0.08 & -0.03 & 2.75 & 1.89 & 2.75 & 1.89 & -4.21 & 3.61 & 0.87 & 0.66 & 4.30 & 3.67 \\ 
\\
\multicolumn{13}{l}{$n = 2500$} \\
 MLE & 0.05 & -0.08 & 2.85 & 2.01 & 2.85 & 2.02 & -4.72 & 3.86 & 0.66 & 0.56 & 4.76 & 3.90 \\ 
 CMLE & 0.10 & -0.10 & 2.32 & 1.61 & 2.32 & 1.61 & -4.27 & 3.64 & 0.58 & 0.47 & 4.30 & 3.67 \\ 
\bottomrule
\end{tabular}
\end{table}

\Cref{simu 1 results} summarises the results of Simulation Study 1, where we compare joint calibrated LTMLE with LTMLE using MLE weights in a simple setting with two follow-up visits and no censoring. 

Under weak confounding with correctly specified models, both methods have similar performance in terms of empirical bias, SE, and RMSE for all sample sizes. This is expected; when confounding is mild and models are correctly specified, the MLE weights achieve satisfactory covariate balance, leading to a relatively stable estimation setting. When transformed covariates are used to induce functional-form misspecification, both methods show increased bias and variability. However, joint calibrated LTMLE achieved smaller RMSEs due to simultaneous reductions in bias and SE relative to LTMLE using MLE weights, particularly with larger sample sizes ($n=1000$ and $n=2500$).

Under strong confounding with correctly specified models, both methods show non-negligible bias, and SEs increase noticeably compared to the weak confounding scenario. This increased variability is likely driven by practical near-positivity violations, which generate highly variable weights even under correct specification. Joint calibrated LTMLE generally achieves similar bias to LTMLE using MLE weights, although in a few cases the absolute biases are slightly larger. Importantly, however, joint calibration generally reduces SEs, resulting in smaller RMSEs across nearly all sample sizes.

An exception occurs for the RMSE of $\widehat\beta_0$ at $n=300$, where strong confounding combined with the small sample size likely leads to increased variability in the joint calibrated weights and less stable estimation. However, the absolute bias of joint calibrated LTMLE remains smaller than that of LTMLE using MLE weights. This indicates a \textit{bias-variance trade-off}: calibration introduces additional moment conditions that improve covariate balance and reduce bias, but may increase weight variability, particularly in small samples under strong confounding. Table~3 in the Supplementary Materials illustrates this phenomenon, showing that when $n=300$ under strong confounding, the joint calibrated weights were substantially more variable, reflected by a larger interquartile range relative to the MLE weights.

Under strong confounding with transformed covariates, the advantages of joint calibration become more pronounced. In this setting, joint calibrated LTMLE achieves smaller biases, SEs, and RMSEs than LTMLE using MLE weights. This finding aligns with the expectation that additional covariate-balancing constraints can mitigate the detrimental effects of misspecification in the treatment assignment and outcome models \cite{yiu_joint_2022}.

Table~4 in the Supplementary Materials presents results for the mean counterfactual outcome under the always-treated strategy at the end of the trial, which show patterns consistent with those observed for the MSM parameters. However, we note that the issue of RMSE being larger for joint-calibrated LTMLE at $n = 300$ for correctly specified models and strong confounding no longer arises when estimating the mean counterfactual outcome under the always-treated strategy at the end of the trial.

Taken together, these findings suggest that joint calibration offers limited gains in settings with weak confounding and correctly specified models, but provides clear improvements in efficiency and robustness when confounding is strong or functional-form misspecification is present. 

\subsubsection{Simulation 2: Increasing estimation complexity by including censoring}

\begin{table}[ht!]
\centering
\caption[Empirical bias, empirical standard error (SE) and root mean squared error (RMSE) for LTMLEs of $(\beta_0, \beta_1)$ in Simulation Study 2 with \textbf{\textit{two follow-up visits}} and \textbf{\textit{censoring}}.]{\small Empirical bias, empirical standard error (SE) and root mean squared error (RMSE) for LTMLEs of $(\beta_0, \beta_1)$ in Simulation Study 2 with \textbf{\textit{two follow-up visits}} and \textbf{\textit{censoring}}. MLE: LTMLE using MLE weights; CMLE: joint calibrated LTMLE.} \label{simu 2 results}
\footnotesize
\begin{tabular}{lcccccccccccc}
\toprule
& \multicolumn{6}{c}{Correct covariates} & \multicolumn{6}{c}{Transformed covariates} \\
\cmidrule(lr){2-7} \cmidrule(lr){8-13}
& \multicolumn{2}{c}{Bias} & \multicolumn{2}{c}{SE} & \multicolumn{2}{c}{RMSE} & \multicolumn{2}{c}{Bias} & \multicolumn{2}{c}{SE} & \multicolumn{2}{c}{RMSE} \\
& \multicolumn{2}{c}{$(\beta_0, \beta_1)$} & \multicolumn{2}{c}{$(\beta_0, \beta_1)$} & \multicolumn{2}{c}{$(\beta_0, \beta_1)$} & \multicolumn{2}{c}{$(\beta_0, \beta_1)$} & \multicolumn{2}{c}{$(\beta_0, \beta_1)$} & \multicolumn{2}{c}{$(\beta_0, \beta_1)$} \\
\midrule
\multicolumn{13}{l}{\textbf{Weak confounding}} \\
\multicolumn{13}{l}{$n = 300$} \\
 MLE & 0.00 & 0.04 & 1.53 & 1.03 & 1.53 & 1.03 & -1.73 & 1.54 & 1.64 & 1.11 & 2.38 & 1.90 \\ 
 CMLE & 0.01 & 0.03 & 1.54 & 1.03 & 1.54 & 1.03 & -1.46 & 1.31 & 1.58 & 1.10 & 2.15 & 1.71 \\ 
\\
\multicolumn{13}{l}{$n = 500$} \\ 
 MLE & -0.03 & 0.02 & 1.09 & 0.78 & 1.09 & 0.78 & -1.91 & 1.63 & 1.36 & 0.89 & 2.34 & 1.86 \\ 
 CMLE & -0.02 & 0.01 & 1.08 & 0.78 & 1.09 & 0.78 & -1.57 & 1.36 & 1.14 & 0.82 & 1.94 & 1.58 \\ 
 \\
\multicolumn{13}{l}{$n = 1000$} \\
 MLE & -0.00 & -0.00 & 0.81 & 0.55 & 0.81 & 0.55 & -2.03 & 1.70 & 1.56 & 0.84 & 2.56 & 1.90 \\ 
 CMLE & -0.00 & -0.00 & 0.81 & 0.54 & 0.81 & 0.54 & -1.58 & 1.38 & 0.85 & 0.58 & 1.80 & 1.50 \\ 
\\
\multicolumn{13}{l}{$n = 2500$} \\
 MLE & 0.01 & -0.01 & 0.51 & 0.35 & 0.51 & 0.35 & -2.50 & 1.97 & 2.78 & 1.29 & 3.74 & 2.36 \\ 
 CMLE & 0.01 & -0.02 & 0.51 & 0.35 & 0.51 & 0.35 & -1.65 & 1.42 & 0.72 & 0.44 & 1.81 & 1.49 \\ 
 \\
\midrule
\multicolumn{13}{l}{\textbf{Strong confounding}}\\
\multicolumn{13}{l}{$n = 300$} \\
 MLE & -0.10 & 0.02 & 3.58 & 2.42 & 3.58 & 2.42 & -4.51 & 3.77 & 1.99 & 1.43 & 4.93 & 4.03 \\ 
 CMLE & 0.08 & -0.05 & 4.28 & 2.91 & 4.28 & 2.91 & -3.45 & 3.10 & 2.12 & 1.42 & 4.05 & 3.41 \\ 
 \\
\multicolumn{13}{l}{$n = 500$} \\
 MLE & 0.12 & -0.18 & 3.49 & 2.43 & 3.49 & 2.43 & -4.67 & 3.90 & 1.48 & 1.13 & 4.89 & 4.06 \\ 
 CMLE & 0.26 & -0.24 & 3.58 & 2.44 & 3.59 & 2.45 & -3.68 & 3.25 & 1.52 & 1.09 & 3.98 & 3.43 \\ 
\\
\multicolumn{13}{l}{$n = 1000$} \\
 MLE & 0.05 & 0.00 & 3.33 & 2.34 & 3.33 & 2.34 & -4.68 & 3.86 & 1.04 & 0.80 & 4.79 & 3.94 \\ 
 CMLE & 0.03 & -0.00 & 2.92 & 2.01 & 2.92 & 2.01 & -3.67 & 3.28 & 1.04 & 0.74 & 3.81 & 3.36 \\ 
\\
\multicolumn{13}{l}{$n = 2500$} \\
 MLE & 0.08 & -0.05 & 2.84 & 2.07 & 2.84 & 2.07 & -4.77 & 3.88 & 0.82 & 0.72 & 4.84 & 3.94 \\ 
 CMLE & 0.11 & -0.09 & 2.38 & 1.70 & 2.38 & 1.71 & -3.76 & 3.34 & 0.70 & 0.51 & 3.82 & 3.38 \\ 
\\
\bottomrule
\end{tabular}
\end{table}

\Cref{simu 2 results} summarises the results of Simulation Study 2, where we increase estimation complexity relative to Study 1 by including censoring in the data-generating mechanism. With the additional censoring, we notice a general pattern of biases and SEs \tb{being larger} for all scenarios and sample sizes compared to Simulation Study 1. This reflects the additional variability introduced by modelling the censoring process and incorporating it into the construction of the weights, which affects both LTMLE using MLE weights and joint calibrated LTMLE.

Under weak confounding with correctly specified models, both methods again demonstrate similar performance in terms of bias, SE and RMSE. With misspecified models, we once again observe increased bias, SE and RMSE for both methods; however, joint calibrated LTMLE achieves smaller RMSE across all sample sizes due to reductions in bias and SE relative to LTMLE using MLE weights. 

Under strong confounding with correctly specified models, the bias-variance trade-off induced by the calibration restrictions becomes more pronounced than in Study 1. In the absence of censoring (Study 1), joint calibrated LTMLE had larger RMSE only at $n = 300$, with better RMSE emerging from $n=500$ and onwards.
However, when censoring is present, the higher variability introduced by the additional calibration restrictions persists for longer: at $n=500$, joint calibrated LTMLE still shows larger RMSE compared to LTMLE using MLE weights due to higher variability as well as larger bias, and only from a sample size of $n=1000$ onwards \tb{do} we begin to observe the anticipated efficiency gains from joint calibration. 

This \tb{finding} highlights that, in the presence of censoring, larger sample sizes \tb{may be} required for the \tb{benefits of improved covariate balance and bias reduction associated with additional calibration restrictions for censoring to outweigh any variance inflation induced by these additional calibration restrictions}. At the same time, larger samples are also necessary to ensure that the initial MLE weights are estimated with sufficient accuracy so that they do not themselves induce excess finite-sample bias. Consistent with Study 1, Table~5 in the Supplementary Materials shows that, with correct model specification, strong confounding and smaller sample sizes ($n=300, 500$), joint calibrated weights have noticeably higher variability, as manifested by the wider interquartile range\tb{, than} the MLE weights.

Under strong confounding with transformed covariates, the advantages of joint calibration are again evident: joint calibrated LTMLE
has smaller biases, SEs, and RMSEs compared to LTMLE using MLE weights across all sample sizes. 

As in Simulation Study 1, the results for the mean counterfactual outcome at the end of the trial under the always-treated strategy in Simulation Study 2 (Table~6 in the Supplementary Materials) show trends that align with those seen for the MSM parameters.

Overall, Simulation Study 2 demonstrates that, while joint calibration continues to provide reductions in RMSE, \tb{under strong confounding and} particularly under functional-form misspecification, the presence of censoring inflates the weight variability and requires a larger sample size at which the gains from joint calibration become apparent.


 \subsubsection{Simulation 3: Evaluating performance under a longer study horizon}

\begin{table}[ht!]
\centering
\caption[Empirical bias, empirical standard error (SE) and root mean squared error (RMSE) for LTMLEs of $(\beta_0, \beta_1)$ in Simulation Study 3 with \textit{\textbf{nine follow-up visits}} and \textit{\textbf{no censoring}}.]{\small Empirical bias, empirical standard error (SE) and root mean squared error (RMSE) for LTMLEs of $(\beta_0, \beta_1)$ in Simulation Study 3 with \textit{\textbf{nine follow-up visits}} and \textit{\textbf{no censoring}}. MLE: LTMLE using MLE weights; CMLE: joint calibrated LTMLE.} \label{simu 3 results}
\footnotesize
\begin{tabular}{lcccccccccccc}
\toprule
& \multicolumn{6}{c}{Correct covariates} & \multicolumn{6}{c}{Transformed covariates} \\
\cmidrule(lr){2-7} \cmidrule(lr){8-13}
& \multicolumn{2}{c}{Bias} & \multicolumn{2}{c}{SE} & \multicolumn{2}{c}{RMSE} & \multicolumn{2}{c}{Bias} & \multicolumn{2}{c}{SE} & \multicolumn{2}{c}{RMSE} \\
& \multicolumn{2}{c}{$(\beta_0, \beta_1)$} & \multicolumn{2}{c}{$(\beta_0, \beta_1)$} & \multicolumn{2}{c}{$(\beta_0, \beta_1)$} & \multicolumn{2}{c}{$(\beta_0, \beta_1)$} & \multicolumn{2}{c}{$(\beta_0, \beta_1)$} & \multicolumn{2}{c}{$(\beta_0, \beta_1)$} \\
\midrule
\multicolumn{13}{l}{\textbf{Weak confounding}} \\
\multicolumn{13}{l}{$n = 300$} \\
MLE & 0.01 & -0.01 & 1.00 & 0.25 & 1.00 & 0.25 & -0.91 & 0.23 & 1.05 & 0.25 & 1.39 & 0.34 \\ 
CMLE & 0.01 & -0.01 & 0.99 & 0.25 & 0.99 & 0.25 & -0.79 & 0.19 & 1.04 & 0.25 & 1.31 & 0.31 \\
 \\
\multicolumn{13}{l}{$n = 500$} \\
MLE & 0.02 & -0.01 & 0.71 & 0.18 & 0.71 & 0.18 & -0.93 & 0.24 & 0.90 & 0.21 & 1.30 & 0.31 \\ 
CMLE & 0.02 & -0.01 & 0.72 & 0.19 & 0.72 & 0.19 & -0.81 & 0.20 & 0.83 & 0.20 & 1.16 & 0.28 \\ 
\\
\multicolumn{13}{l}{$n = 1000$} \\
 MLE & 0.00 & 0.00 & 0.51 & 0.13 & 0.51 & 0.13 & -1.03 & 0.26 & 0.83 & 0.18 & 1.32 & 0.32 \\ 
 CMLE & 0.00 & 0.00 & 0.51 & 0.13 & 0.51 & 0.13 & -0.88 & 0.22 & 0.65 & 0.17 & 1.09 & 0.28 \\ 
\\
\multicolumn{13}{l}{$n = 2500$} \\
 MLE & -0.01 & -0.00 & 0.31 & 0.08 & 0.31 & 0.08 & -1.22 & 0.29 & 1.11 & 0.17 & 1.65 & 0.33 \\ 
 CMLE & -0.00 & -0.00 & 0.31 & 0.08 & 0.31 & 0.08 & -0.96 & 0.23 & 0.69 & 0.14 & 1.19 & 0.27 \\ 
 \\
\midrule
\multicolumn{13}{l}{\textbf{Strong confounding}} \\
\multicolumn{13}{l}{$n = 300$} \\
MLE & 0.06 & 0.00 & 2.11 & 0.55 & 2.11 & 0.55 & -3.84 & 0.76 & 2.01 & 0.47 & 4.33 & 0.89 \\ 
 CMLE & 0.09 & -0.00 & 1.93 & 0.50 & 1.93 & 0.50 & -2.70 & 0.52 & 1.59 & 0.41 & 3.13 & 0.66 \\ 
\\
\multicolumn{13}{l}{$n = 500$} \\
MLE & -0.01 & 0.02 & 1.96 & 0.49 & 1.96 & 0.49 & -3.93 & 0.81 & 1.83 & 0.43 & 4.34 & 0.92 \\ 
CMLE & 0.00 & 0.00 & 1.71 & 0.44 & 1.71 & 0.44 & -2.72 & 0.55 & 1.41 & 0.37 & 3.07 & 0.66 \\ 
\\
\multicolumn{13}{l}{$n = 1000$} \\
MLE & 0.04 & -0.01 & 1.75 & 0.43 & 1.75 & 0.43 & -4.32 & 0.86 & 1.57 & 0.44 & 4.59 & 0.97 \\ 
CMLE & 0.05 & -0.01 & 1.53 & 0.38 & 1.53 & 0.39 & -3.04 & 0.60 & 1.18 & 0.37 & 3.26 & 0.71 \\ 
\\
\multicolumn{13}{l}{$n = 2500$} \\
MLE & -0.04 & 0.00 & 1.28 & 0.37 & 1.28 & 0.37 & -4.71 & 0.94 & 1.37 & 0.44 & 4.90 & 1.04 \\ 
 CMLE & -0.02 & -0.00 & 1.12 & 0.33 & 1.12 & 0.33 & -3.31 & 0.68 & 0.95 & 0.35 & 3.44 & 0.76 \\
\\
\bottomrule
\end{tabular}
\end{table}

 \Cref{simu 3 results} summarises the results from Simulation Study 3, which extends a longer study horizon to nine follow-up visits in order to reflect practical challenges common in epidemiological studies \cite{schomaker_using_2019}. Censoring was not included, so that the impact of a longer study horizon on treatment weight estimation and subsequent calibration could be examined in isolation.
 
 Across nearly all scenarios, both empirical bias and SE are smaller than in Simulation Study 1. This improvement is likely due to the increased number of observation times: with more follow-up visits, the treatment assignment model is informed by a larger set of data points, leading to more accurate estimation of the treatment mechanism and, consequently, more stable MLE weights. 
 
 \tb{Comparing Study 3's results to Study 1, which had a shorter study horizon, we notice that although} the calibration restrictions are applied sequentially from the trial baseline, we do not observe that the variability introduced at earlier visits propagates over time and degrades performance. 
 Instead, the longer study horizon appears primarily to improve the initial weight estimation rather than amplify instability in finite samples, which results in improved overall performance for joint calibrated LTMLE as well. 

Similarly to Study 1, under weak confounding with correctly specified models, both methods again demonstrate nearly identical performance. Under strong confounding with correct specification, we no longer observe the large variability of joint calibrated LTMLE in small sample sizes ($n = 300$) as in Study 1, likely due to the improved initial weight estimation as discussed above. Joint calibrated LTMLE consistently achieved negligible bias and smaller SE than LTMLE using MLE weights, leading to lower RMSE across all sample sizes.

Under both weak and strong confounding with functional-form misspecification, joint calibrated LTMLE again outperforms LTMLE using MLE weights as in Studies 1 and 2, with smaller biases, SEs, and RMSEs across all sample sizes. 

As in the previous simulation studies, the results for the mean counterfactual outcome at the end of the trial under the always-treated strategy show similar patterns with those observed for the MSM parameters (Table~7 in the Supplementary Materials).

Overall, extending the study horizon improves efficiency for both methods by improving the estimation of the initial MLE weights through increased information. Joint calibrated LTMLE consistently performs as well as or better than LTMLE using MLE weights in all scenarios, and provides clear advantages under functional-form misspecification, highlighting that calibration remains beneficial even when the MLE weights are relatively stable. 

\subsubsection{Simulation 4: Evaluating the performance of the 95\% CIs using the three bootstrap methods}

\begin{table}[ht!]
\centering
\caption[Empirical coverage, average CI width and average computation time (in seconds) of confidence intervals of $(\beta_0, \beta_1)$ in Simulation Study 4 with \textbf{\textit{two follow-up visits}} and \textbf{\textit{no censoring}}.]{\small Empirical coverage, average CI width and average computation time (in seconds) of confidence intervals of $(\beta_0, \beta_1)$ in Simulation Study 4 with \textbf{\textit{two follow-up visits}} and \textbf{\textit{no censoring}}. \textit{MLE Full boot.}: \tb{Full nonparametric bootstrap CIs with LTMLE using MLE weights, re-estimating the MLE weights for
each bootstrap sample}; \textit{MLE Modified.boot.}: Bootstrap CIs \tb{based on} \textit{modified} LTMLE using MLE weights, fixing the MLE weights and initial predicted outcomes $\bm{\widehat Q}_k^t$ at the values estimated from the original data; \textit{CMLE Full boot.}: \tb{Full nonparametric bootstrap} CIs with joint calibrated LTMLE, re-estimating the initial weights and re-calibrating them within each bootstrap sample; \textit{CMLE Modified boot.}: Bootstrap CIs \tb{based on} \textit{modified} LTMLE using calibrated weights, fixing the calibrated weights and initial predicted outcomes $\bm{\widehat Q}_k^t$ at the values estimated from the original data; \textit{CMLE Modified boot. with cali.}: Bootstrap CIs \tb{based on} \textit{modified} LTMLE \tb{with calibration performed within each bootstrap sample}, fixing the initial weights and initial predicted outcomes $\bm{\widehat Q}_k^t$ at the values estimated from the original data and re-calibrating the weights within each bootstrap sample.} \label{simu 4 results}
{\footnotesize
\begin{tabular}{lcccccccccc}
\toprule
& \multicolumn{5}{c}{\textbf{Weak confounding}} & \multicolumn{5}{c}{\textbf{Strong confounding}} \\
\cmidrule(lr){2-6} \cmidrule(lr){7-11}
& \multicolumn{2}{c}{Coverage} & \multicolumn{2}{c}{Width} & Computation & \multicolumn{2}{c}{Coverage} & \multicolumn{2}{c}{Width} & \multicolumn{1}{c}{Computation} \\
& \multicolumn{2}{c}{$(\beta_0, \beta_1)$} & \multicolumn{2}{c}{$(\beta_0, \beta_1)$} & time & \multicolumn{2}{c}{$(\beta_0, \beta_1)$} & \multicolumn{2}{c}{$(\beta_0, \beta_1)$} & time\\
\midrule
\multicolumn{11}{l}{$n = 300$} \\
MLE Full boot. & 0.93 & 0.93 & 5.31 & 3.58 & 40.74 & 0.94 & 0.94 & 9.35 & 6.86 & 45.47 \\ 
 MLE Modified.boot. & 0.93 & 0.93 & 5.28 & 3.54 & 17.80 & 0.90 & 0.89 & 8.19 & 5.87 & 19.34 \\ 
 CMLE Full boot. & 0.93 & 0.93 & 5.30 & 3.56 & 176.96 & 0.95 & 0.96 & 12.12 & 8.29 & 297.62 \\ 
 CMLE Modified boot. & 0.93 & 0.93 & 5.28 & 3.54 & 39.84 & 0.92 & 0.92 & 10.48 & 7.05 & 52.81 \\ 
 CMLE Modified boot. & 0.93 & 0.93 & 5.27 & 3.53 & 137.26 & 0.95 & 0.96 & 11.62 & 7.79 & 244.19 \\
 with cali & & & & & & & & & & \\
 \\
\multicolumn{6}{l}{$n = 500$} \\
MLE Full boot. & 0.94 & 0.95 & 4.10 & 2.76 & 45.63 & 0.93 & 0.93 & 8.30 & 6.03 & 48.38 \\ 
 MLE Modified.boot. & 0.95 & 0.95 & 4.09 & 2.75 & 20.05 & 0.91 & 0.91 & 7.55 & 5.39 & 20.72 \\ 
 CMLE Full boot. & 0.95 & 0.95 & 4.08 & 2.75 & 212.07 & 0.93 & 0.94 & 9.50 & 6.45 & 260.28 \\ 
 CMLE Modified boot. & 0.95 & 0.94 & 4.09 & 2.74 & 45.39 & 0.91 & 0.90 & 9.03 & 6.06 & 47.34 \\ 
 CMLE Modified boot. & 0.95 & 0.94 & 4.07 & 2.74 & 169.28 & 0.93 & 0.93 & 9.10 & 6.09 & 213.93 \\ 
 with cali & & & & & & & & & & \\
\\
\multicolumn{6}{l}{$n = 1000$} \\
MLE Full boot. & 0.94 & 0.95 & 2.90 & 1.94 & 63.30 & 0.92 & 0.93 & 7.25 & 5.34 & 67.59 \\ 
 MLE Modified.boot. & 0.94 & 0.95 & 2.90 & 1.94 & 28.56 & 0.92 & 0.91 & 6.86 & 5.02 & 29.67 \\ 
 CMLE Full boot. & 0.94 & 0.95 & 2.89 & 1.94 & 330.31 & 0.92 & 0.92 & 7.46 & 5.18 & 357.26 \\ 
 CMLE Modified boot. & 0.94 & 0.96 & 2.90 & 1.94 & 63.17 & 0.92 & 0.92 & 7.70 & 5.29 & 60.23 \\ 
 CMLE Modified boot. & 0.93 & 0.96 & 2.89 & 1.93 & 271.04 & 0.91 & 0.92 & 7.22 & 4.97 & 298.22 \\ 
 with cali & & & & & & & & & & \\
\\
\multicolumn{6}{l}{$n = 2500$} \\
MLE Full boot. & 0.94 & 0.94 & 1.82 & 1.23 & 135.27 & 0.93 & 0.91 & 6.12 & 4.47 & 137.56 \\ 
 MLE Modified.boot. & 0.94 & 0.94 & 1.82 & 1.23 & 65.26 & 0.93 & 0.91 & 5.98 & 4.34 & 64.18 \\ 
 CMLE Full boot. & 0.94 & 0.94 & 1.82 & 1.23 & 782.36 & 0.93 & 0.92 & 6.00 & 4.18 & 780.28 \\ 
 CMLE Modified boot. & 0.94 & 0.94 & 1.82 & 1.23 & 145.50 & 0.93 & 0.92 & 6.50 & 4.48 & 136.19 \\ 
 CMLE Modified boot. & 0.94 & 0.94 & 1.82 & 1.23 & 662.88 & 0.93 & 0.91 & 5.89 & 4.08 & 666.46 \\
 with cali & & & & & & & & & & \\
 \bottomrule
\end{tabular}}
\end{table}

\Cref{simu 4 results} summarises the results from Simulation Study 4, which compares the bootstrap CI methods discussed in \Cref{confidence intervals} when either MLE weights or joint calibrated weights are used. The data-generating mechanism is the same as in Simulation Study 1. The Monte Carlo standard errors of the estimated coverage rates, computed as $\text{SE}_{MC} = \sqrt{\frac{\widehat{\text{Coverage}}(1-\widehat{\text{Coverage}})}{N_{\text{sim}}}},$ ranged between 0.6\% and 1\% across all scenarios \cite{morris_using_2019}.

Under weak confounding, all CI construction methods perform similarly across all sample sizes. Empirical coverage remains close to the nominal 95\% level, and the average CI widths are nearly indistinguishable, indicating no systematic differences in accuracy or efficiency between methods in this setting. 

Under strong confounding, the CI coverage is slightly reduced overall, particularly for smaller sample sizes $n = 300,500$ and for the bootstrap CIs based on modified LTMLE with fixed weights \tb{and initial outcome predictions} (``MLE Modified boot.'' and ``CMLE Modified boot.''). This is likely driven by two factors: (1) large finite-sample bias in the point estimates under strong confounding (as seen in Simulation Study 1), and (2) the CI methods with fixed weights being unable to correct 
more chance covariate imbalances arising from strong confounding within the bootstrap samples. When
weights are held fixed, any imbalance in a given bootstrap sample is not re-adjusted, which can induce estimation error and reduce coverage.

In contrast, bootstrap CIs that re-estimate weights in each bootstrap sample show slightly better coverage. For \tb{full nonparametric} bootstrap CIs with LTMLE using MLE weights applied to each bootstrap sample (``MLE Full boot.''), re-estimation partially improves the imbalance despite no calibration applied. 
For CI methods that calibrate the weights in each bootstrap sample, either the full
re-estimation procedure (``CMLE Full boot.'') or modified LTMLE with
re-calibration in each bootstrap sample (``CMLE Modified boot. with
cali.''), coverage is consistently close to nominal. 
Their similar performance aligns with the findings in \citet{su_sensitivity_2022}, who also found that fixing the initial MLE weights has minimal impact on bootstrap CIs once joint calibration is applied. Notably, for $n = 300$, these CIs have better coverage compared to the bootstrap CIs using LTMLE using MLE weights, highlighting the benefits of joint calibration under strong confounding. For larger sample sizes $n = 1000$ and $2500$, all CI methods demonstrate similar performance in terms of coverage and average CI width.

Focusing on the bootstrap CIs with joint calibration, the biggest difference among them lies in computational burden. The \tb{full nonparametric} bootstrap CIs re-estimating and re-calibrating weights in every bootstrap sample(``CMLE Full boot.'') is the most computationally intensive. In contrast, the bootstrap CIs based on modified LTMLE with fixed calibrated weights (``CMLE Modified boot.'') requires noticeably less computation time on average, 
while the bootstrap CIs based on modified LTMLE with fixed initial weights but recalibrating them within each bootstrap sample (``CMLE Modified boot. with cali.'') offer an intermediate computational burden. When computational efficiency is a practical concern, the latter method may provide an attractive compromise by offering close-to-nominal coverage and better computational efficiency relative to CIs based on full \tb{nonparametric} bootstrap.

 \subsection{Summary}
Across Simulation Studies 1-3, joint calibrated LTMLE performed comparably to LTMLE using MLE weights under weak confounding with correctly specified models. Under strong confounding and correct model specification, joint calibrated LTMLE generally achieved lower RMSE through reduced variability, demonstrating the efficiency gains that can be obtained from improved covariate balance via calibration. Some exceptions occurred in small samples, where additional calibration restrictions introduced more variability while reducing bias, highlighting a bias-variance trade-off in small samples. 

Under functional-form misspecification of confounding covariates, joint calibrated LTMLE had noticeably smaller bias and variability across all scenarios and sample sizes, demonstrating that the joint calibrated weights add robustness to LTMLE by mitigating the detrimental effects of misspecification. 

Simulation Study 4 showed that the two bootstrap CI methods incorporating joint calibration within each bootstrap sample provided empirical coverage close to the nominal level across all sample sizes. In contrast, the bootstrap CIs based on modified LTMLE with fixed calibrated weights were more sensitive to strong confounding, particularly in small samples. Given that the full \tb{nonparametric} bootstrap CIs with joint calibrated LTMLE and the bootstrap CIs with modified LTMLE and re-calibration in each bootstrap sample achieved similar coverage and interval widths, the latter may be preferred in practice due to its better computational efficiency.

\section{Application to the HERS data}\label{application}

In this section, we apply the proposed joint calibrated LTMLE to the HERS data. Our objective is to estimate the per-protocol effect of HAART use on CD4 cell count over time by characterising the mean counterfactual CD4 cell counts under the always-treated and never-treated strategies using the working MSM in~\Cref{HERS msm}. We estimate the model parameters using four approaches:
\begin{enumerate}
 \item[(1)] Standard LTMLE \tb{with} stabilised MLE weights, \tb{using parametric models for the stabilised weights and the initial predicted outcomes $\bm{\widehat Q}_k^t$};
 \item[(2)] Joint calibrated LTMLE with parametric models for the initial stabilised MLE weights and the initial predicted outcomes $\bm{\widehat Q}_k^t$;
 \item[(3)] \tb{Standard LTMLE with stabilised weights, using data-adaptive estimation via SuperLearner for the stabilised weights and the initial predicted outcomes $\bm{\widehat Q}_k^t$.}
 \item[(4)] Joint calibrated LTMLE with data-adaptive estimation via SuperLearner for the initial stabilised weights and the initial predicted outcomes $\bm{\widehat Q}_k^t$.
\end{enumerate}

We include the analysis based on SuperLearner to reflect its widespread use in the LTMLE literature \cite{schomaker_using_2019} and its integration within existing implementations of LTMLE (e.g., the \texttt{ltmle} package). Importantly, SuperLearner is fully compatible with the proposed joint calibration approach, as discussed further in \Cref{discussion}. As in the simulation studies, we implemented the LTMLE algorithm manually in \texttt{R} to incorporate joint calibrated weights, and used the \texttt{SuperLearner} package \cite{polley_superlearner_2025} to implement SuperLearner.

\subsection{Model specifications}
 We set $h(a,t,\bm V)$ equal to the numerator of stabilised MLE weights, where $\bm V$ is described below. 
\paragraph{Treatment models.} The denominator of the stabilised inverse probability of treatment weights was estimated via a logistic regression model for the treatment process. The model included the previous visit's HAART treatment status, ethnicity and study site. Because HAART use was self-reported for the interval between the previous and current visits, we also included CD4 cell count (after square-root-transformation and standardisation) and HIV viral load (after $\log_{10}$-transformation and standardisation) measured at the previous two visits, as well as HIV symptoms \tb{(coded as a categorical variable with 6 symptom levels)} at the previous visit. For the numerator of the stabilised treatment weights, a logistic regression model including only the previous visit's HAART status and the baseline CD4 cell count stratum variable $G$ \tb{(which is also included as a covariate in the MSM, see Equation~\eqref{HERS msm})} was specified.

\paragraph{Censoring models.}
The denominator of the stabilised inverse probability of censoring weights was estimated using a logistic regression model for the censoring process. The model 
included the current and previous visits' CD4 cell count, HIV viral load and HIV symptoms, as well as ethnicity and study site. CD4 cell count and HIV viral load included in the censoring models were also transformed as described for the treatment process models. 
For the numerator of the stabilised censoring weights, we specified a logistic regression model including only the baseline CD4 cell count stratum variable $G$.

\paragraph{Joint calibration.} The joint calibrated weights were obtained by calibrating the stabilised MLE weights using the method described in \Cref{implementation calibration}. For artificial censoring due to treatment protocol non-adherence, the calibration restrictions included the same covariates as in the denominator model for treatment assignment (excluding the previous visit's HAART status), together with an intercept term. For censoring due to loss to follow-up, the calibration restrictions included the same covariates as in the denominator model for the censoring process, along with an intercept term.

\paragraph{Outcome regression.} The initial predicted outcomes $\bm{\widehat Q}_k^t$ in the LTMLE algorithm were estimated using quasi-binomial logistic regression models that included the following covariates: the current and previous two visits' HAART status; the transformed CD4 cell count at the previous two visits; the transformed HIV viral load and HIV symptoms at the current and previous two visits; ethnicity and study site.
\paragraph{SuperLearner implementation.} 
For SuperLearner implementation, we included the same sets of covariates as stated above for the corresponding models, and specified the following candidate learners in the \texttt{SuperLearner} package: the arithmetic mean (\texttt{SL.mean}), generalised linear models (GLMs) with all main effect terms (\texttt{SL.glm}), GLMs including pairwise interactions (\texttt{SL.glm.interaction}), GLMs chosen by stepwise variable selection using AIC (\texttt{SL.step}), Bayesian GLMs (\texttt{SL.bayesglm}), and generalised additive models (\texttt{SL.gam}). 

\paragraph{Bootstrap confidence intervals.} We constructed 95\% bootstrap CIs of the MSM parameters and per-protocol effects over time (defined as the differences in mean counterfactual outcomes over time) using 500 bootstrap samples. We considered two approaches: (1) a full nonparametric bootstrap for LTMLE using MLE weights, re-estimating the MLE weights within each bootstrap sample, and (2) bootstrap based on modified LTMLE, in which the initial MLE weights and initial predicted outcomes $\bm{\widehat Q}_k^t$ were fixed at those values estimated from the original data and the calibration step was repeated within each bootstrap sample. We chose these two bootstrap CI approaches because the sample size of the trial-eligible cohort in the HERS data was close to $n =500$, for which Simulation Study 4 indicated that these two methods achieved better coverage. \tb{For the two standard LTMLE approaches, we calculated CIs using the first CI approach. For the two joint calibrated LTMLE approaches, we employed the second CI approach.}

\subsection{Results}
\begin{table}[ht]
\centering
\caption[Estimates and 95\% CIs of the MSM coefficients for the cumulative effect of sustained HAART treatment, stratified by initial CD4 cell count at visit 7 in the HERS data.]{Estimates and 95\% CIs of the MSM coefficients for the cumulative effect of sustained HAART treatment, stratified by initial CD4 cell count at visit 7 in the HERS data. MLE: LTMLE using MLE weights estimated via parametric models; CMLE: joint calibrated LTMLE with parametric models; MLE-SL: \tb{LTMLE using weights estimated via SuperLearner; CMLE-SL: joint calibrated LTMLE with SuperLearner.}} \label{HERS msm results}
\begin{tabular}{llll}
 \toprule
 & \multicolumn{3}{c}{Strata by CD4 cell count at visit 7}\\ 
 \cmidrule(lr){2-4} 
 &\multicolumn{1}{c}{$ < 200$} & \multicolumn{1}{c}{$200-500$} & \multicolumn{1}{c}{$> 500$} \\ 
 \midrule
MLE & 24.46 (9.08, 43.14) & 12.13 (-12.67, 38.98) & -6.43 (-36.29, 28.80) \\
 CMLE & 22.13 (4.07, 45.71) & 13.96 (-7.59, 38.97) & -13.85 (-47.83, 22.79) \\
 MLE-SL & 29.70 (13.48, 47.87) & 27.69 (-6.01, 56.85) & -47.03 (-84.00, 0.10) \\ 
 CMLE-SL & 35.13 (14.64, 51.69) & 32.08 (-0.31, 59.70) & -35.06 (-77.40, 8.10) \\ 
 \bottomrule
\end{tabular}
\end{table}

\Cref{HERS msm results} presents the estimated MSM coefficients and corresponding 95\% CIs obtained \tb{from the four approaches considered}. 
\Cref{HERS ATE plot} presents the estimated per-protocol treatment effect (i.e. the difference in mean counterfactual CD4 cell count under the always-treated versus never-treated strategy) over visits 8-12, stratified by initial CD4 cell count groups at visit 7. 

\begin{figure}[ht!]
 \centering
 \includegraphics[width=0.85\linewidth]{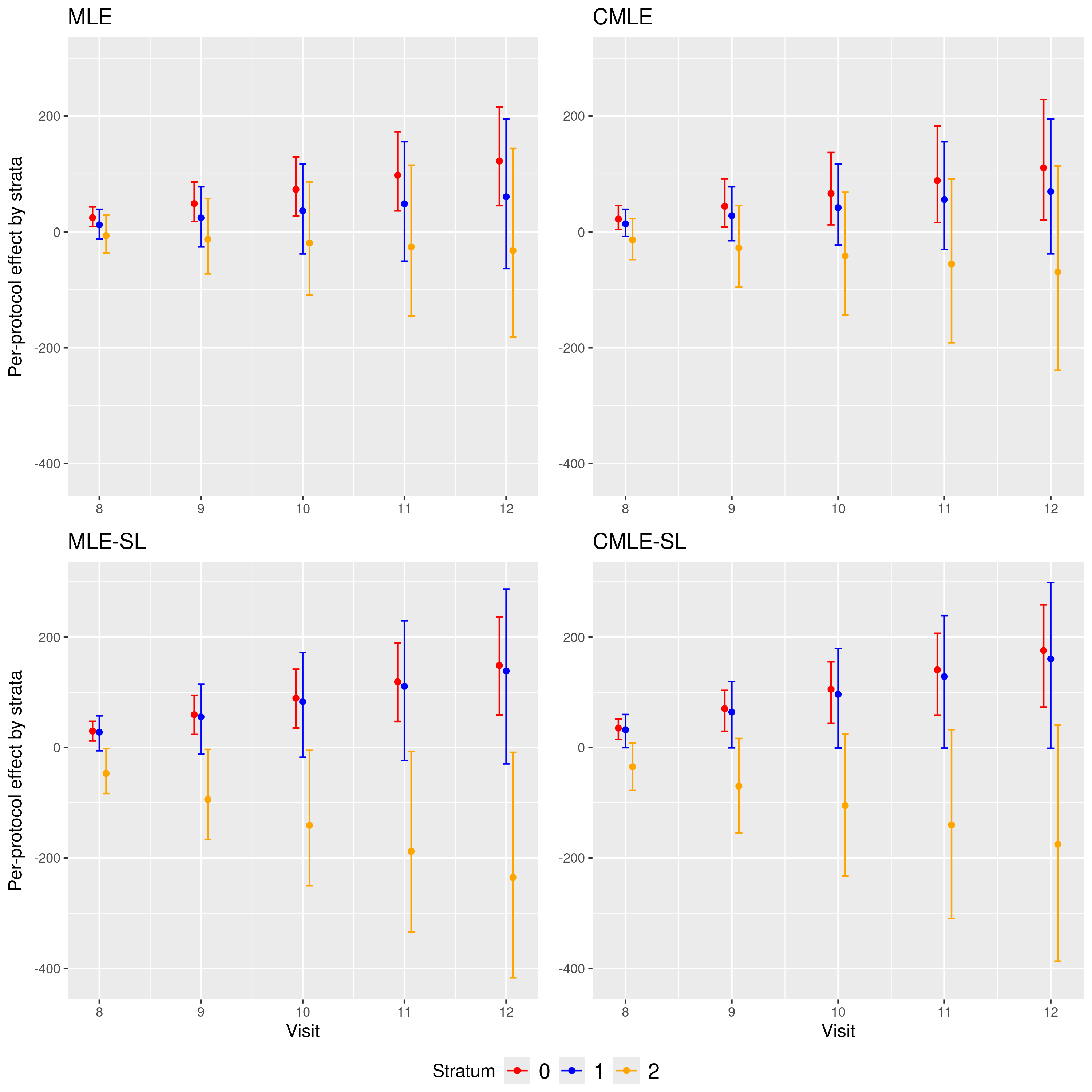}
 \caption[The estimates and 95\% CIs of per-protocol effect of HAART treatment on CD4 cell count over visits 8-12, stratified by initial CD4 cell count at visit 7 in the HERS data example.]{The estimates and 95\% CIs of per-protocol effect of HAART treatment on CD4 cell count over visits 8-12, stratified by initial CD4 cell count at visit 7 in the HERS data example. MLE: LTMLE using MLE weights estimated via parametric models; CMLE: joint calibrated LTMLE with parametric models; \tb{MLE-SL: LTMLE using MLE weights estimated by SuperLearner;} CMLE-SL: joint calibrated LTMLE with SuperLearner; Stratum 0: initial CD4 cell count below 200; Stratum 1: initial CD4 cell count between 200 and 500; Stratum 2: initial CD4 cell count above 500.} \label{HERS ATE plot}
\end{figure}

Although most CIs include 0 and are therefore not statistically significant at the 5\% level, the estimated effects are generally consistent with previous studies \cite{ko_estimating_2003, yiu_joint_2022}, suggesting that cumulative exposure to HAART may be beneficial for patients with an initial CD4 cell count below 500. 
\Cref{HERS ATE plot} shows positive per-protocol treatment \tb{effects} across visits 8-12 for patients with initial CD4 cell counts under 500, while the estimated effects are negative but \tb{non-significant} across visits for patients with initial CD4 cell counts above 500. 

We note that, for the always-treated strategy, the calibration procedure failed to converge from visit 10 onwards when applied to the initial weights obtained using both parametric regression models and SuperLearner. Therefore, from visit 10 onwards, the joint calibrated weights for the always-treated strategy were set equal to the corresponding initial weights. This non-convergence issue is likely due to the small number of patients adhering to the always-treated strategy at later visits, especially beyond visit 10
(see \Cref{HERS data summary} in \Cref{HERSintro}). In this case, the \texttt{nleqslv} function in the \texttt{nleqslv} package could not find a solution to the convex minimisation problem in~\Cref{convex minimization} that satisfied the calibration restrictions. \tred{This provides evidence of practical violations of the positivity assumption at later follow-up visits. In such cases, the data offer insufficient support to enforce the desired covariate balance restrictions, and estimation necessarily relies more heavily on modelling assumptions.}

For visits 8 and 9, where the calibration procedure converged for both the always-treated and never-treated strategies, we provide a table of baseline characteristics of patients before weighting at visit 8 in Table~8 of the Supplementary Materials, along with Love plots assessing covariate balance at visits 8 and 9, in Figures~\ref{balance_8} and~\ref{balance_9} (generated using the \texttt{cobalt} package \cite{greifer_cobalt_2026}).

\tred{Figures~\ref{balance_8} and~\ref{balance_9} show covariate balance diagnostics by comparing standardised mean differences (SMDs) of covariates weighted by MLE weights estimated via parametric models, MLE weights estimated via SuperLearner, and their joint calibrated counterparts.}

\tred{In Figure~\ref{balance_8}, we assess \tp{the balance of baseline covariates} at visit 8 by comparing the weighted covariate distributions among patients who initiated the always-treated or never-treated strategies with the total trial population at baseline. Weighting patients who\tp{, at trial baseline,} initiated the always-treated or never-treated strategy, aims to recover the covariate mean of the entire baseline population, corresponding to the hypothetical scenarios in which everyone initiated the always-treated or never-treated strategies at trial baseline. The SMDs without weighting (`Unadjusted') in Figure~\ref{balance_8} show substantial imbalance between the baseline population and those initiating treatment ($A_0=1$) or not ($A_0=0$), with treated patients having lower CD4 counts and higher HIV viral load. Both MLE weights estimated via parametric models and \tp{those estimated using SuperLearner} reduce baseline covariate imbalance relative to the unweighted sample, but neither achieves exact balance. In particular, SuperLearner-based weights do not improve covariate balance beyond that achieved by MLE weights with parametric models in this application, illustrating that flexible estimation of nuisance models does not guarantee improved finite-sample balance. In contrast, joint calibrated weights, based on initial weights with either parametric models or SuperLearner, achieve near-exact balance across all covariates.}

\trd{This assessment extends to visit 9 in} Figure~\ref{balance_9}, \tp{where the SMDs between time-invariant and time-varying covariates at visits 8 and 9 after weighting are shown. The weighting at visit 9 \trd{aims to recover} the covariate distributions that would be observed when all patients adhered to the two treatment strategies up to visit 9. As in Figure~\ref{balance_8}, both uncalibrated weights estimated via parametric models and SuperLearner improve balance relative to the unweighted sample, but neither consistently achieves close balance across all covariates. After calibration, both approaches again achieve near-exact covariate balance.}

\tred{\tp{In summary, Figures~\ref{balance_8} and \ref{balance_9} show that inverse} probability weights based on either maximum likelihood or SuperLearner estimation reduce covariate imbalance relative to the unweighted population, but do not consistently achieve exact covariate balance. In contrast, joint calibrated weights achieve near-exact covariate balance regardless of whether they are based on initial weights estimated via parametric models or SuperLearner, indicating that calibration is the key mechanism driving the improvements in covariate balance observed across visits 8 and 9.}

\tred{Overall, the HERS application provides empirical support for the use of calibration, showing that both uncalibrated MLE weights estimated via parametric models and SuperLearner\tp{-based weights} may exhibit residual covariate imbalance in finite samples, which is effectively addressed by calibration. In combination with our simulation results, which demonstrate that improved covariate balance can translate into improved treatment effect estimation, these findings support joint calibration as a practical tool that may enhance finite-sample performance of LTMLE \tp{for parameter estimation of MSMs in TTE applications}.}

\begin{figure}[h!]
 \centering
 \includegraphics[width=0.9\linewidth]{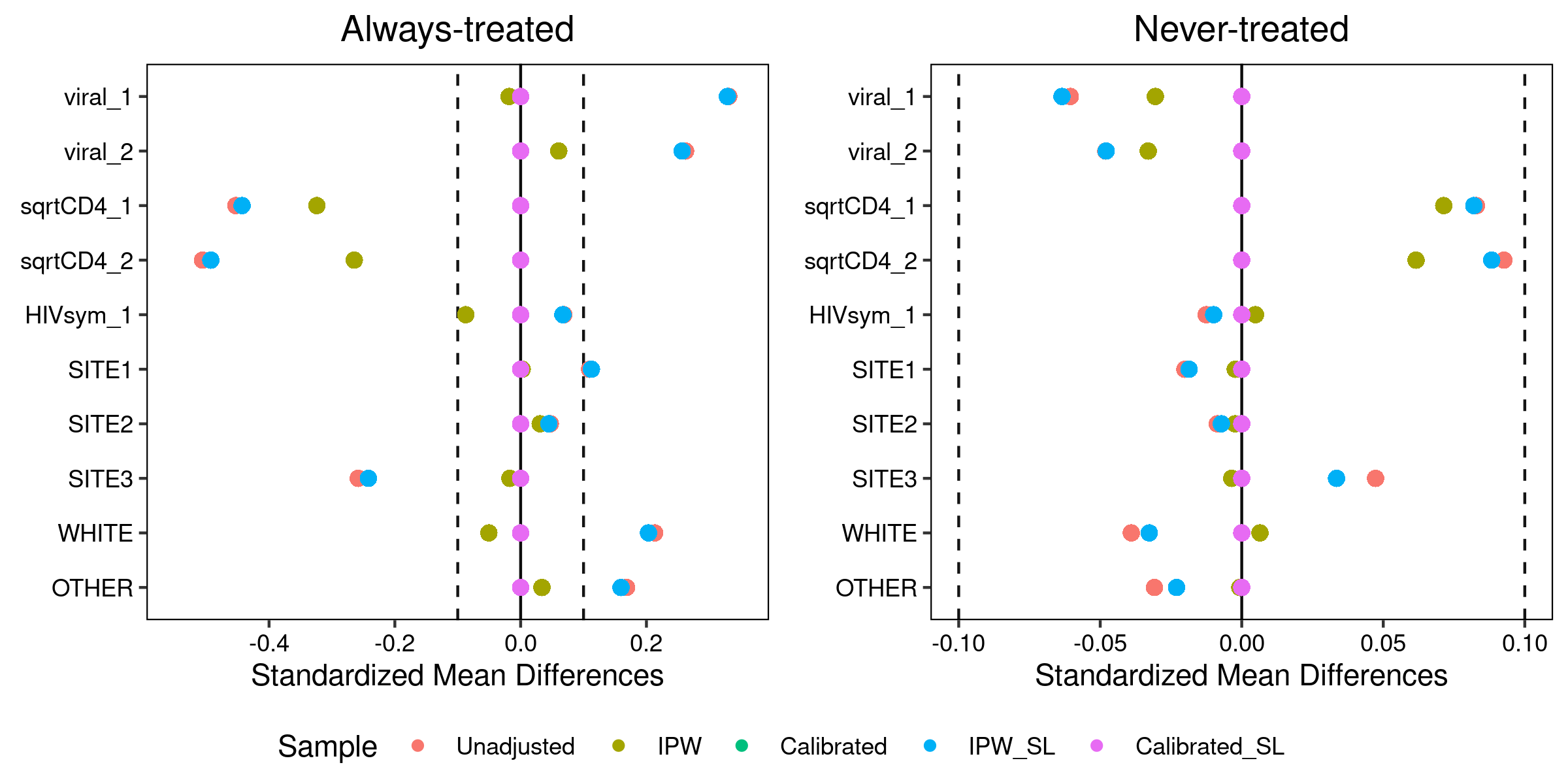}
 \caption[Love plots of covariate balance at visit 8 in the HERS data.]{Love plots of covariate balance at visit 8 in the HERS data using no weights (`Unadjusted'), MLE weights estimated by parametric models (`IPW'), joint calibrated weights with parametric models for initial weights (`Calibrated'), weights with SuperLearner (`IPW\_SL') and joint calibrated weights with SuperLearner (`Calibrated\_SL'). Time-varying covariates: viral\_1 \& viral\_2: HIV viral load at previous two visits; sqrtCD4\_1 \& sqrtCD4\_2: CD4 cell count (after square root transformation and standardisation) at previous two visits; HIVsym\_1: HIV symptoms at the previous visit. Time-invariant covariates: SITE1, SITE2, SITE3: binary variables indicating study sites with Site 0 as the reference group; WHITE, OTHER: binary variables indicating patient race, with Black as the reference group.} \label{balance_8}
\end{figure}

\begin{figure}[h!]
 \centering
 \includegraphics[width=0.9\linewidth]{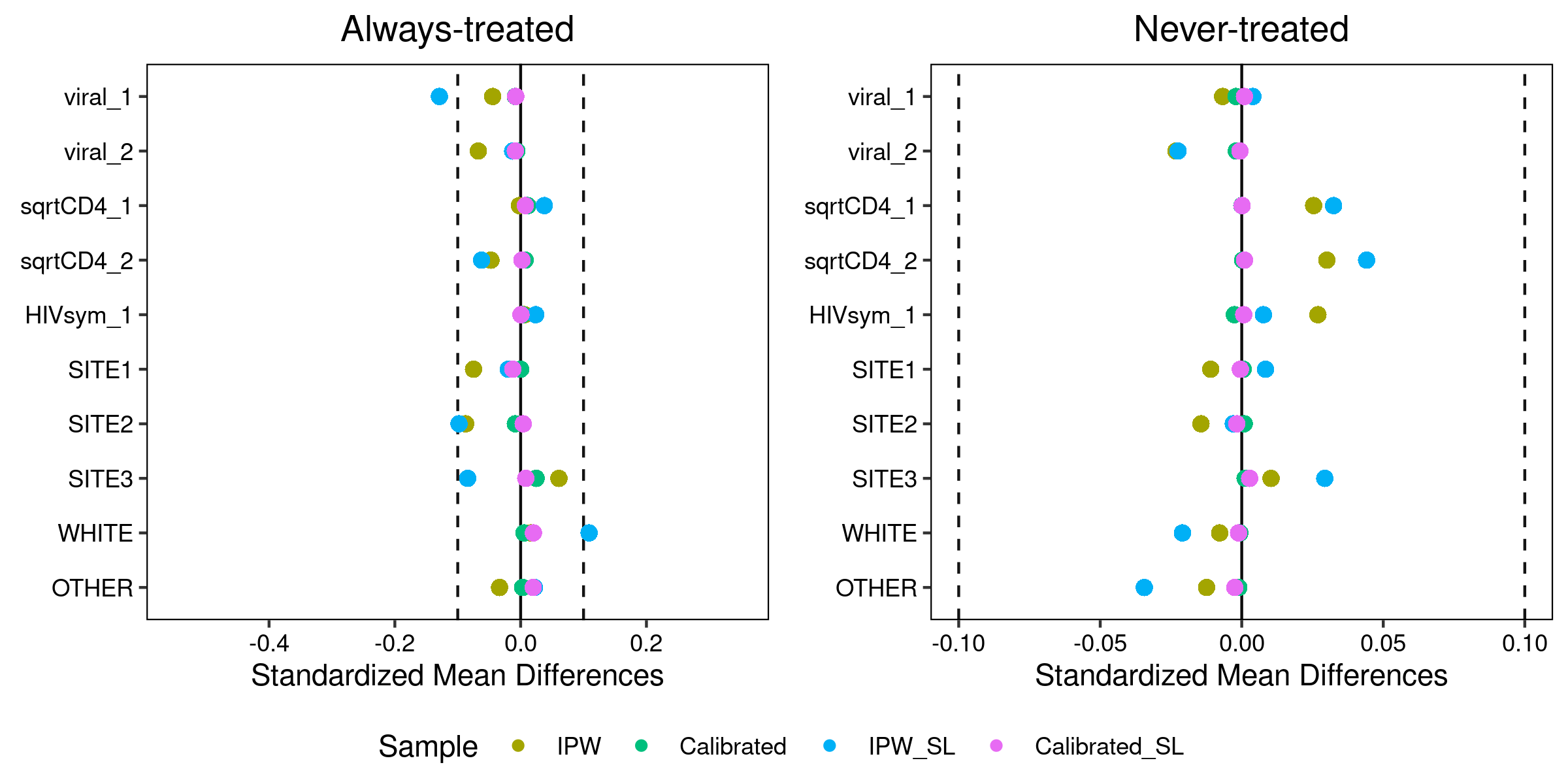}
 \caption[Love plots of covariate balance at visit 9 in the HERS data.]{Love plots of covariate balance at visit 9 in the HERS data using MLE weights estimated by parametric models (`IPW'), joint calibrated weights with parametric models for initial weights (`Calibrated'), weights with SuperLearner (`IPW\_SL') and joint calibrated weights with SuperLearner (`Calibrated\_SL'). Time-varying covariates: viral\_1 \& viral\_2: HIV viral load at previous two visits; sqrtCD4\_1 \& sqrtCD4\_2: CD4 cell count (after square root transformation and standardisation) at previous two visits; HIVsym\_1: HIV symptoms at the previous visit. Time-invariant covariates: SITE1, SITE2, SITE3: binary variables indicating study sites with Site 0 as the reference group; WHITE, OTHER: binary variables indicating patient race, with Black as the reference group.} \label{balance_9}
\end{figure}

\section{Conclusion and discussion}\label{discussion}

In this article, we adapted the joint calibrated weights approach of \citet{yiu_joint_2022} to the TTE framework and combined it with the LTMLE estimator of \citet{petersen_targeted_2014} to develop the joint calibrated LTMLE estimator for per-protocol effects of binary treatments on longitudinal continuous outcomes, characterised by a working MSM.

Through a series of simulation studies, we evaluated the finite-sample performance of joint calibrated LTMLE relative to standard LTMLE using MLE weights. The results showed that joint calibration strengthens the finite-sample performance of LTMLE across a range of realistic scenarios in TTE. Under correct model specification, joint calibrated LTMLE performed comparably to standard LTMLE and improved efficiency under strong confounding. Although small increases in variance were observed in some small-sample settings, these cases were \tp{infrequent}. Under functional-form model misspecification, the benefits of joint calibration were more pronounced: joint calibrated LTMLE consistently achieved lower RMSE, indicating improved robustness when treatment, censoring or outcome models were incorrectly specified. 

For inference, bootstrap CI methods that recalibrated weights within each bootstrap sample provided reliable coverage across sample sizes, whereas approaches that fixed calibrated weights \tb{had poorer coverage} under strong confounding. Overall, these findings support joint calibration as a practical and effective enhancement to LTMLE, particularly when confounding is strong or \tp{nuisance models are misspecified}. However, the HERS data analysis highlighted an important finite-sample limitation: when few patients adhere to certain treatment strategies, the calibration approach may fail to converge, \tb{meaning that one needs to revert to using the initial MLE weights at later time points}.

This work focused on continuous outcomes with two treatment strategies of interest. The LTMLE estimator \tb{can also be used with} survival outcomes \trd{and/or more than two} treatment strategies \cite{petersen_targeted_2014}, and extending our joint calibrated LTMLE to these settings would be of interest. \trd{Survival outcomes settings} may exacerbate \tred{the lack of support for} certain treatment strategies, \tp{because individuals are no longer observed after experiencing the event, leading to progressively less information on treatment assignment at later follow-up times and thus possibly increased risk of model misspecification \cite{schomaker_using_2019}. It would therefore be of interest to investigate the extent to which joint calibration could improve LTMLE's performance in such settings.} 

\tb{Joint calibrated weights have been shown to improve the performance of IPW-based estimators for general MSMs \cite{yiu_joint_2022}. Based on the improvements we observed in this work when combining joint calibrated weights with LTMLE for MSM \tp{parameter estimation} in TTE, we anticipate that replacing the MLE weights with our joint calibrated weights may yield similar or potentially greater performance gains for IPW-based estimators for TTE. This expectation is motivated by the fact that IPW-based estimators rely entirely on the weighting for causal effect estimation, making their performance particularly sensitive to improvements in weight stability and covariate balance.}

Although we did not address high-dimensional settings, where the number of confounders may be large, several strategies can help \tb{estimation in such settings}. Our implementation imposed calibration restrictions \tred{sequentially} at each trial visit. Alternatively, restrictions may be aggregated across visits to reduce the number of moment conditions to be solved, as discussed by \citet{yiu_joint_2022} and \citet{su_sensitivity_2022}. Another practical strategy is to restrict calibration to covariates most strongly predictive of the outcome, as recommended by \citet{yiu_joint_2022}, thereby reducing the dimensionality of the calibration problem. Finally, one may consider imposing `soft' calibration restrictions that allow a controlled degree of imbalance for those covariates \tp{that are less predictive of} the outcome \cite{wang_minimal_2020}.

A further avenue for improving joint calibrated LTMLE is the incorporation of data-adaptive methods. As illustrated in the HERS data analysis, SuperLearner can be used to estimate the initial weights and the initial outcome predictions $\widehat Q_k^t$. Data-adaptive methods may \tp{reduce the risk of model misspecification} when estimating nuisance parameters. When coupled with calibration to enhance \tred{covariate balance and} weight stability, this has the potential to further improve the finite-sample performance of joint calibrated LTMLE \cite{ertefaie_nonparametric_2023}. \tred{The HERS data application showed that SuperLearner-based weight estimation does not guarantee improved covariate balance in finite samples relative to parametric models. This highlights that increased flexibility in nuisance model estimation does not necessarily translate into better balancing properties without explicit calibration.} Investigating the combined impact of joint calibration and data-adaptive estimation of nuisance models in complex longitudinal settings therefore remains an important direction for future research \cite{schomaker_using_2019}.

\bibliography{references}

\end{document}


\maketitle

\section{Additional articles in covariate-balancing weights literature}

The following table provides a comprehensive overview of the additional articles identified in the covariate-balancing weights literature. 

{\footnotesize
\begin{longtable}{p{0.15\linewidth} p{0.35\linewidth} p{0.35\linewidth}}
    \caption{Summary of key references on covariate balancing weights methods.} \label{CBW papers table} \\
    \toprule
    Paper & Settings & Methods \\
    \midrule
    \endfirsthead

    \caption[]{Summary of key references (continued)} \\
    \toprule
    Paper & Settings & Methods \\
    \midrule
    \endhead

    \multicolumn{3}{r}{\textit{Continued on next page...}} \\
    \endfoot

    \bottomrule
    \endlastfoot

Graham et al. (2012) \cite{graham_inverse_2012} & Missing-at-random, propensity score, point intervention, eventual outcome (i.e., a single outcome measured at a study end) & Inverse probability tilting (IPT): a method-of-moments estimator of the propensity score model parameter. If the unconditional moments used to estimate the propensity score parameter are appropriately chosen, IPT  (i) is locally efficient and (ii) remains consistent even if the propensity score model is misspecified.\\
\midrule
Hainmueller (2012) \cite{hainmueller_entropy_2012} & Point intervention, binary treatment, eventual outcome & Entropy Balancing weights: convex optimisation-based reweighting estimator that directly enforces exact balance of specified covariate moments between treatment groups by minimising entropy divergence from base weights subject to linear balance constraints; yields unique weights without explicit modelling of the treatment assignment mechanism.\\
\midrule
Imai and Ratkovic (2014) \cite{imai_covariate_2014} & Point intervention, binary and multivalued treatment, eventual outcome & Covariate Balancing Propensity Score (CBPS) estimator: Generalised method of moments estimator of the treatment assignment mechanism model parameter based on moments conditions for covariate balance. \\
\midrule
Imai and Ratkovic (2015) \cite{imai_robust_2015} & MSMs, IPTW, longitudinal data, time-varying confounders, binary time-dependent treatments & CBPS method applied to longitudinal data: Generalised method of moments estimator of the MSM parameter using covariate balance moment conditions. R package \texttt{CBPS} \cite{fong_cbps_2025} developed to implement CBPS in multiple settings, including for point interventions (Imai and Ratkovic (2014) \cite{imai_covariate_2014}) and MSMs. \\
\midrule
Zubizaretta (2015) \cite{zubizarreta_stable_2015}& Singular time point, missing outcome data & Stable balancing weight: design-based approach, directly find optimal minimum-variance covariate balancing weights subject to a penalised empirical difference between the weighted covariate distribution of respondents and that of the full  sample. Can be done as a convex optimisation problem. R package: \texttt{sbw} \cite{zubizarreta_sbw_2025}\\
\midrule
Han (2016) \cite{han_intrinsic_2016} & Longitudinal data, eventual outcome, dependent censoring/drop-out & Multiply robust method to estimate the mean eventual outcome in the presence of drop-out where they impose calibration restrictions for the cumulative probability of drop-out (inverse of the inverse probability of censoring weights) in the form of moment conditions involving  functions of time-varying covariates.\\
\midrule 
Chan et al. (2016) \cite{chan_globally_2016} & Horvitz-Thompson estimator, eventual outcome, point intervention, binary treatment & Empirical Balancing Calibration Weighting (EBCW): a nonparametric approach to calibrate a set of misspecified uniform weights (as opposed to model-estimated weights) to obtain covariate balancing weights by solving a constrained minimization problem, based on minimising a distance measure between final and misspecified weights subject to moment constraints.\\
\midrule
Huffman and Van Gameren (2018) \cite{huffman_covariate_2018} & Time-varying continuous treatments, MSMs, time-varying confounding, longitudinal outcome & Extension of the CBPS method to MSMs and time-varying continuous treatment combined together to minimise the association between time-varying covariates and the treatment, by including the correlation-breaking point condition in the GMM: checks that the histories of both exposure and the time-varying covariates to be uncorrelated in the weighted data.\\
\midrule
Li et al. (2018) \cite{li_balancing_2018} & Eventual outcome, point intervention, binary treatment & Overlap weights: weights are constructed to emphasise individuals with the most covariate overlap between treatment groups. These weights are bounded, so they yield less extreme weights compared to inverse probability weights. Overlap weights based on a logistic propensity score model achieve exact finite-sample balance in the means of included covariates and reduce the influence of extreme weights arising from limited overlap. \\
\midrule
Yiu and Su (2018) \cite{yiu_covariate_2018} & Point intervention: binary, ordinal or continuous treatment; eventual outcome; extension to longitudinal data & Covariate association eliminating weights are obtained via a convex optimisation problem to obtain minimum-variance weights subject to the weighted score equation of a parametric model of the treatment assignment given covariates is solved under the restriction that coefficients corresponding to covariates are set to zero. Weights are therefore optimised so that after weighting the data by covariate association eliminating weights, there is no association between treatment and covariates. \\
\midrule
Wang and Zubizaretta (2020) \cite{wang_minimal_2020} & Singular time point, censoring due to non-response (missing at random) & Minimal dispersion approximating weights: class of weights of minimum dispersion that approximately balance the covariates. Minimal weights can be viewed as linked to shrinkage estimation of the propensity score (probability of response given covariates). The minimal weights require choosing the degree of approximate balance $\delta$ of covariates, and a tuning algorithm for $\delta$ is provided by the authors. Finding the minimal weights is an unconstrained optimisation problem. R package \texttt{optweight} \cite{greifer_optweight_2026} combines stable balancing weights of Zubizarreta (2015) \cite{zubizarreta_sbw_2025} with minimal dispersion approximating weights and extends the method to multi-category, continuous, and multivariate treatments and provides a simple user interface and compatibility with the \texttt{cobalt} package \cite{greifer_cobalt_2026} for balance assessment.\\
\midrule
Zhou and Wodtke (2020) \cite{zhou_residual_2020}& Longitudinal data, MSM, IPTW, time-varying continuous treatment, time-varying confounding affected by treatment history& Residual balancing weights: an extension of Hainmueller (2012) \cite{hainmueller_entropy_2012}. Residual Balancing (RBW) constructs weights by modelling the conditional means of post‑treatment confounders, then balancing the residuals of these models rather than modelling the treatment distribution. This avoids the instability of IPW, which requires modelling conditional treatment probabilities and performs poorly with continuous treatments.
Also applicable to mediation analyses involving post‑treatment confounding. R package: \texttt{rbw} \cite{zhou_rbw_2022}. \\
\midrule
Ning et al. (2020) \cite{ning_robust_2020} & Point intervention, binary treatment, eventual outcome & Extension of the CBPS approach for high dimensional settings where the number of confounders might surpass the sample size. This is achieved broadly via a penalised estimation approach. The average treatment effect is estimated via covariate balancing propensity score and Horvitz-Thompson estimator. This method is also included in the \texttt{CBPS} package.\\
\midrule
Martinet (2020) \cite{martinet_balancing_2020} & Point intervention, eventual outcomes, continuous treatment & Covariate Balance via Discrepancy Minimization: optimal weights for covariate balance are obtained via a discrepancy minimisation algorithm. The algorithm derives weights that minimize a discrepancy notion between source and target data that incorporates the hypothesis class and the loss chosen for the regression to estimate the dose-response function of the potential outcomes.\\
\midrule
Tan (2020) \cite{tan_regularized_2020} & Point intervention, eventual outcome & Provides theoretical bounds for MSE of  estimators with inverse probability weights obtained via calibrated propensity score models, even under model misspecification. They present regularised calibrated estimator of the propensity score model parameters via lasso penalty and an algorithm for computing this estimator that is applicable to high-dimensional data. \\
\midrule
Kallus and Santacatterina (2021) \cite{kallus_optimal_2021} & IPTW, MSMs, longitudinal data, binary time-varying treatment, eventual outcome & Kernel Optimal Weighting (KOW): provides weights for fitting an MSM that balance time-dependent confounders while controlling for precision. Uses a quadratic optimisation problem over weights to directly minimize covariate imbalance while penalising extreme weights.\\
\midrule
Avagyan and Vansteelandt (2021) \cite{avagyan_stable_2021}& IPTW, MSMs, longitudinal data, time-varying exposure, eventual outcome &  Stable inverse probability weighting (IPW): estimating‑equation calibration of propensity models at each time point to prevent highly volatile weights and stabilize IPW estimation for time‑varying exposures; with extensions to MSMs. \\
\midrule
Yiu and Su (2022) \cite{yiu_joint_2022} & MSM, IPTW, IPCW, longitudinal data, time-varying confounders, time-varying treatment (binary, continuous and ordinal) & Joint calibrated inverse probability of treatment and censoring weights: base on Yiu and Su (2018) \cite{yiu_covariate_2018}, jointly solve a set of linear calibration restrictions  for the treatment and censoring weights, where the calibration restrictions are obtained based on moments conditions for the observed confounding covariates. \\
\midrule
Fan et al. (2023) \cite{fan_optimal_2023} & Point intervention, eventual outcome, binary treatment & Extending on the CBPS approach,  robustness under alternative balancing conditions is investigated, and optimal balancing functions are proposed. A theoretical asymptotic distribution under local misspecification of the propensity score model is proposed. The authors developed an `optimal CBPS' methodology for the inverse probability of treatment weighting that is doubly robust as a consistent average treatment effect estimator is obtained when either the propensity score or outcome model is correctly specified. This method is also included in the \texttt{CBPS} package. \\
\midrule
Santacatterina (2023) \cite{santacatterina_robust_2023} & Time-to-event outcome, point intervention, binary and continuous treatments & Aim is to estimate the marginal hazard ratio of the Cox proportional hazard model for a time-to-event outcome. Robust orthogonality weights (ROW) are proposed, based on Yiu and Su (2018) \cite{yiu_covariate_2018} and Zubizaretta (2015) \cite{zubizarreta_stable_2015}. The method focuses on the correlation between treatment and covariates as a measure of covariate balance and solves optimal weights with minimum variance subject to minimal treatment-covariates correlation. Weighting by ROW orthogonalizes covariate and treatment variables, thus eliminating associations between them.\\
\midrule
Li, De los Angeles Resa, Zubizarreta (2025) \cite{li_adaptive_2025} & Longitudinal data, binary time-varying treatment, continuous eventual outcome, IPW & The method builds on \citet{imai_robust_2015} by deriving sufficient balance conditions for identification in longitudinal treatment settings and then enforcing them via orthogonalisation of time‑varying covariates. Instead of balancing the raw covariates, the procedure balances only the innovation components, i.e., the parts orthogonal to past covariate and treatment histories,  thus isolating new information at each time point. This reduces dimensionality, improves robustness, and achieves efficiency comparable to g‑computation. Weights are chosen to minimise variance subject to these orthogonalized balance constraints, improving stability under propensity model misspecification.\\
\midrule 
Viviano and Bradic (2026)\cite{viviano_dynamic_2026} & Longitudinal data, binary time-varying treatment, time-varying outcome, dynamic treatment regimes & Dynamic Covariate Balancing (DCB) constructs weights sequentially over time, using local projections of potential outcomes onto past histories. At each period, the method forms weights that balance current and lagged covariates as well as treatment histories. Weights are obtained by solving a series of quadratic optimisation problems that also penalise weight variance to ensure stability. The method scales to high‑dimensional settings and is implemented in the \texttt{DynBalancing} R package.\\
\end{longtable}
}

\newpage
\section{Target trial  protocol for the HERS data example}

\begin{table}[ht!]
\centering
\caption{Protocol of a target trial to evaluate the sustained
effect of HAART on CD4 cell count over time}
\small
\begin{tabular}{p{0.3\linewidth} p{0.6\linewidth}}
\toprule
\textit{Eligibility criteria} & HIV-infected women with no prior exposure to HAART.
\\
\\
\textit{Treatment strategies} &The intervention strategy consists of initiating and maintaining HAART throughout follow-up till the end of the trial.  
The comparator strategy consists of refraining from initiating HAART during follow-up till the end of the trial (participants may receive other antiviral therapies). \\
\\
\textit{Assignment procedures} & Patients are randomly assigned to one of the treatment strategies at baseline and are aware of the strategy to which they have been assigned. \\
\\
\textit{Follow-up period} & Follow-up begins at baseline (treatment randomisation) and continues for 4 scheduled visits occurring every 6 months, or until loss to follow-up, whichever occurs first.\\
\\
\textit{Outcome} & CD4 cell count measured at each follow-up visit. \\
\\
\textit{Causal contrast of interest} & Per-protocol effect of sustained HAART use on CD4 cell count over time stratified by baseline CD4 cell count categories: less than 200 units, between 200 and 500 units, and above 500 units.\\
\\
\textit{Statistical Methods} & The per-protocol effect will be estimated using \textit{joint calibrated LTMLE} to fit a working MSM summarising the mean counterfactual CD4 cell trajectories stratified by baseline CD4 cell count categories. Joint calibrated weights, constructed by calibrating initial inverse probability weights estimated via maximum likelihood,  address the bias induced by treatment protocol non-adherence and dependent censoring due to loss to follow-up.\\
 \bottomrule
\end{tabular}

\label{appendix B: HERS protocol}
\end{table}

\section{Additional details on stabilised inverse probability weights}
\subsection{Definition and estimation} \label{appendix B: stabilsied weights}
The stabilised inverse probability of treatment weights are defined as 
\begin{align}\label{SIPTW}
 SW^A_{a,t} = \frac{\prod_{j = 0}^t \Pr( A_{j}= a\mid\overline A_{ j-1}=\overline a, \boldsymbol V, C_{j-1} = 0)}{\prod_{j = 0}^t \Pr( A_{j}= a\mid\overline A_{ j-1}=\overline a, \boldsymbol V, \overline{\boldsymbol L}_{j}, C_{j-1} = 0)}, ~~~~~~~~~~a \in \{1,0\}.
\end{align}

The stabilised inverse probability of censoring weights are defined as 

\begin{align}\label{SIPCW}
 SW^C_{a,t} = \frac{\prod_{j = 0}^{t-1}\Pr(C_{j} =0 \mid C_{j-1} = 0,\overline A_{j}= \overline a, \boldsymbol V )}{\prod_{j = 0}^{t-1}\Pr(C_{j} =0 \mid C_{j-1} = 0,\overline A_{j}= \overline a, \boldsymbol V, \overline{\boldsymbol L}_{j})}, ~~~~~~~~~~a \in \{1,0\}.
\end{align}
with $SW^C_{a,0} = 1$.

Similar to the probabilities  $\Pr(A_{j}=a\mid\overline A_{j-1}=\overline a, \boldsymbol V, \overline{\boldsymbol L}_{j}, C_{j-1} = 0)$, $\Pr(C_{j} =0\mid\overline C_{j-1} = 0,\overline A_{j} = \overline a, \boldsymbol V, \overline{\boldsymbol L}_{j})$ in the denominators of \eqref{SIPTW} and \eqref{SIPCW}, $\Pr(A_{j}=a\mid\overline A_{j-1}=\overline a, \boldsymbol V, C_{j-1} = 0)$  and $\Pr(C_{j} =0\mid\overline C_{j-1} = 0,\overline A_{j} = \overline a, \boldsymbol V ), j = 0,\ldots,T$  in the numerators of \eqref{SIPTW} and \eqref{SIPCW} are often  estimated by  $\Pr(A_{j}=a\mid\overline A_{j-1}=\overline a, \boldsymbol V, C_{j-1} = 0; \hat {\boldsymbol \eta})$ and $\Pr(C_{j} =0\mid\overline C_{j-1} = 0,\overline A_{j} = \overline a, \boldsymbol V ; \hat {\boldsymbol \mu})$ \tb{with $\overline A_{j-1}$ set at $\overline a$}, where $\hat {\boldsymbol \eta}$ and $\hat {\boldsymbol \mu}$ are the maximum likelihood estimates of ${\boldsymbol \eta}$ and ${\boldsymbol \mu}$ in pooled logistic regression models $\Pr(A_{j}=a\mid\overline A_{j-1}, \boldsymbol V, C_{j-1} = 0;  {\boldsymbol \eta})$ and $\Pr(C_{j} =0\mid\overline C_{j-1} = 0,\overline A_{j}, \boldsymbol V ; {\boldsymbol \mu})$, respectively.

Combining both weights, the stabilised inverse probability of treatment and censoring weight for each patient at time $t$ estimated via maximum likelihood is
\begin{align} \label{stabilised MLE Weights}
SW^{AC}_{a,t}(\hat{\bm \theta}) = SW^A_{a,t}(\hat{\boldsymbol \alpha},\hat{\boldsymbol \eta})\times SW^C_{a,t}(\hat{\boldsymbol \gamma}, \hat{\boldsymbol \mu}), ~~~~~~~~~~ a \in \{1,0\}
\end{align}
where $\hat{\bm \theta}$ is the set of parameter estimates $\hat{\boldsymbol \alpha},\hat{\boldsymbol \eta},\hat{\boldsymbol \gamma}, \hat{\boldsymbol \mu}$.

\subsection{Joint calibrated LTMLE algorithm for stabilised weights} \label{appendix B: stabilised joint calibrated LTMLE}
\begin{enumerate}
 \item Estimate stabilised inverse probability of treatment and censoring weights for both treatment strategies via maximum likelihood. Define $h(a,t,\bm V) = \prod_{j = 0}^t \Pr( A_{j}= a\mid\overline A_{ j-1}=\overline a, \boldsymbol V, C_{j-1} = 0; \hat{\boldsymbol \eta})\times \prod_{j = 0}^{t-1}\Pr(C_{j} =0 \mid C_{j-1} = 0,\overline A_{j}= \overline a, \boldsymbol V; \hat{\boldsymbol \mu})$, which corresponds to the numerator of the stabilised inverse probability weights.
 \item Jointly calibrate the stabilised MLE weights following the calibration algorithm in Algorithm~1 of the main text. This gives joint calibrated weights $\bm{W}^{\star}_{a,t}(\hat{\bm\lambda}^{(a)}_t)$ for $a \in \{1,0\}$ and $t=1, \ldots, T$.
 \item Follow Steps~2--5 of the LTMLE algorithm in Section~3.3 of the main text, replacing the weights $h(a,t,\bm V)\times \boldsymbol{W}_{a,t}^{AC}(\hat{\bm \alpha}, \hat{\bm \gamma})$ in Step~3(c) with the joint calibrated weights.
 \end{enumerate}

\section{Additional details and results  from the simulation studies} \label{appendix B: additional simu results}
\subsection{Outcome model specification details}

In Simulation Studies 1-4 in Section~4 of the main text, LTMLE is implemented using working models for predicted outcomes $ \hat Q_k^t$ that include the correct set of time-varying covariates and treatment variables. We have $k = t,..., 0$, $t = 0,...,T$ with $T = 2$ in Simulation Studies 1,2 and 4, and $T = 9$ in Simulation Study 3. In the \texttt{glm()} function in R, the models can be specified as follows:

 For $k = t$,
 \begin{align*}
     \text{logit}&(Y_t) \sim A_t + A_{t-1} + X_{1,t} + X_{2,t} + X_{3,t} + X_{4,t} + X_{1,t-1} +  X_{2,t-1}+  X_{3,t-1}+  X_{4,t-1},\\
     &\qquad\qquad\qquad\qquad\qquad\qquad\qquad\qquad\qquad\qquad\qquad\qquad\qquad\qquad\qquad\qquad t >0,\\
     \text{logit}&(Y_0) \sim A_0 + X_{1,0} + X_{2,0} + X_{3,0} + X_{4,0} , ~~~ t = 0.
 \end{align*} 
 
For $k = t-1$,
 \begin{align*}
     \text{logit}(Q_{k+1}^{1,t*}) \sim \sum_{j = 0}^k A_j + X_{1,k} + X_{2,k} + X_{3,k} + X_{4,k}, \\
      \text{logit}(Q_{k+1}^{0,t*}) \sim \sum_{j = 0}^k A_j + X_{1,k} + X_{2,k} + X_{3,k} + X_{4,k}. 
 \end{align*} 

For $k = t-2$,
 \begin{align*}
     \text{logit}(Q_{k+1}^{1,t*}) \sim \sum_{j = 0}^k A_j, \\
      \text{logit}(Q_{k+1}^{0,t*}) \sim \sum_{j = 0}^k A_j.  
 \end{align*}

For $k < t-2$,
 \begin{align*}
     \text{logit}(Q_{k+1}^{1,t*}) \sim 1, \\
      \text{logit}(Q_{k+1}^{0,t*}) \sim 1. \\
      \text{(Intercept-only models)}
 \end{align*} 

\subsection{Additional simulation results}

\begin{table}[ht]
\centering
\caption[Summary statistics for the estimated MLE and calibrated weights from one simulated dataset with different sample sizes and correctly specified models for Simulation Study 1.]{\small Summary statistics for the estimated MLE and calibrated weights from one simulated dataset with different sample sizes and correctly specified models for Simulation Study 1. MLE: MLE weights; CMLE: Joint calibrated weights.} \label{appendix B: simu 1 weights}
\begin{tabular}{lcccccc}
\toprule
& Minimum & 1st quantile & Median &  Mean & 3rd quantile & Maximum \\
\midrule
\multicolumn{7}{l}{\textbf{Weak confounding}} \\
\multicolumn{7}{l}{$n = 300$} \\
 MLE & 1.07 & 1.73 & 2.26 & 2.66 & 3.06 & 12.75 \\ 
  CMLE & 0.88 & 1.68 & 2.27 & 2.70 & 3.20 & 9.88 \\ 
  \\
\multicolumn{7}{l}{$n = 500$} \\
  MLE & 1.10 & 1.70 & 2.19 & 2.59 & 3.00 & 17.61 \\ 
  CMLE & 1.05 & 1.68 & 2.21 & 2.62 & 3.04 & 19.74 \\ 
\\
\multicolumn{7}{l}{$n = 1000$} \\
 MLE & 1.08 & 1.73 & 2.22 & 2.54 & 2.94 & 16.52 \\ 
  CMLE & 1.04 & 1.71 & 2.18 & 2.53 & 2.94 & 16.89 \\ 
\\
\multicolumn{7}{l}{$n = 2500$} \\
 MLE & 1.08 & 1.72 & 2.21 & 2.58 & 3.00 & 16.02 \\ 
  CMLE & 1.05 & 1.72 & 2.21 & 2.58 & 3.01 & 15.83 \\ 
\midrule
\multicolumn{7}{l}{\textbf{Strong confounding}} \\
\multicolumn{7}{l}{$n = 300$} \\
MLE & 1.00 & 1.00 & 1.04 & 1.71 & 1.26 & 72.33 \\ 
  CMLE & 0.00 & 0.53 & 1.21 & 2.49 & 2.27 & 102.00 \\ 
  \\
\multicolumn{7}{l}{$n = 500$} \\
MLE & 1.00 & 1.00 & 1.02 & 1.69 & 1.21 & 194.01 \\ 
  CMLE & 0.00 & 0.40 & 1.13 & 2.54 & 2.32 & 146.39 \\ 
\\
\multicolumn{7}{l}{$n = 1000$} \\
MLE & 1.00 & 1.00 & 1.02 & 2.19 & 1.17 & 545.86 \\ 
  CMLE & 0.01 & 0.54 & 1.06 & 2.60 & 1.92 & 311.92 \\ 
\\
\multicolumn{7}{l}{$n = 2500$} \\
MLE & 1.00 & 1.00 & 1.03 & 1.95 & 1.26 & 508.16 \\ 
  CMLE & 0.02 & 0.76 & 1.23 & 2.58 & 1.92 & 475.39 \\ 
\bottomrule
\end{tabular}
\end{table}

\begin{table}[ht]
\centering
\caption[Empirical bias, empirical standard deviation (SD) and root mean squared error (RMSE) for LTMLEs of the mean counterfactual outcome $E(Y^{\bar 1}_2)$ under the always-treated strategy at the end of the trial in Simulation Study 1 with \textbf{\textit{two follow-up visits}} and \textbf{\textit{no dependent censoring}}.]{\small Empirical bias, empirical standard deviation (SD) and root mean squared error (RMSE) for LTMLEs of the mean counterfactual outcome $E(Y^{\bar 1}_2)$ under the always-treated strategy at the end of the trial in Simulation Study 1 with \textbf{\textit{two follow-up visits}} and \textbf{\textit{no dependent censoring}}. MLE: LTMLE with MLE weights; CMLE: joint calibrated LTMLE.} \label{appendix B: simu 1 mean}
\begin{tabular}{lcccccc}
\toprule
& \multicolumn{3}{c}{Correct covariates} & \multicolumn{3}{c}{Transformed covariates} \\
\cmidrule(lr){2-4} \cmidrule(lr){5-7}
& Bias & SD & RMSE &  Bias & SD & RMSE \\
& $E(Y^{\bar 1}_2)$ & $E(Y^{\bar 1}_2)$ & $E(Y^{\bar 1}_2)$ & $E(Y^{\bar 1}_2)$ & $E(Y^{\bar 1}_2)$& $E(Y^{\bar 1}_2)$  \\
\midrule
\multicolumn{7}{l}{\textbf{Weak confounding}} \\
\multicolumn{7}{l}{$n = 300$} \\
  MLE  & 0.14 & 2.19 & 2.20 & 2.92 & 2.32 & 3.73 \\ 
  CMLE  & 0.14 & 2.18 & 2.19 & 2.82 & 2.23 & 3.59 \\ 
  \\
\multicolumn{7}{l}{$n = 500$} \\
  MLE  & -0.02 & 1.66 & 1.66 & 2.93 & 1.82 & 3.45 \\ 
  CMLE  & -0.03 & 1.66 & 1.66 & 2.79 & 1.77 & 3.30 \\ 
\\
\multicolumn{7}{l}{$n = 1000$} \\
  MLE  & -0.02 & 1.18 & 1.18 & 3.10 & 1.77 & 3.57 \\ 
  CMLE  & -0.02 & 1.18 & 1.18 & 2.89 & 1.41 & 3.22 \\ 
\\
\multicolumn{7}{l}{$n = 2500$} \\
  MLE  & -0.02 & 0.75 & 0.75 & 3.35 & 2.25 & 4.03 \\ 
  CMLE  & -0.03 & 0.75 & 0.75 & 3.04 & 1.31 & 3.31 \\ 
\midrule
\multicolumn{7}{l}{\textbf{Strong confounding}} \\
\multicolumn{7}{l}{$n = 300$} \\
MLE  & 0.06 & 5.47 & 5.47 & 6.81 & 3.21 & 7.52 \\ 
  CMLE  & -0.10 & 5.40 & 5.40 & 6.44 & 2.89 & 7.06 \\ 
  \\
\multicolumn{7}{l}{$n = 500$} \\
MLE  & -0.38 & 5.24 & 5.25 & 6.99 & 2.45 & 7.41 \\ 
  CMLE  & -0.25 & 4.91 & 4.91 & 6.61 & 2.34 & 7.02 \\ 
\\
\multicolumn{7}{l}{$n = 1000$} \\
MLE  & 0.03 & 5.49 & 5.49 & 6.88 & 1.75 & 7.10 \\ 
  CMLE  & 0.00 & 4.42 & 4.42 & 6.63 & 1.62 & 6.83 \\ 
\\
\multicolumn{7}{l}{$n = 2500$} \\
MLE  & -0.20 & 4.60 & 4.61 & 6.85 & 1.38 & 6.99 \\ 
  CMLE  & -0.19 & 3.70 & 3.70 & 6.65 & 1.15 & 6.75 \\ 
\bottomrule
\end{tabular}
\end{table}

\begin{table}[ht]
\centering
\caption[Summary statistics for the estimated MLE and calibrated weights from one simulated dataset with different sample sizes and correctly specified models for Simulation Study 2.]{\small Summary statistics for the estimated MLE and calibrated weights from one simulated dataset with different sample sizes and correctly specified models for Simulation Study 2. MLE: MLE weights; CMLE: Joint calibrated weights.} \label{appendix B: simu 2 weights}
\begin{tabular}{lcccccc}
\toprule
& Minimum & 1st quantile & Median &  Mean & 3rd quantile & Maximum \\
\midrule
\multicolumn{7}{l}{\textbf{Weak confounding}} \\
\multicolumn{7}{l}{$n = 300$} \\
 MLE & 1.07 & 1.73 & 2.31 & 2.85 & 3.29 & 19.73 \\ 
  CMLE & 0.65 & 1.60 & 2.07 & 2.60 & 3.01 & 12.03 \\ 
  \\
\multicolumn{7}{l}{$n = 500$} \\
  MLE & 1.10 & 1.72 & 2.31 & 2.80 & 3.24 & 25.56 \\ 
  CMLE & 0.77 & 1.61 & 2.13 & 2.59 & 3.00 & 16.28 \\  
\\
\multicolumn{7}{l}{$n = 1000$} \\
 MLE & 1.08 & 1.75 & 2.30 & 2.71 & 3.19 & 18.33 \\ 
  CMLE & 1.04 & 1.68 & 2.11 & 2.49 & 2.87 & 21.15 \\ 
\\
\multicolumn{7}{l}{$n = 2500$} \\
 MLE & 1.08 & 1.75 & 2.32 & 2.76 & 3.24 & 20.38 \\ 
  CMLE & 1.05 & 1.68 & 2.15 & 2.52 & 2.95 & 14.71 \\ 
\midrule
\multicolumn{7}{l}{\textbf{Strong confounding}} \\
\multicolumn{7}{l}{$n = 300$} \\
MLE & 1.00 & 1.01 & 1.14 & 1.83 & 1.43 & 80.42 \\ 
  CMLE & 0.00 & 0.07 & 0.52 & 2.48 & 1.70 & 71.34 \\ 
  \\
\multicolumn{7}{l}{$n = 500$} \\
MLE & 1.00 & 1.01 & 1.10 & 1.79 & 1.35 & 194.01 \\ 
  CMLE & 0.00 & 0.05 & 0.38 & 2.30 & 1.24 & 146.39 \\ 
\\
\multicolumn{7}{l}{$n = 1000$} \\
MLE & 1.00 & 1.01 & 1.10 & 1.73 & 1.29 & 56.55 \\ 
  CMLE & 0.00 & 0.23 & 0.65 & 2.47 & 1.51 & 141.35 \\ 
\\
\multicolumn{7}{l}{$n = 2500$} \\
MLE & 1.00 & 1.01 & 1.12 & 2.04 & 1.40 & 545.16 \\ 
  CMLE & 0.01 & 0.53 & 0.93 & 2.43 & 1.65 & 372.19 \\ 
\bottomrule
\end{tabular}
\end{table}

\begin{table}[ht]
\centering
\caption[Empirical bias, empirical standard deviation (SD) and root mean squared error (RMSE) for LTMLEs of the mean counterfactual outcome $E(Y^{\bar 1}_2)$ under the always-treated strategy at the end of the trial in Simulation Study 2 with \textbf{\textit{two follow-up visits}} and \textbf{\textit{dependent censoring}}.]{\small  Empirical bias, empirical standard deviation (SD) and root mean squared error (RMSE) for LTMLEs of the mean counterfactual outcome $E(Y^{\bar 1}_2)$ under the always-treated strategy at the end of the trial in Simulation Study 2 with \textbf{\textit{two follow-up visits}} and \textbf{\textit{dependent censoring}}. MLE: LTMLE with MLE weights; CMLE: joint calibrated LTMLE.} \label{appendix B: simu 2 mean}
\begin{tabular}{lcccccc}
\toprule
& \multicolumn{3}{c}{Correct covariates} & \multicolumn{3}{c}{Transformed covariates} \\
\cmidrule(lr){2-4} \cmidrule(lr){5-7}
& Bias & SD & RMSE &  Bias & SD & RMSE \\
& $E(Y^{\bar 1}_2)$ & $E(Y^{\bar 1}_2)$ & $E(Y^{\bar 1}_2)$ & $E(Y^{\bar 1}_2)$ & $E(Y^{\bar 1}_2)$& $E(Y^{\bar 1}_2)$  \\
\midrule
\multicolumn{7}{l}{\textbf{Weak confounding}} \\
\multicolumn{7}{l}{$n = 300$} \\
MLE  & 0.13 & 2.35 & 2.35 & 2.88 & 2.48 & 3.80 \\ 
  CMLE  & 0.12 & 2.35 & 2.36 & 2.48 & 2.46 & 3.49 \\ 
  \\
\multicolumn{7}{l}{$n = 500$} \\
MLE  & 0.02 & 1.83 & 1.83 & 2.97 & 1.99 & 3.58 \\ 
  CMLE  & 0.00 & 1.83 & 1.83 & 2.50 & 1.89 & 3.13 \\ 
\\
\multicolumn{7}{l}{$n = 1000$} \\
MLE  & -0.00 & 1.28 & 1.28 & 3.08 & 1.83 & 3.58 \\ 
  CMLE  & -0.02 & 1.28 & 1.28 & 2.56 & 1.35 & 2.89 \\ 
\\
\multicolumn{7}{l}{$n = 2500$} \\
MLE  & -0.03 & 0.82 & 0.82 & 3.42 & 2.42 & 4.19 \\ 
  CMLE  & -0.04 & 0.82 & 0.82 & 2.61 & 0.95 & 2.77 \\  
  \midrule
  \multicolumn{7}{l}{\textbf{Strong confounding}} \\
\multicolumn{7}{l}{$n = 300$} \\
MLE  & -0.05 & 5.51 & 5.51 & 6.79 & 3.32 & 7.56 \\ 
  CMLE  & -0.07 & 6.72 & 6.72 & 5.84 & 3.13 & 6.62 \\ 
  \\
\multicolumn{7}{l}{$n = 500$} \\
MLE  & -0.41 & 5.51 & 5.53 & 7.05 & 2.68 & 7.54 \\ 
  CMLE  & -0.46 & 5.75 & 5.77 & 6.07 & 2.48 & 6.56 \\ 
\\
\multicolumn{7}{l}{$n = 1000$} \\
MLE  & 0.06 & 5.51 & 5.51 & 6.91 & 1.92 & 7.17 \\ 
  CMLE  & 0.03 & 4.65 & 4.65 & 6.17 & 1.73 & 6.41 \\  
\\
\multicolumn{7}{l}{$n = 2500$} \\
MLE  & -0.07 & 4.78 & 4.79 & 6.86 & 1.89 & 7.11 \\ 
  CMLE  & -0.15 & 3.97 & 3.97 & 6.27 & 1.18 & 6.38 \\ 
\bottomrule
\end{tabular}
\end{table}

\begin{table}[ht]
\centering
\caption[Empirical bias, empirical standard deviation (SD) and root mean squared error (RMSE) for the mean counterfactual outcome $E(Y^{\bar 1}_9)$ under the always-treated strategy at the end of the trial in Simulation Study 3 with \textit{\textbf{nine follow-up visits }} and \textbf{\textit{no dependent censoring}}.]{\small Empirical bias, empirical standard deviation (SD) and root mean squared error (RMSE) for the mean counterfactual outcome $E(Y^{\bar 1}_9)$ under the always-treated strategy at the end of the trial in Simulation Study 3 with \textit{\textbf{nine follow-up visits }} and \textbf{\textit{no dependent censoring}}. MLE: LTMLE with MLE weights; CMLE: joint calibrated LTMLE.} \label{appendix B: simu 3 mean}
\begin{tabular}{lcccccc}
\toprule
& \multicolumn{3}{c}{Correct covariates} & \multicolumn{3}{c}{Transformed covariates} \\
\cmidrule(lr){2-4} \cmidrule(lr){5-7}
& Bias & SD & RMSE &  Bias & SD & RMSE \\
& $E(Y^{\bar 1}_9)$ & $E(Y^{\bar 1}_9)$ & $E(Y^{\bar 1}_9)$ & $E(Y^{\bar 1}_9)$ & $E(Y^{\bar 1}_9)$& $E(Y^{\bar 1}_9)$  \\
\midrule
\multicolumn{7}{l}{\textbf{Weak confounding}} \\
\multicolumn{7}{l}{$n = 300$} \\
MLE  & -0.02 & 0.75 & 0.75 & -0.23 & 0.76 & 0.80 \\ 
  CMLE  & -0.02 & 0.75 & 0.75 & -0.22 & 0.75 & 0.78 \\ 
  \\
\multicolumn{7}{l}{$n = 500$} \\
MLE  & 0.01 & 0.58 & 0.58 & -0.22 & 0.65 & 0.69 \\ 
  CMLE  & 0.00 & 0.58 & 0.58 & -0.20 & 0.62 & 0.65 \\ 
\\
\multicolumn{7}{l}{$n = 1000$} \\
MLE  & 0.01 & 0.40 & 0.40 & -0.26 & 0.60 & 0.65 \\ 
  CMLE  & 0.01 & 0.40 & 0.40 & -0.21 & 0.50 & 0.54 \\ 
\\
\multicolumn{7}{l}{$n = 2500$} \\
MLE  & -0.01 & 0.25 & 0.25 & -0.35 & 0.77 & 0.85 \\ 
  CMLE  & -0.01 & 0.25 & 0.25 & -0.25 & 0.51 & 0.57 \\ 
\midrule
\multicolumn{7}{l}{\textbf{Strong confounding}} \\
\multicolumn{7}{l}{$n = 300$} \\
MLE  & 0.06 & 1.70 & 1.70 & -1.61 & 1.46 & 2.17 \\ 
  CMLE  & 0.07 & 1.54 & 1.54 & -1.15 & 1.25 & 1.70 \\ 
  \\
\multicolumn{7}{l}{$n = 500$} \\
MLE  & 0.04 & 1.63 & 1.63 & -1.56 & 1.40 & 2.09 \\ 
  CMLE  & -0.01 & 1.41 & 1.41 & -1.09 & 1.13 & 1.57 \\ 
\\
\multicolumn{7}{l}{$n = 1000$} \\
MLE  & 0.02 & 1.33 & 1.33 & -1.78 & 1.32 & 2.22 \\ 
  CMLE  & 0.01 & 1.19 & 1.19 & -1.23 & 0.99 & 1.58 \\ 
\\
\multicolumn{7}{l}{$n = 2500$} \\
MLE  & -0.04 & 1.06 & 1.06 & -1.94 & 1.29 & 2.33 \\ 
  CMLE  & -0.03 & 0.91 & 0.91 & -1.29 & 0.94 & 1.59 \\ 
\bottomrule
\end{tabular}
\end{table}

\section{Additional results for the HERS data analysis} \label{appendix B: HERS additional results}

\begin{table}[ht]
\centering
\caption[Baseline characteristics of HIV-positive patients in the emulated trial using the HERS data.]{Baseline characteristics of HIV-positive patients in the emulated trial using the HERS data. Patients are stratified according to whether they are about to initiate HAART at the visit ($A_0 = 1$) or not ($A_0 = 0$). For categorical variables, values are presented as counts and percentages; for continuous variables, values are reported as medians with interquartile ranges.} \label{appendix B: HERS stats}
{ 
\begin{tabular}{lllll}
  \toprule
Variable & Level  & $A_0 = 1$ & $A_0 = 0$ & Total \\ 
  \hline
 HIV viral load &  & 4365 (1172.5 - 17440) & 1980 (310 - 9290) & 2250 (380 - 10630) \\ 
 \\
 HIV symptom level &&& \\
  & 0 & 51 (67.1\%) & 277 (66.7\%) & 328 (66.8\%) \\  
  & 1 & 11 (14.5\%) & 82 (19.8\%) & 93 (18.9\%) \\ 
 & 2 & 7 (9.2\%) & 31 (7.5\%) & 38 (7.7\%) \\  
  & 3 & 6 (7.9\%) & 19 (4.6\%) & 25 (5.1\%) \\
  & 4 & 1 (1.3\%) & 6 (1.4\%)  & 7 (1.4\%) \\ 
& 5 & 0 (0\%) & 0 (0\%) & 0 (0\%) \\
  \\
  Site  & & & \\
  & 0 & 24 (31.6\%)& 107 (25.6\%) & 131 (26.7\%) \\
  & 1 & 21 (27.6\%) & 92 (22.2\%)& 113 (23\%) \\
   & 2 & 19 (25\%) & 94 (22.7\%) & 113 (23\%) \\ 
    & 3 & 12 (15.8\%) & 122 (29.4\%) & 134 (27.3\%) \\ 
    \\
CD4 cell count &  & 268.6 (133.6 -- 395.9) & 375.4 (234.1 -- 548.6) & 351 (213.2 -- 521.3) \\ 
  & $< 200$ & 28 (36.8\%) & 88 (21.2\%) &116 (23.6\%) \\ 
  & $[200,500]$ & 40 (52.6\%) & 198 (47.7\%) & 238 (48.5\%) \\ 
  & $ > 500$ & 8 (10.5\%) & 129 (31.1\%) &  137 (27.9\%) \\ 
  \\
  Ethinicity & & & \\
   & White & 22 (28.9\%) & 78 (18.8\%)  & 100 (20.4\%) \\  
   & Black & 36 (47.3\%) & 270 (57.4\%) & 306 (62.3\%) \\
   & Other & 18 (23.7\%) & 67 (16.1\%)  & 85 (17.3\%) \\  
 \midrule
 Total number of patients & & 76 (15.5\%) &415 (84.5\%) & 491\\
   \bottomrule
\end{tabular}}
\end{table}
\bibliography{references}